\documentclass{aa}

\usepackage{amsmath}
\usepackage{graphicx}
\usepackage{url}
\usepackage{ulem}
\usepackage{txfonts}
\usepackage{hyperref}

\def\msun{{\rm ~M}_{\odot}}
\def\rsun{{\rm ~R}_{\odot}}
\def\zsun{{\rm ~Z}_{\odot}}
\def\gpy{{\rm ~Gpc}^{-3} {\rm ~yr}^{-1}}
\def\kms{{\rm ~km} {\rm ~s}^{-1}}
\def\mpy{{\rm ~M}_{\odot} {\rm ~yr}^{-1}}
\def\sfr{{\rm ~M}_{\odot} {\rm ~Mpc}^{-3} {\rm ~yr}^{-1}}

\usepackage{color}

\begin{document}

   \title{The evolutionary roads leading to low effective spins, high black 
          hole masses, and O1/O2 rates of LIGO/Virgo binary black holes}

  \titlerunning{BH-BH effective spins, masses and rates}

   \author{K. Belczynski\inst{1}\thanks{chrisbelczynski@gmail.com}
          \and
          J. Klencki\inst{2}
          \and
          C.E. Fields\inst{3}
          \and
          A. Olejak\inst{4,5}
          \and
          E. Berti\inst{6}
          \and 
          G. Meynet\inst{7}
          \and
          C.L. Fryer\inst{8}
          \and
          D.E. Holz\inst{9}
          \and
          R.~O'Shaughnessy\inst{10}
          \and
          D.A. Brown\inst{11}
          \and
          T. Bulik\inst{4}
          \and
          S.C. Leung\inst{12,13}
          \and
          K. Nomoto\inst{12}
          \and
          P. Madau\inst{14}
          \and
          R. Hirschi\inst{15,12}
          \and
          E. Kaiser\inst{15}
          \and 
          S. Jones\inst{8}
          \and
          S.~Mondal\inst{1}
          \and
          M. Chruslinska\inst{2}
          \and
          P. Drozda\inst{4}
          \and
          D. Gerosa\inst{16}
          \and
          Z. Doctor\inst{9,22}
          \and
          M. Giersz\inst{1}
          \and
          S. Ekstrom\inst{7}
          \and
          C. Georgy\inst{7}
          \and
          A. Askar\inst{17}
          \and
          V. Baibhav\inst{6}
          \and
          D. Wysocki\inst{10}
          \and
          T. Natan\inst{18} 
          \and
          W.M. Farr\inst{19}
          \and
          G. Wiktorowicz\inst{20}
          \and
          M. Coleman Miller\inst{21}
          \and
          B. Farr\inst{22}
          \and 
          J.-P. Lasota\inst{1,23}
   }

   \institute{Nicolaus Copernicus Astronomical Center, Polish Academy of Sciences,
           ul. Bartycka 18, 00-716 Warsaw, Poland
         \and
           Department of Astrophysics/IMAPP, Radboud University, P.O. Box 9010, 6500 GL
           Nijmegen, The Netherlands
         \and
           Department of Physics and Astronomy, Michigan State University,
           East Lansing, MI 48824, USA
         \and
           Astronomical Observatory, Warsaw University, Al. Ujazdowskie 4,
           00-478 Warsaw, Poland
         \and
           Center for Theoretical Physics, Polish Academy of Sciences, Al.
           Lotnikow 32/46, 02-668 Warsaw, Poland
         \and
           Department of Physics and Astronomy, Johns Hopkins University,
           Baltimore, MD 21218 USA
         \and
           Department of Astronomy, University of Geneva, Chemin des
           Maillettes 51, 1290 Versoix, Switzerland
         \and
           CCS-2, MSD409, Los Alamos National Laboratory, Los Alamos, NM 87545, USA
         \and 
           University of Chicago, Chicago, IL 60637, USA  
         \and
           Center for Computational Relativity and Gravitation, Rochester Institute of
           Technology, Rochester, New York 14623, USA
         \and
           Department of Physics, Syracuse University, Syracuse NY 13224
         \and
           Kavli Institute for the Physics and Mathematics of the Universe (WPI),
           The University of Tokyo, Kashiwa, Chiba 277-8583, Japan
         \and
           TAPIR, Walter Burke Institute for Theoretical Physics, Mailcode 350-17, 
           Caltech, Pasadena, CA 91125, USA
         \and
           Department of Astronomy \& Astrophysics, University of California, 
           1156 High Street, Santa Cruz, CA 95064, USA
         \and
           Lennard-Jones Laboratories, Keele University, Keele ST5 5BG, UK 
         \and
           School of Physics \& Astronomy \& Institute for Gravitational
           Wave Astronomy, University of Birmingham, Birmingham, UK
         \and 
           Lund Observatory, Department of Astronomy, and Theoretical Physics, 
           Lund University, Box 43, SE-221 00 Lund, Sweden         
         \and
           Physics Department, Kenyon College, 201 North College RD, Gambier, 
           OH 43022, USA
         \and
           Department of Physics and Astronomy, Stony Brook University,
           Stony Brook, NY 11794, USA
         \and
           National Astronomical Observatories \& University of the Chinese 
           Academy of Sciences, Beijing, China 
         \and
           Department of Astronomy and Joint Space-Science Institute University 
           of Maryland, College Park, MD 20742-2421, USA         
         \and 
           Department of Physics, University of Oregon, Eugene, OR 97403, USA
         \and
           Institut d'Astrophysique de Paris, CNRS et Sorbonne Universit\'e,
           UMR 7095, 98bis Bd Arago, 75014 Paris, France
   }

  \abstract
{
All ten LIGO/Virgo binary black hole (BH-BH) coalescences reported from the O1/O2 runs 
have near zero effective spins. There are only three potential explanations of this 
fact. If the BH spin magnitudes are large then {\em (i)} either both BH spin vectors 
must be nearly in the orbital plane or {\em (ii)} the spin angular momenta of the BHs 
must be oppositely directed and similar in magnitude.  Or, {\em (iii)} the BH spin 
magnitudes are small.
We test the third hypothesis within the framework of the classical isolated binary 
evolution scenario of the BH-BH merger formation. We test three models of angular 
momentum transport in massive stars: a mildly efficient transport by meridional currents 
(as employed in the Geneva code), an efficient transport by the Tayler-Spruit magnetic dynamo 
(as implemented in the MESA code), and a very-efficient transport (as proposed by Fuller 
et al.) to calculate natal BH spins. We allow for binary evolution to increase the BH 
spins through accretion and account for the potential spin-up of stars through tidal 
interactions. Additionally, we update the calculations of the stellar-origin BH masses, 
include revisions to the history of star formation and to the chemical evolution across 
cosmic time. 
We find that we can match simultaneously the observed BH-BH merger rate density,
BH masses, and effective spins. Models with efficient angular momentum transport 
are favored. The updated stellar-mass weighted gas-phase metallicity evolution
now used in our models appears to be a key in better reproducing the LIGO/Virgo 
merger rate estimate. Mass losses during the pair-instability pulsation supernova 
phase are likely overestimated if the merger GW170729 hosts a BH more massive than 
$50\msun$. We also estimate rate of BH-NS mergers from recent LIGO/Virgo observations.
If in fact angular momentum transport in massive stars is efficient, then any 
(electromagnetic or gravitational wave) observation of a rapidly spinning BH would 
indicate either a very effective tidal spin up--of the progenitor star (homogeneous 
evolution, high-mass X-ray binary formation through case A mass transfer, or a spin-up 
of a Wolf-Rayet star in a close binary by a close companion), or significant mass 
accretion by the hole, or a BH formation through the merger of two or more BHs (in a dense 
stellar cluster). 
Our updated models of BH-BH, BH-NS and NS-NS mergers are now publicly available at 
\url{www.syntheticuniverse.org} under the tab "StarTrack models vs. Gravitational Wave 
Observations".
}


\maketitle

\section{Introduction}

The LIGO/Virgo Collaboration has reported the detection of ten binary black hole 
(BH-BH) mergers during the O1/O2 observations: GW150914, GW151012, GW151226, GW170104, 
GW170608, GW170729, GW170809, GW170814, GW170818 and GW170823~\citep{LIGO2019a,LIGO2019b}. 
We list the basic properties of these mergers in Table~\ref{tab.ligodata}. 

In our analysis we use only the LIGO/Virgo--reported events but three additional
BH-BH mergers, GW170121, GW170304, and GW170727, were detected at high 
confidence ($P_\mathrm{astro} > 0.98$\footnote{LIGO/Virgo uses $>0.95$ 
probability to report events of astrophysical origin}) by \cite{Venumadhav2019}. 
They have low effective spins consistent with the triggers reported by the 
LIGO/Virgo collaboration. The three other events reported by \cite{Venumadhav2019} 
are most likely not of astrophysical origin. In particular GW170403, with a
highly negative effective spin, was reported as the least secure event. Two more 
events were reported with high effective spins:  one with 
$\chi_\mathrm{eff}=0.7^{+ 0.2}_{-0.3}$ \citep{Venumadhav2019} and the second with 
$\chi_\mathrm{eff}=0.81^{+0.15}_{-0.21}$ \citep{Zackay2019}, however, these 
triggers have only a $56\%$ and $71\%$ chance of being of astrophysical origin, 
respectively.

\begin{table*}
\caption{LIGO/Virgo O1/O2 BH-BH mergers\tablefootmark{a}.}

\centering
\begin{tabular}{c| c c c c c}
\hline\hline
Event & $\chi_{\rm eff}$ & $M_{\rm tot}$ & $M_1$ & $M_2$ & $z$ \\
\hline
\hline
&&&&&\\
{\bf GW150914} &        -0.01 &        62.8 &        35.6 &        30.6 &        0.09 \\
               & [-0.14,0.11] & [59.9,65.9] & [32.6,40.4] & [26.2,33.6] & [0.06,0.12] \\
&&&&&\\
{\bf GW151012} &         0.04 &        36.3 &        23.3 &        13.6 &        0.21 \\
               & [-0.15,0.32] & [33.1,45.2] & [17.8,37.3] &  [8.8,17.7] & [0.12,0.30] \\
&&&&&\\
{\bf GW151226} &         0.18 &        20.6 &        13.7 &         7.7 &        0.09 \\
               &  [0.06,0.38] & [19.3,25.7] & [10.5,22.5] &   [5.1,9.9] & [0.05,0.13] \\
&&&&&\\
{\bf GW170104} &        -0.04 &        50.4 &        31.0 &        20.1 &        0.19 \\
               & [-0.24,0.13] & [46.7,55.0] & [25.4,38.2] & [15.6,25.0] & [0.11,0.26] \\
&&&&&\\
{\bf GW170608} &         0.03 &        17.7 &        10.9 &         7.6 &        0.07 \\
               & [-0.04,0.22] & [17.1,20.7] &  [9.2,16.2] &   [5.5,8.9] & [0.05,0.09] \\
&&&&&\\
{\bf GW170729} &         0.36 &        86.3 &        50.6 &        34.3 &        0.48 \\
               &  [0.11,0.57] &  [75.3,100] & [40.4,67.2] & [24.2,43.4] & [0.28,0.67] \\
&&&&&\\
{\bf GW170809} &         0.07 &        57.9 &        35.2 &        23.8 &        0.20 \\
               & [-0.09,0.23] & [54.3,63.2] & [29.2,43.5] & [18.7,29.0] & [0.13,0.25] \\
&&&&&\\
{\bf GW170814} &         0.07 &        54.1 &        30.7 &        25.3 &        0.12 \\
               & [-0.04,0.19] & [51.7,57.3] & [27.7,36.4] & [21.2,28.2] & [0.08,0.15] \\
&&&&&\\
{\bf GW170818} &        -0.09 &        61.3 &        35.5 &        26.8 &        0.20 \\
               & [-0.29,0.09] & [57.7,65.8] & [30.8,43.0] & [21.6,31.1] &v[0.13,0.27] \\
&&&&&\\
{\bf GW170823} &         0.08 &        69.1 &        39.6 &        29.4 &        0.34 \\
               & [-0.14,0.28] & [62.3,78.4] & [33.0,49.6] & [22.3,28.6] & [0.20,0.47] \\
\hline
\hline

\end{tabular}
\tablefoot{\\
\tablefootmark{a}{For all parameters average and $90\%$ confidence limits are listed, 
$\chi_{\rm eff}$: effective spin parameter, 
$M_{\rm tot}$: total intrinsic BH-BH binary mass, 
$M_1$: primary BH intrinsic mass, 
$M_2$: secondary BH intrinsic mass,
$z$: event redshift.}
}
\label{tab.ligodata}
\end{table*}

The majority of the reported mergers contain ``heavy'' BHs with component masses $>20\msun$ 
and are consistent with being formed by isolated binary evolution of stars with 
metallicities $\lesssim 10\%\zsun$ and initial masses in the range $40\mbox{--}100\,\msun$, 
while lower mass BHs may have formed from lower mass stars or at higher metallicity
~\citep{Belczynski2010b,Belczynski2016b}. A typical evolution involves the 
interaction of stars in a binary through mass transfer and a common envelope 
phase~\citep{Tutukov1993,Lipunov1997,Belczynski2002,Voss2003,Dominik2012,
Belczynski2016a,Eldridge2016,Woosley2016,Stevenson2017,Kruckow2018,Hainich2018,
Marchant2018,Spera2019} and its outcome was predicted to produce the first LIGO/Virgo 
sources~\citep{Belczynski2010a}.

The dynamical formation scenario of BH-BH mergers is an evolutionary channel alternative 
to isolated binary evolution that could operate in globular clusters, nuclear 
clusters, or open/young clusters~\citep{Miller2002a,Miller2002b,PortegiesZwart2004,
Gultekin2004,Gultekin2006,OLeary2007,Sadowski2008,Downing2010,Benacquista2013,Bae2014,
Chatterjee2016,Mapelli2016,Hurley2016,Rodriguez2016a,VanLandingham2016,Askar2017,
ArcaSedda2017,Samsing2018,Morawski2018,Banerjee2018,DiCarlo2019,Zevin2019,Rodriguez2018a,
Perna2019}. The dynamical channel can also explain the range of BH masses observed by 
LIGO/Virgo. 

However, these two basic scenarios, dynamical and isolated, predict different 
spin-orbit misalignment distributions, with dynamical formation generating nearly 
isotropic distributions~\citep{2010CQGra..27k4007M,Rodriguez2016c}, while binary 
evolution favors distributions that show mostly small to moderate misalignments, 
with only a small fraction of mergers reaching high misalignments~\citep{Wysocki2018,
Gerosa2018}. 
A hybrid BH-BH merger formation channel that involves isolated triple (dynamically 
interacting) star systems~\citep{Antonini2017} favors BH spin vectors that are 
located in the BH-BH orbital plane, with $\sim90^\circ$ misalignment with respect to 
the orbital angular momentum if the tertiary dominates the angular momentum of
the system~\citep{Antonini2018}.

The spin orientations of the BHs can therefore provide important information 
about their formation~\citep{Farr2017,Farr2018,Vitale2017b}. 
However, the effect of spin is sub-dominant in gravitational waveforms, so spins are
more difficult to measure than masses. The waveform is most sensitive to the binary's
effective spin
\begin{equation}
\chi_{\rm eff}\equiv {M_{\rm BH1} a_{\rm spin1} \cos\Theta_1 + M_{\rm BH2}
a_{\rm spin2} \cos\Theta_2 \over M_{\rm BH1}+M_{\rm BH2}},
\label{eq.xeff}
\end{equation}
with $a_{\rm spin1,2}$ being the BH spin magnitudes and $\Theta_{1,2}$ being the 
angles between the BH spins and the orbital angular momentum.
The dimensionless BH spin magnitude is defined as
\begin{equation}
a_{\rm spin} \equiv {c J_{\rm BH} \over G M_{\rm BH}^2},
\label{eq.bhmag}
\end{equation}
where $c$ is the speed of light, $G$ is the gravitational constant, and 
$M_{\rm BH}$ and $J_{\rm BH}$ are respectively the mass and angular momentum 
of the BH. 

All ten of the BH-BH binaries observed by LIGO/Virgo are consistent with effective 
spins near zero. Indeed, within the $90\%$ confidence levels reported by LIGO/Virgo 
all mergers are consistent with $-0.1 \lesssim \chi_{\rm eff} \lesssim 0.1$ 
(see Tab.~\ref{tab.ligodata}). If the BH spin magnitudes are large, then 
either {\em (i)} both black hole spin vectors are close to the orbital plane, so 
that $\cos\Theta_{1,2}\approx 0$, or {\em (ii)} the black hole angular momenta 
are nearly equal in magnitude but close to oppositely directed. Alternatively, 
one can obtain small values of the effective spin parameter if {\em (iii)} BH 
spin magnitudes are small. 

In this study we investigate the third possibility, i.e., that the spin magnitudes of BHs in 
BH-BH mergers detected by LIGO/Virgo are small. Our study is limited to the
classical isolated binary evolution BH-BH formation scenario. We test several 
models of natal BH spins to predict the effective spin parameters of BH-BH mergers 
in the local Universe and to compare them with LIGO/Virgo observations. Compared to
\citet{Belczynski2016c,Belczynski2016b,Belczynski2016a}, we incorporated updated 
mass loss by pair-instability pulsation supernovae, revised our model of accretion 
onto BHs in close binaries, allowed for effective tidal spin-up of Wolf-Rayet stars, 
and adopted a new model for the evolution of metallicity and the star formation 
rate across cosmic time.

To guide the reader through our article we will briefly summarize the most important 
ingredients and results of this study. Based on single stellar evolution calculations 
we introduce three BH natal spin models (see Sec.~\ref{sec.spins}, and we argue that 
these models can be reasonably used in binary evolution as implemented in our calculations 
(see Sec.~\ref{sec.sinbin}). We show that initial star rotation does not play a significant 
role, while angular momentum transport plays crucial role in setting the BH natal spin
(see Sec.~\ref{sec.ang_mom_transport}). In binary evolution, accretion/mass transfer 
(see Sec.~\ref{sec.bhacc}) does not play significant role, while tidal interactions 
(see Sec.~\ref{sec.tides}) may increase BH spins for about $20-30\%$ of the BH-BH mergers 
(see Sec.~\ref{sec.xeff}). Merger rates of NS-NS, BH-NS and BH-BH binaries are sensitive 
to assumptions on the common envelope and natal kicks, but also depend strongly on the 
assumed cosmic chemical evolution model while the change of merger rates with redshift 
depends mostly on the cosmic star-formation history (see Sec.~\ref{sec.rates}).

\section{Model}
\label{sec.model}

\subsection{Natal Black Hole Spin}
\label{sec.spins}

\begin{figure}
\hspace*{-0.3cm}
\includegraphics[width=9.2cm]{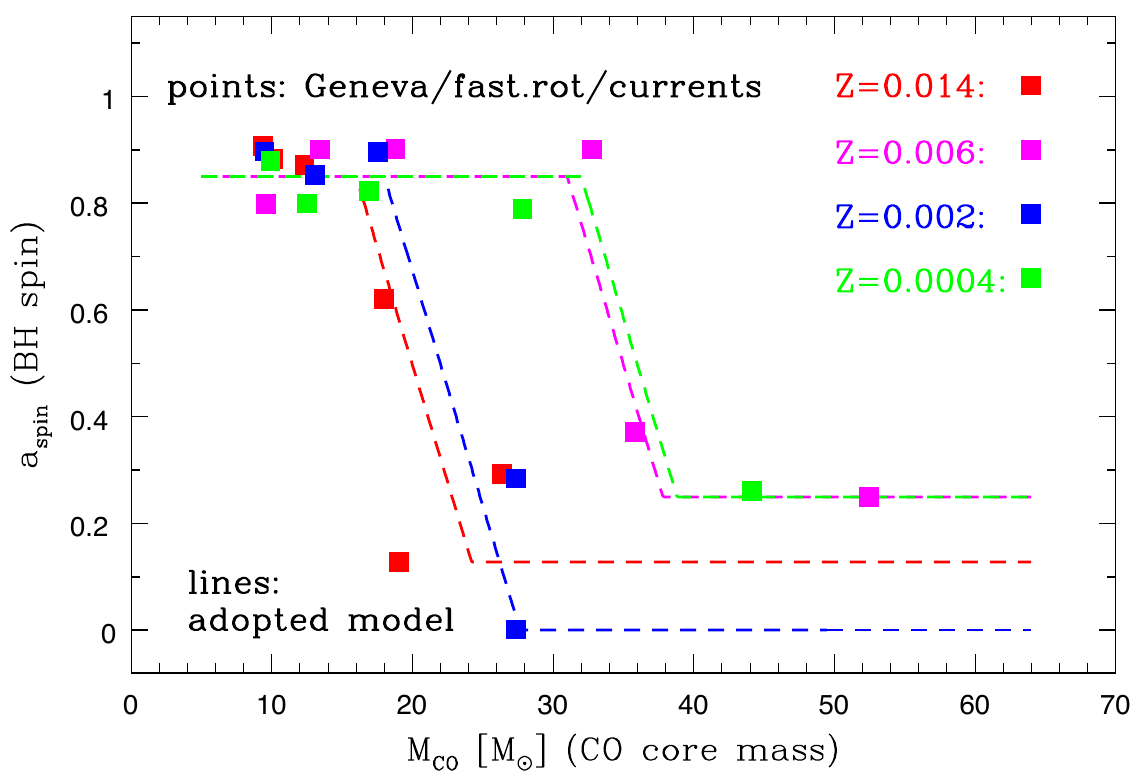}
\caption{
Magnitude of natal BH spin as a function of the CO core mass of the collapsing 
star for the {\tt Geneva} stellar models with $40\%$ critical initial velocity 
and mild angular momentum transport by meridional currents. 
Lines mark our adopted model for natal BH spins and its dependence on metallicity 
(see eq.~\ref{eq.bhspin1}). The star's CO core mass may be used as a proxy for the 
BH mass (see Appendix~\ref{sec.COBH}).
}
\label{fig.bhspin1}
\end{figure}

\begin{figure}
\hspace*{-0.3cm}
\includegraphics[width=9.2cm]{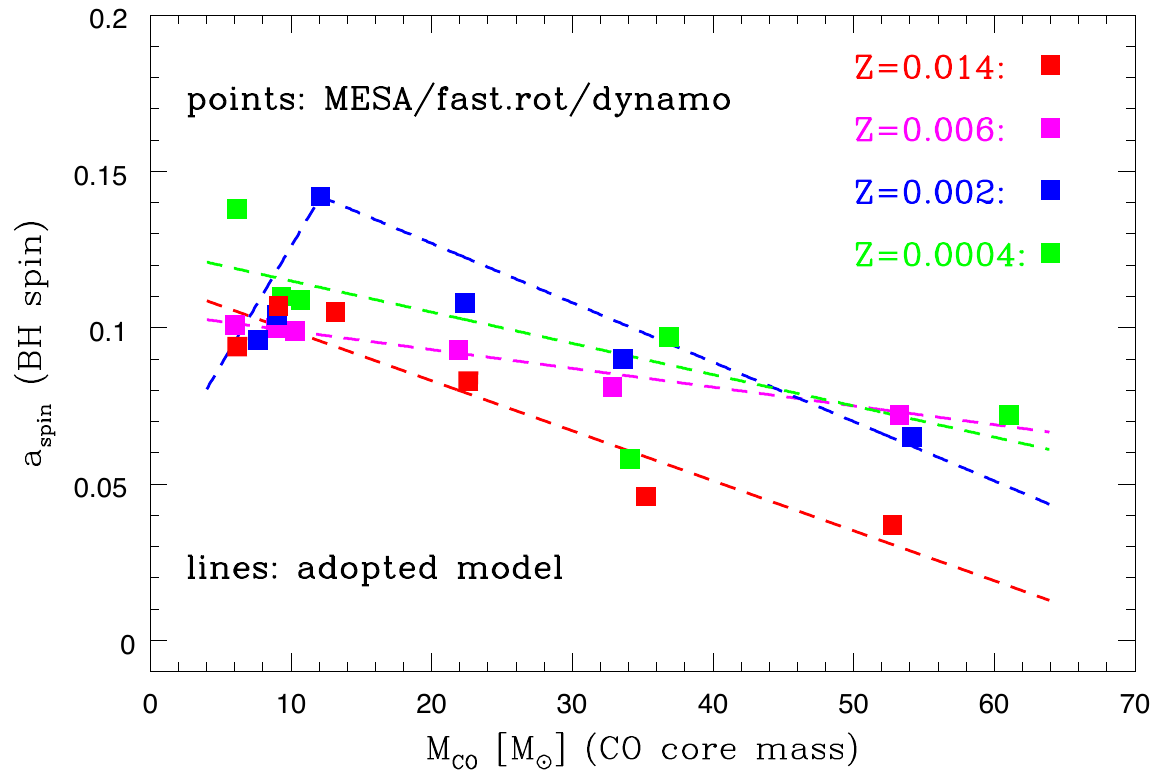}
\caption{
Magnitude of natal BH spin as a function of the CO core mass of the collapsing 
star for the {\tt MESA} stellar models with $40\%$ critical initial velocity 
and the Tayler-Spruit magnetic dynamo (efficient) angular momentum transport. 
Lines mark our adopted model for natal BH spins and its dependence on metallicity 
(see eq.~\ref{eq.bhspin2}). 
}
\label{fig.bhspin2}
\end{figure}

Various rotating star models differ in terms of the physics of rotation, and 
in particular in the efficiency of angular momentum transport. Here we test 
three different models of angular momentum transport for massive stars. 
Moderate angular momentum transport is adopted along with Geneva stellar models 
\citep{Eggenberger2008,Ekstrom2012} that are based on \cite{Zahn1992} theory, 
in which angular momentum is mainly transported by meridional currents: the
shellular model.
Efficient angular momentum transport is adopted with standard MESA stellar
models (calculated in this study), which use an efficient transport mechanism
mediated by the action of the so-called ``Tayler-Spruit magnetic dynamo'' in 
the radiative zone~\citep{Spruit1999,Spruit2002}. Super-efficient angular
momentum transport is based on the Tayler-Spruit dynamo modified to include
stronger magnetic field generation that leads to more efficient angular
momentum transport~\citep{Fuller2019a,Ma2019,Fuller2019b}. 

To test the validity of each of these models through gravitational wave 
astrophysics, we compare their predictions with the LIGO/Virgo 
effective--spin estimates. For the Geneva (moderate) and MESA (efficient) angular 
momentum transport we use models for a wide range of metallicity: 
$Z=0.014, 0.006, 0.002, 0.0004$ (see Appendix~\ref{sec.Geneva} and ~\ref{sec.MESA}).   
We assume that stars on the zero age main sequence (ZAMS) have an equatorial 
velocity equal to $40\%$ of the critical velocity, defined as the velocity at 
which the centrifugal acceleration completely balances gravity.\footnote{\texttt{GENEC} 
and \texttt{MESA} use two slightly different definitions of the critical velocity. 
\cite{Maeder2000} discusses the two definitions. \texttt{MESA} models use their 
eq.\,(3.12), whereas \texttt{GENEC} models use the minimum between the values 
obtained from their eq.\,(3.14) and (3.19), respectively.}
To test the importance of the initial rotation rates we ran several models with different 
assumptions about this initial condition, see Section~\ref{sec.ang_mom_transport} and Figures
~\ref{fig.jrot_endHe} and \ref{fig.jrot_endHe_MESA}. We find that the final core 
rotation rate is almost independent of the initial rotation rate but depends 
strongly on the angular momentum transport process included in the simulation 
(i.e. non-magnetic or magnetic). 
For the ``Fuller'' model (super efficient) of angular momentum transport we assume
that the natal BH spin depends neither on initial stellar rotation nor on
metallicity, since in each case almost all angular momentum is removed from the
stellar core ~\citep{Fuller2019a,Ma2019,Fuller2019b}. 

The stellar models provide the angular momentum in each zone corresponding to a
spherical shell in the star.  We assume that angular momentum is conserved in the
collapse phase, and calculate the angular momentum of the compact remnant by
summing the angular momentum of the zones with enclosed mass lower than the
compact remnant mass. 

From the amount of angular momentum contained in the collapsing core, we calculate 
the dimensionless BH spin magnitude $a_{\rm spin}$. However, we limit the spin 
magnitude to $0.9$: $a_{\rm spin}={\rm min}(a_{\rm spin},0.9)$ to account for any 
potential processes that might remove some angular momentum during BH formation. 
We arbitrarily picked $0.9$ as a maximum for the spin rate. The difficulty in 
predicting the fastest spin values is that they require high-angular momentum 
disk-forming infalls. There are several spurious mechanisms which can remove angular 
momentum from these accretion disks but for now we do not have means to perform 
predictive  calculations of this effect. Disk winds are one example, however, the 
actual amount of material and angular momentum that is removed by such winds can 
vary dramatically~\citep{Vlahakis2001,Surman2006,Janiuk2017}. Therefore, we have 
used $a_{\rm spin}=0.9$ to denote any value between $0.9-1.0$, depending on these 
accretion loss mechanisms. 

In Figures~\ref{fig.bhspin1} and ~\ref{fig.bhspin2} we show the BH spins as a function 
of the progenitor's CO core mass for the Geneva and MESA models. We approximate the 
natal BH spin by simple fits. These rough fits are meant to reproduce the general trends 
in the data and are used in our population synthesis calculations. We list the actual 
data in Tables~\ref{tab.spin1} and ~\ref{tab.spin2}, so other fits can be attempted 
if desired. 

For the Geneva models the natal BH spin may be approximated by:
\begin{equation}
a_{\rm spin}= \left\{ \begin{array}{ll}
  0.85               & M_{\rm CO} \leq m_1\\
  a M_{\rm CO} + b   & m_1 < M_{\rm CO} < m_2\\
  a_{\rm low}        & M_{\rm CO} \geq m_2
\end{array}
\right.
\label{eq.bhspin1}
\end{equation}
with $a=-0.088$ for all models;
$b=2.258$, $m_1=16.0\msun$, $m_2=24.2\msun$, $a_{\rm low}=0.13$ for $Z=0.014$;
$b=3.578$, $m_1=31.0\msun$, $m_2=37.8\msun$, $a_{\rm low}=0.25$ for $Z=0.006$;
$b=2.434$, $m_1=18.0\msun$, $m_2=27.7\msun$, $a_{\rm low}=0.0$ for $Z=0.002$;
and $b=3.666$, $m_1=32.0\msun$, $m_2=38.8\msun$, $a_{\rm low}=0.25$ for $Z=0.0004$.
Note that progenitor stars with CO cores less massive than $\sim 20\msun$ (low- to 
intermediate-mass BHs) tend to produce high-spin BHs ($a_{\rm spin} \sim 0.8$--$0.9$), 
but higher mass stars (massive BHs) tend to produce low-spin BHs 
($a_{\rm spin} \sim 0$--$0.3$). This general trend is easily understood. Stellar 
winds during the evolution of a massive star can carry away a considerable amount 
of angular momentum~\citep{Meynet2015}. For the most massive stars, this mass 
loss is extensive, efficiently removing angular momentum and producing low-spin 
BHs. The data points also show a non-monotonic dependence on metallicity, which is
the result of a complex and metallicity-dependent interplay between the strength 
of stellar winds, the extent of the H-burning shell, and the model for the efficiency 
of element diffusion within the meridional current (see Appendix~\ref{sec.Geneva}).

For MESA models the natal BH spin may be approximated by:
\begin{equation}
a_{\rm spin}= \left\{ \begin{array}{ll}
  a_1 M_{\rm CO} + b_1  & M_{\rm CO} \leq m_1\\
  a_2 M_{\rm CO} + b_2  & M_{\rm CO} > m_1
\end{array}
\right.
\label{eq.bhspin2}
\end{equation}
with 
$a_1=-0.0016$, $b_1=0.115$, $m_1=\infty$ for $Z=0.014$;
$a_1=-0.0006$, $b_1=0.105$, $m_1=\infty$ for $Z=0.006$;
$a_1=0.0076$, $b_1=0.050$, $a_2=-0.0019$, $b_2=0.165$, $m_1=12.09\msun$ for $Z=0.002$;
and $a_1=-0.0010$, $b_1=0.125$, $m_1=\infty$ for $Z=0.0004$. 
The MESA models include the Tayler-Spruit magnetic dynamo and thus models of 
all masses and at all metallicities end up with low BH spin magnitudes in the 
range $0.05 \lesssim a \lesssim 0.15$. There is a mild tendency for lower 
metallicity and lower initial mass models to end up with slightly higher BH 
spin magnitudes but the dependence is much weaker than in the Geneva models, 
which do not include magnetic field related transport of angular momentum.

Finally, for the Fuller model that employs super-efficient angular momentum
transport, we adopt a single value of BH natal spin for stars of all masses 
and all metallicities: 
\begin{equation}
a_{\rm spin}= 0.01. 
\label{eq.bhspin3}
\end{equation}
Note that single stellar models (of different mass and metallicity) presented by 
\cite{Fuller2019b} produce very low spins for BHs in range: $a=0.003-0.035$ with 
a typical value of $a \sim 0.01$.

In our binary population synthesis calculations we employ the presented-above natal 
BH spin estimates obtained from single stellar evolution (see Sec.~\ref{sec.calcul}).
The discussion of binary interactions that can change these natal BH spins (that are 
and are not taken into account) is given in Section~\ref{sec.sinbin}.

\subsection{Black Hole Spin Misalignment}
\label{sec.misa}

To calculate the effective spin of BH-BH mergers we need to know the misalignment
angles of the two BH spin vectors with respect to the orbital angular momentum 
vector: $\Theta_1$ and $\Theta_2$ (see eq.~\ref{eq.xeff}). 

In our estimate of these angles we ignore the potential effects of tides and mass 
transfer/accretion: see \cite{Gerosa2013} and \cite{Wysocki2018} for alternatives 
and further discussion. However, in our calculations, we take into account the binary 
components spin precession. 
We assume that the two stars are born with spins that are fully 
aligned with the ZAMS binary orbital  angular momentum (${\bf L_0}$). At each BH 
formation, the natal kick may change the orbit and its orientation in space. After the 
first BH formation the new orbital angular momentum is ${\bf L_1}$, while after the 
second BH formation it is ${\bf L_2}$. 
After the first BH formation, if there was a natal kick, the binary component spins 
$\bf S_1$ and $\bf S_2$ are not aligned with the binary angular momentum vector. 
We assume that after the first BH formation both the BH spin and the companion star 
spin are small compared to the binary orbital angular momentum. With this assumption 
the total angular momentum of the binary is equal to the orbital angular momentum. 
Therefore, both the binary component spins  precess around ${\bf L_1}$. The precession 
periods for the BH and the companion star spin are different. In this 
approximation the spin of the BH (or the star) can be described as a sum of two 
components $\bf{S_{1/2}}= \bf{S_{1/2}^\parallel} +  \bf{S_{1/2}^\perp}$, where 
$\bf{S_{1/2}^\parallel}$ is the spin component parallel to the orbital angular 
momentum and $\bf{S_{1/2}^\perp}$ is the spin component perpendicular to the 
orbital angular momentum. During precession the former is constant and the latter 
has a constant value but rotates in the plane perpendicular to the direction of 
the angular orbital momentum. The precession period of each spin is different
~\citep{HamiltonSarazin82}. Thus at the time of the second BH formation we choose 
random positions of each component vector on its precessing trajectory and they 
become $\bf S_1'$ and $\bf S_2'$ . The natal kick (if any) changes again the 
binary orbit orientation in space, which becomes now $\bf L_2$. We follow all 
these changes to calculate the effective spin
\begin{equation}
\chi_{eff} = {M_{\rm BH1}{\bf S_1'} + M_{\rm BH2}{\bf S_2'}\  \over 
M_{\rm BH1}+ M_{\rm BH2}} \cdot {{\bf L_2} \over |{\bf L_2}|}
\end{equation}
Note that although (after the second BH formation) misaligned BH spins are subject to 
precession, the value of $\chi_{\rm eff}$ is expected to remain constant during 
the long inspiral towards the LIGO/Virgo band \citep{Gerosa2015}.

\subsection{Black Hole Masses}
\label{sec.bh_mass}

In our calculation of BH masses we use formulas assuming both the rapid and delayed 
supernova engines~\citep{Fryer2012}. The rapid development of the engine naturally 
creates a mass gap between NSs and BHs (dearth of compact objects in the range 
$2-5\msun$), while the delayed engine does produce compact objects in this range
~\citep{Belczynski2012a}. For single-star evolution, the delayed model minimum BH 
mass is $2.5\msun$, while it is $5\msun$ for the rapid model. Note that binary 
evolution may create light BHs ($\sim 2.5-5\msun$) by accretion induced collapse of 
a NS to a BH independent of supernova model~\citep{Belczynski2004b}. 

The maximum mass of a BH is set by the maximum mass of a star and wind mass loss rates 
and depends sensitively on the potential explosive mass--loss during the final stages 
of the star's life. We adopt a rather conservative maximum initial mass for the stars: 
$M_{\rm ZAMS}<150\msun$, although there seems to be evidence that stars with
mass of $\sim 200-300\msun$ may exists even in the local Universe~\citep{Crowther2010}.
For stellar winds we adopt formulas based on theoretical predictions of radiation 
driven mass loss~\citep{Vink2001} with inclusion of Luminous Blue Variable mass 
loss~\citep{Belczynski2010b}. These wind mass loss rates may be overestimated 
by as much as an order of magnitude~\citep{Oskinova2011,Ramachandran2019}. 
Therefore we allow (conservatively) for the reduction of stellar wind mass loss 
rates to $f_{\rm wind}=0.3$ of their currently adopted values. 

We also allow for pair-instability supernovas (PSNs) to entirely disrupt 
massive stars (stars with He core mass in range $M_{\rm He}=65-135\msun$)
and leave no NS/BH remnant. For somewhat lower-mass stars (stars with He
core mass in range $M_{\rm He}\approx40-65\msun$) we allow for pair-instability
pulsation supernovas (PPSNs). This process may remove the outer layers of a star, 
but does not lead to its disruption and allows for the BH formation in core collapse. 
We adopt three models for PPSN mass loss. In the first model we adopt strong 
PPSNs that are assumed to always remove the entire star mass above the  inner 
$45.0\msun$~\citep{Belczynski2016c} for stars with $M_{\rm He}=45-65\msun$.
Therefore, the post PPSN star mass is:  
\begin{equation}
M_{\rm star}/\msun= 45.0 \ \ \ \ 45.0 \leq M_{\rm He} < 65.0\msun. 
\label{eq.ppsn1}
\end{equation} 
In the second model we adopt recent PPSN calculations that allow for as much 
as $51.2\msun$ of the star to remain bound after a PPSN~\citep{Leung2019}. In this 
moderate PPSN model, the explosive mass loss depends on the He core mass. The
post PPSN star mass is:
\begin{equation}
M_{\rm star}/\msun= \left\{ \begin{array}{ll}
  0.65 M_{\rm He} + 12.2  & 40.0 \leq M_{\rm He} < 60.0\msun\\
  51.2  & 60.0 \leq M_{\rm He} < 62.5\msun\\ 
  -14.3 M_{\rm He} + 938.0  & 62.5 \leq M_{\rm He} \leq 65.0\msun.
\end{array}
\right.
\label{eq.ppsn2}
\end{equation}
Finally, we adopt a weak PPSN model that allows only for $50\%$ of the mass loss 
calculated by ~\cite{Leung2019}. In this scheme, the post PPSN star mass may 
reach $55.6\msun$, and we approximate the post PPSN star mass with: 
\begin{equation}
M_{\rm star}/\msun= \left\{ \begin{array}{ll}
  0.83 M_{\rm He} + 6.0  & 40.0 \leq M_{\rm He} < 60.0\msun\\
  55.6  & 60.0 \leq M_{\rm He} < 62.5\msun\\ 
  -14.3 M_{\rm He} + 938.1  & 62.5 \leq M_{\rm He} \leq 65.0\msun.
\end{array}
\right.
\label{eq.ppsn3}
\end{equation}

\begin{figure}
\hspace*{-0.3cm}
\includegraphics[width=9.2cm]{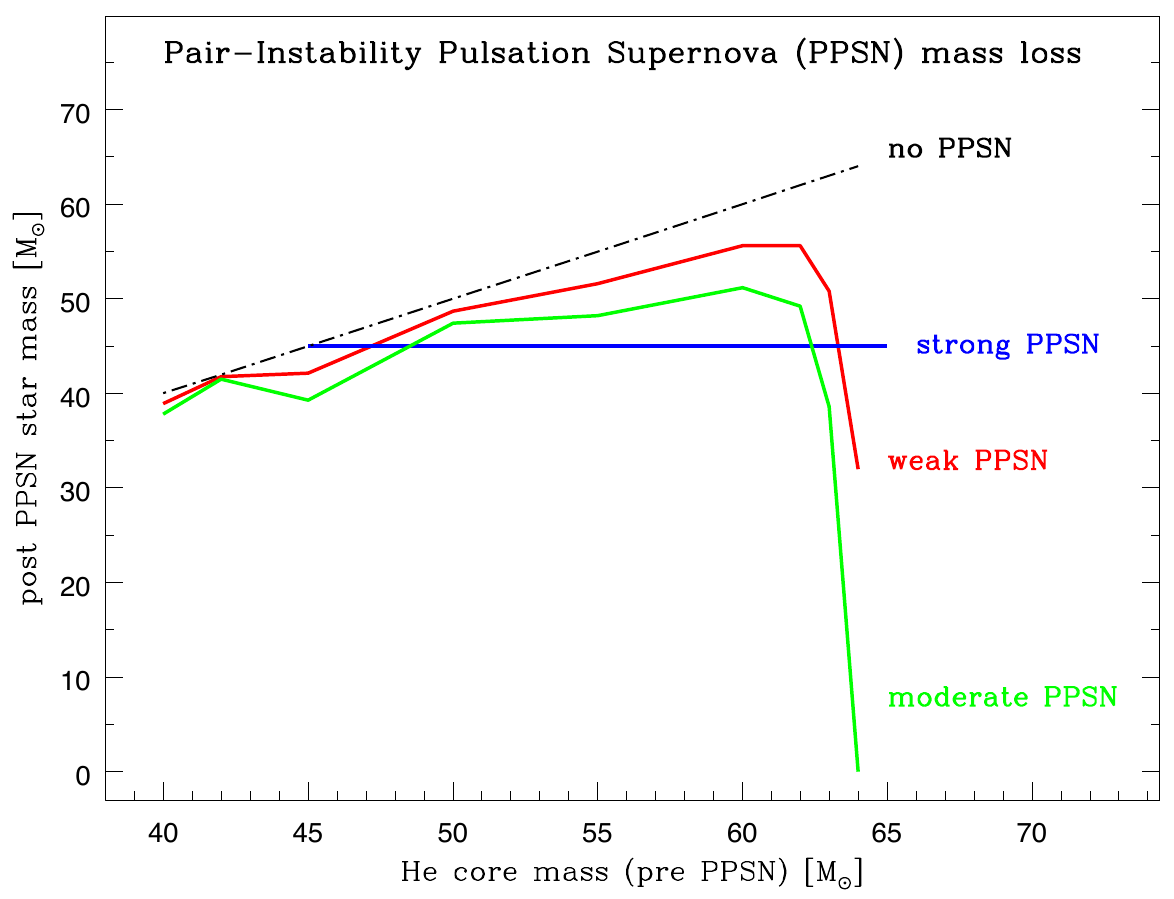}
\caption{Adopted models for pair-instability pulsation supernova mass loss. 
For a given He core mass we show the mass of a star after PPSN mass loss.
Moderate PPSN mass loss is adopted directly from~\cite{Leung2019}, while its 
modified ($50\%$ reduced mass loss) version is presented as weak PPSN model. 
Strong PPSN are adopted from~\cite{Belczynski2016c}.
}
\label{fig.ppsn}
\end{figure}

All three models are shown in Figure~\ref{fig.ppsn}. For the two models based 
on calculations from~\citet{Leung2019} we note a steep decrease in post PPSN 
star mass for He core masses above $\sim 60\msun$. The masses of these He stars 
are very close to the boundary mass between the PPSN and the PSN. As the mass 
of the He star increases, the central temperature at the bounce increases and 
oxygen burning becomes more explosive and causes a larger amount of mass ejection. 
Eventually explosive oxygen burning in the $\sim 65\msun$ He star produces 
large enough nuclear energy to disrupt the star completely with no BH remnant,
i.e., induces PSN. The amount of mass ejection increases steeply as the He 
star mass increases from $\sim 60\msun$ to $\sim 65\msun$, because the oxygen 
burning rate is very sensitive to the temperature. Thus the remnant BH mass 
decreases steeply as the He star mass approaches $65\msun$.

In the PPSN mass regime massive BHs are formed, and according to our scheme
~\citep{Fryer2012} these BHs form through collapse of the entire star into a BH 
(direct BH formation). However we allow for some mass loss in neutrinos. In 
the original formulas we have adopted a high neutrino fractional mass loss 
\citep[$f_{\rm neu}=0.1$; see e.g.][]{Belczynski2016b}, while here we also allow 
also for much smaller mass loss in some models ($f_{\rm neu}=0.01$). The final BH 
mass, in case of the direct BH formation, is calculated from 
\begin{equation}
M_{\rm BH}= (1 - f_{\rm neu}) M_{\rm star}, 
\label{eq.bhmass}
\end{equation}
where $M_{\rm star}$ denotes star's mass just prior the core collapse. 

\subsection{Black Hole Accretion Model}
\label{sec.bhacc}

\begin{figure}
\hspace*{-0.3cm}
\includegraphics[width=9.2cm]{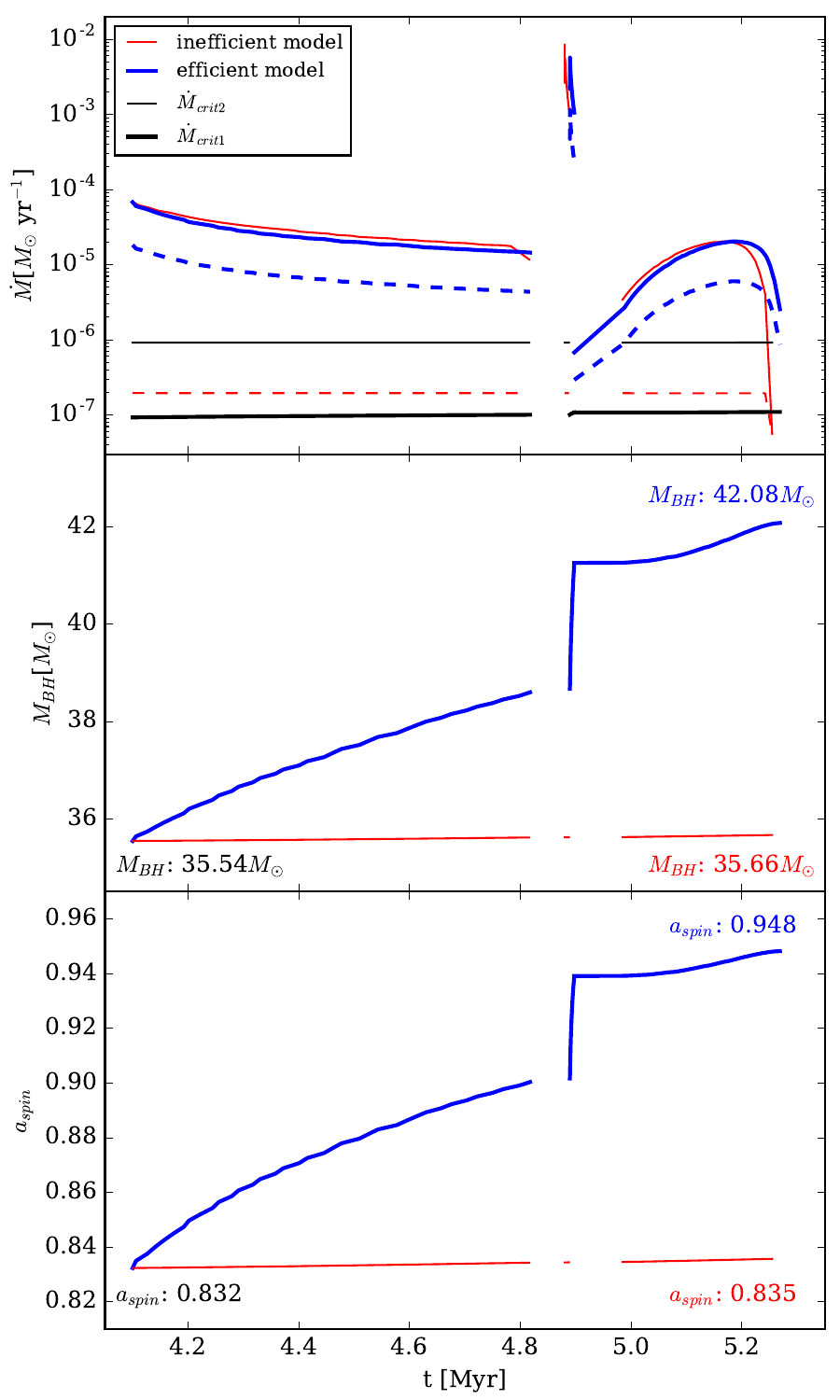}
\caption{
    The figure shows the comparison of the two black hole accretion models employed 
    in the {\tt StarTrack} code. Presented are the evolution of the mass transfer rate from 
    a donor star ($\dot M_{\rm don}$; top panel, solid lines), BH mass accumulation 
    rate ($\dot M_{\rm BH}$; top panel, dashed lines), the BH mass ($M_{\rm BH}$; 
    middle panel), and the BH spin ($a_{\rm spin}$; bottom panel) during the RLOF 
    stable mass transfer phases. Critical mass transfer rates (above which mass 
    ejection from a system is expected) are provided for reference. During the first 
    part of the RLOF ($t\approx 4-4.8$ Myr) the donor is a MS star, during the short-duration 
    peak it is a HG star ($t\approx 4.9$ Myr), and for the remaining RLOF it is a 
    helium-core burning star. See Sec.~\ref{sec.bhacc} for a description of the full 
    evolutionary path and both accretion models.}
\label{fig.bhacc}
\end{figure}

Mass accretion may increase the BH mass and spin after its formation. Here we test 
two models of accretion from a stable Roche lobe overflow (RLOF; not common
envelope [CE]) mass-transfer or from stellar winds. 
The first model (efficient BH accretion) is based on the results of global, axisymmetric 
simulations of accretion disks with $\alpha P$ viscosity, disk winds and photon trapping 
performed by \cite{Ohsuga2007}. \cite{Belczynski2008b} obtained a fit to the results of 
\cite{Ohsuga2007}:\\

$\log \left( {\dot M_{\rm BH} \over \dot M_{\rm crit1}} \right)  =$\\
\begin{equation} 
    \left\{ \begin{array}{ll} 
        \log \left( {|\dot M_{\rm don}| \over \dot M_{\rm crit1}} \right) & |\dot M_{\rm don}| \leq \dot M_{\rm crit1}\\
        0.544 \log \left( {|\dot M_{\rm don}| \over \dot M_{\rm crit1}} \right) & \dot M_{\rm crit1} < |\dot M_{\rm don}| \leq 10 \times \dot M_{\rm crit1}\\
        0.934 \log \left( {|\dot M_{\rm don}| \over \dot M_{\rm crit1}} \right) - 0.380 & |\dot M_{\rm don}| > 10 \times \dot M_{\rm crit1}\\ 
    \end{array} \right.,
    \label{eq.eff_bhacc}
\end{equation} 
\noindent where $\dot M_{\rm BH}$ is the mass accumulation rate onto the BH, 
$\dot M_{\rm crit1}=2.6 \times 10^{-8} (M_{\rm BH}/10\msun) \mpy$ is the critical 
mass transfer rate (obtained from numerical simulations) above which the accretion 
onto the BH is not fully efficient, and $\dot M_{\rm don}$ is the mass transfer 
rate from the donor to the accretion disk around the BH. Note that above the 
critical mass transfer rate some mass transferred from the donor is lost/ejected 
from the system. 

The second model (inefficient BH accretion) uses the analytical prescription of 
\cite{Shakura1973}. In this model, disk winds play a significant role effectively 
limiting the mass accumulation for super-Eddington mass transfer rates by limiting 
the local accretion rate to its Eddington value. This model is supported by the 
numerical results of strong outflow from the super-critical accretion disk 
\citep{Abolmasov2009,Sadowski2016}, as well as by observations of many ultra 
luminous X-ray sources in centers of unusually large and bright emission nebulae 
which are powered by the accumulated kinetic energy of the outflow 
\citep{Pakull2003,King2004}, and also supported by the fact that the neutron star 
in Cygnus X-2 while being subject to high mass transfer rate ($\sim 10^{-5}\mpy$) 
has ejected most of the mass transferred from the donor star~\citep{King1999}.

In this model we assume that the mass accumulation rate onto the BH is given by
\begin{equation} 
    \dot M_{\rm BH}=f_1 f_2 \dot M_{\rm don},
    \label{eq.ineff_bhacc}
\end{equation} 
\noindent where $(1-f_1)$ is the fraction of the mass transferred to the disk
from the donor which is lost in wind from the outer part of the accretion disk. 
It is difficult to determine the wind mass loss rate from the outer part of the 
disk \citep[see][]{Sadowski2016}, and we treat $f_1$ as a parameter in our model. 
For the current calculations we adopt $f_1=1$, i.e., no wind mass loss in the outer 
part of the disk, which maximizes accumulation of mass onto the BH. 
Similarly, $(1-f_2)$ represents the fraction of mass which is lost from the inner 
part of the accretion disk. The value of $f_2$ depends on the mass transfer rate as:
\begin{equation}
f_2=
\begin{cases}
1, & \text{for\ }\dot M_{\rm don}\leq\dot M_{\rm crit2}\\
\frac{R_{\rm ISCO}}{R_{\rm sph}}, & \text{for\ } \dot M_{\rm don}>\dot M_{\rm crit2}
\end{cases}
\end{equation}
\noindent where the critical mass transfer rate is 
$\dot M_{\rm crit2} = \dot M_{\rm Edd}= 4.375 \times10^{-8} (1+X)^{-1} (M_{\rm BH}/\msun)\mpy$, 
where hydrogen mass fraction $X$ is $0.7$ for H-rich donor stars and $0.0$ for 
H-deficient donor stars, $R_{\rm ISCO}$ is the innermost stable circular orbit 
radius, and $R_{\rm sph}$ is the spherisation radius, where the disk's height 
becomes comparable to the radius. $R_{\rm ISCO}$ depends on the spin value of 
the accreting BH and varies between $0.5R_{\rm S}$ (for a maximally prograde 
spinning BH) to $4.5R_{\rm S}$ (for a maximally retrograde spinning BH), 
where $R_{\rm S}=\frac{2GM_{\rm BH}}{c^2}$ is the Schwarzchild radius of a BH. 
$R_{\rm sph}$ is given by \citep{Shakura1973}
\begin{equation}
    R_{\rm sph}=\frac{27}{4}\frac{\dot M_{\rm don}}{\dot M_{\rm Edd}}R_{\rm S}.
\end{equation}
The above two equations are a simplified description of the change of the
accretion mode onto compact objects. In particular, they lead to a jump in
mass accretion rate onto a compact object when mass transfer from a donor star  
is equal to the critical mass transfer rate ($M_{\rm don}=\dot M_{\rm crit2}$). 
This simplification does not influence the results of our evolutionary calculations. 

In our calculations we always assume a prograde BH spin, and the initial BH spin 
magnitude is adopted from a given stellar model (see Sec.~\ref{sec.spins}) and 
then increased by accretion as detailed in \cite{Belczynski2008b}.
At sub-Eddington accretion rates ($\dot M_{\rm don}\leq\dot M_{\rm Edd}$), 
we assume that there is no mass loss from the inner part of the disk ($f_2=1$).

The mass accumulation onto a BH in the inefficient BH accretion model is always 
Eddington-limited ($\dot M_{\rm BH}<\dot M_{\rm Edd}$), whereas in the
efficient BH accretion model it increases monotonically with the mass transfer rate 
and may significantly exceed $\dot M_{\rm Edd}$. These two models produce the same 
accumulation on a BH for mass transfer rates below 
$\dot M_{\rm crit1} = 0.06\times\dot M_{\rm Edd}$ where both models give 
exactly the same prescription ($\dot M_{\rm BH}= |\dot M_{\rm don}|$, note that
$\dot M_{\rm crit1} = 0.06 \dot M_{\rm crit2}$). 

As a result, the evolution of BH binaries which go through a phase of stable mass 
transfer with high (super-critical) mass transfer rate may be considerably different 
under different assumptions about the BH accumulation efficiency. However, the 
dominant formation channel for BH-BH mergers \citep[see][and also Fig.~\ref{fig.evol1} 
and Fig.~\ref{fig.evol2}]{Belczynski2016b} contains no such phase. Along the way 
to formation of most BH-BH mergers in our models, accretion of matter by a BH takes 
place only during a short-lived CE event or as result of capturing a fraction of a 
stellar wind from the companion. Accretion during CE can be best related to the 
Bondi-Hoyle accretion  and has been found to be rather insignificant (less than 
$\sim 1\msun$ for a typical $30\msun$ BH, see Sec.~\ref{sec.Binary} for a detailed 
discussion of recent calculations of accretion during CE). The wind-fed accretion 
during the subsequent phase of a compact BH -- Wolf-Rayet (BH-WR) binary evolution 
is even less significant, partially due to small wind-mass loss rates from low 
metallicity systems (which are the progenitors of most BH-BH mergers). 

Having said that, the BH accretion models presented in this section can play an 
important role in certain sub-dominant channels for the BH-BH merger formation. In 
Fig.~\ref{fig.bhacc} we showcase the time evolution of the mass transfer and accretion 
rates as well as the BH spin and mass in a system in which the BH accretion is 
particularly significant. The system began its evolution as two ZAMS stars with masses 
of $84.6\msun$ and $48\msun$ formed at a very low metallicity $Z=0.0002$ in a binary 
with separation of about $1900\rsun$. At the age of $3.8$ Myr the primary, now a 
$66\msun$  helium-core burning star with a radius of $800\rsun$, goes into RLOF and 
initiates a CE phase. As a result, the primary becomes a $36 \msun$ WR star and the 
binary separation decreases down to $30\rsun$. With the orbit already being quite 
compact, the companion MS star ($\sim 47\msun$) initiates another RLOF and starts 
stable mass transfer back onto the WR primary. The WR star grows to about $39\msun$, 
before collapsing directly into a $35.54\msun$ BH at the age of $4$ Myr (with no natal 
kick, $10\%$ of mass being lost in neutrinos, and the natal spin of $a_{\rm init}=0.832$ 
adopted from Geneva BH natal spin model). Soon thereafter, the companion MS initiates 
a RLOF again and the first phase of stable, super-critical mass transfer onto the 
newly-formed BH begins ($\dot{M}_{\rm don} \sim$ a few $\times\;10^{-5}\mpy$).
At that moment the system may be potentially observable as an ultraluminous X-ray 
source \citep[see e.g.,][for the recent analysis of these objects]{Wiktorowicz2019}. 
The mass transfer continues up until about $t=4.8$ Myr (Fig.~\ref{fig.bhacc}), at which 
point the donor is at the very end of its MS evolution and contracts a bit, causing a 
temporary detachment. The mass transfer starts again when the companion begins to expand 
on its HG at the age of about $t=5.9$ Myr having mass of $28\msun$. The rate is high 
($\sim 10^{-3}\mpy$), but the phase is short-lived ($7.7$ kyr). Finally, the last 
phase of mass transfer occurs when the companion is a slowly expanding core helium
burning star and terminates at about $5.25$ Myr.

The net result of the subsequent stable mass transfer phases shown in 
Fig.~\ref{fig.bhacc} is an increase of the binary separation, which at the point 
of the final detachment is about $160\rsun$ (similar in both accretion models). 
At that point the donor star has been stripped almost down to its helium core and 
has a remaining mass of only about $10\msun$.

Since accumulation of mass onto the BH (dashed lines in the top panel of 
Fig.~\ref{fig.bhacc}) is different in the two models for super-critical mass 
transfer rates, the final BH mass is also different: $42.08\msun$ for the 
efficient BH accretion model and $35.66\msun$ for the inefficient BH accretion 
model (Fig. \ref{fig.bhacc}, middle panel). We also note that the BH spin, was 
increased to $0.948$ for the efficient model and to $0.835$ for the inefficient 
model. 

This particular binary ends its evolution at $t \approx 5.6$ Myr forming a BH-BH 
with masses of $35.7\msun$ and $6.2\msun$ for the inefficient model, or $42.1\msun$ 
and $6.0\msun$ for the efficient model. The delay time to merger is about 3 Myr. The 
second BH is formed through a supernova explosion with a natal kick. As the separation 
before the supernova explosion was relatively large in both accretion models 
($\sim 160\rsun$), the formation of a binary that could merge within the Hubble time 
was only possible thanks to the preferentially oriented natal kick, which decreased 
the separation down to $\sim90\rsun$ and, more importantly, induced a high eccentricity 
of $e>0.988$ (from a $e=0.0$ pre-supernova value). Such an influence of a natal kick on 
the binary orbit is very rare \citep{Andrews2019}, so the above example is an extreme 
example of a BH-BH merger formation.  In the vast majority of systems, the impact of 
the accretion model on the BH-BH formation is much smaller.

\subsection{Tidal Interactions}
\label{sec.tides}

In this section, we discuss the efficiency of tidal spin-up in close binaries. 
Our standard implementation of tides follows from \cite{Hurley2002} and is based 
on the standard \cite{Zahn1977,Zahn1989} theory as updated by more recent calibrations 
\citep{Claret2007}. These prescriptions, which are laid out in
\cite{Belczynski2008a},
result in rather weak tidal interactions between stars in binary systems.
This is particularly true for very massive stars (e.g., progenitors of BHs) that
evolve so fast that the tides do not affect significantly their rotation. 

In our classical BH-BH formation scenario ~\citep{Belczynski2016b}, the only
evolutionary phase at which tides could spin-up either of the BH progenitors
happens after the CE phase when the initially wide binary ($a\gtrsim1000\rsun$) is 
reduced to a rather tight orbit ($a\lesssim 100\rsun$). Such a configuration will 
consist of at least one stripped stellar core (WR star): BH-WR, WR-BH or WR-WR 
binaries. 

Recently \cite{Kushnir2016,Zaldarriaga2017,Kushnir2017,Hotokezaka2017,Qin2018} 
investigated the strength and treatment of tides and they argued that WR stars 
may potentially be significantly spun up if placed in very close binaries with 
massive companions (e.g., immediate progenitors of BH-BH mergers).

\begin{figure}
\hspace*{-0.3cm}
\includegraphics[width=9.2cm]{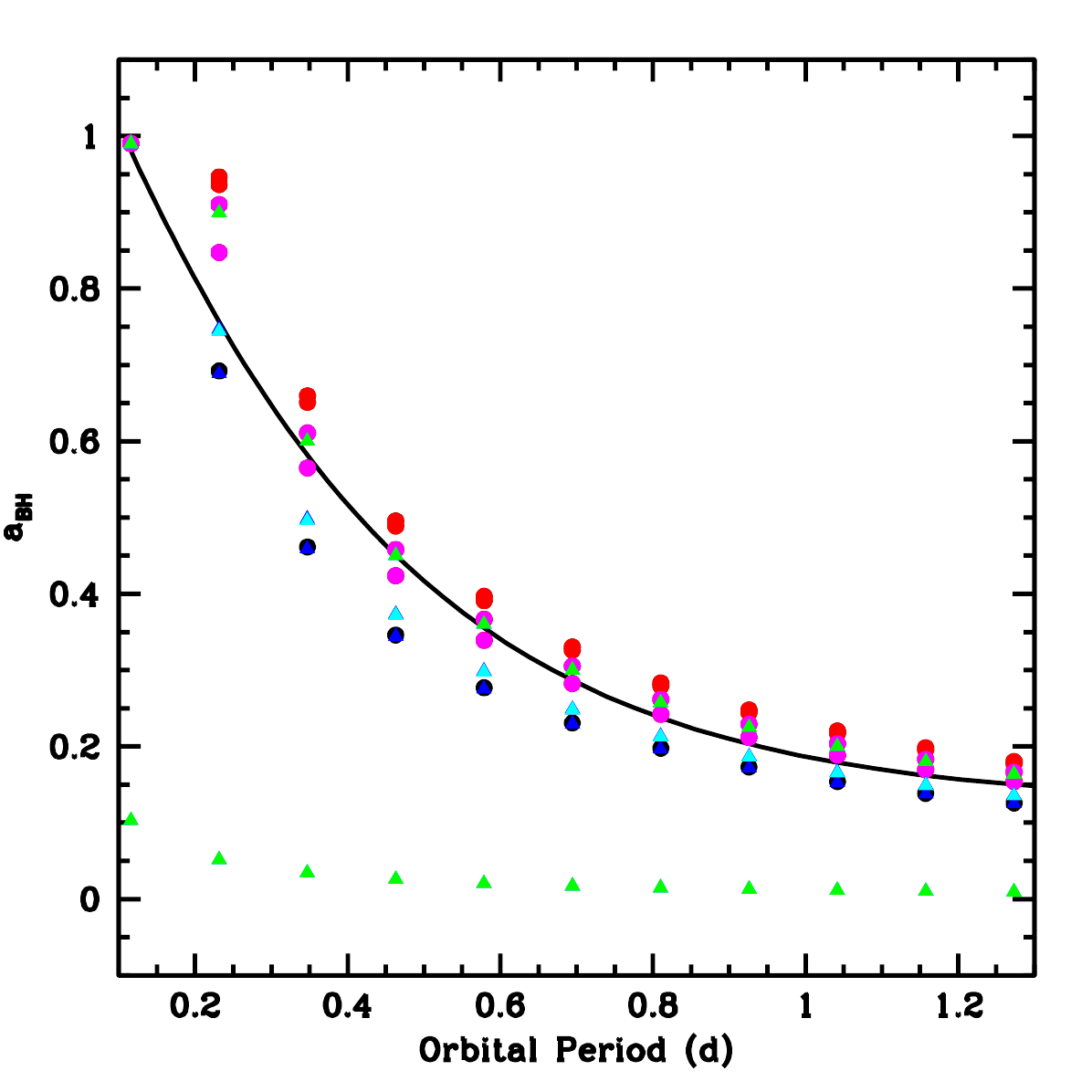}
\caption{
Black hole spin magnitudes as a function of the orbital period for our $Z=0.014$ 
(circles) and  $Z=0.0004$ (triangles) MESA models. The color coding corresponds to the 
black hole remnants with masses: $M_{\rm BH}<15\msun$ (black/blue), $15\msun<M_{\rm BH}<30\msun$
(cyan/red), $M_{\rm BH}> 30\msun$ (green/magenta). Binary systems with short orbital 
periods: $0.1-1.3$d and WR stars will produce BHs with broad range of spin magnitudes: 
$0.15-1$. Systems with orbital periods below $0.1$d form black holes from tidally locked 
WR stars, and the BH spins are maximal. Binaries with an orbital period above $1.3$d produce 
BHs with spins below $0.2$ and typically are not tidally locked at all. A set of the 
most massive, lowest-metallicity stars have such dense cores that tidal spin-up does not 
dramatically increase their angular momentum and these stars require orbital periods of 
less than $0.1$d to have spin values above $0.1$ (lower set of green triangles). 
However, we ignore this fact, and let these cores to be spun-up to estimate the maximal 
effect of tidal interactions in our models (higher set of green triangles). The black 
curve shows our fit to the BH spin magnitude as a function of orbital period.
}
\label{fig.tidal}
\end{figure}

To determine the upper limit to the spins produced by tidal locking, we ignore 
the  orbital evolution and assume that the entire star is tidally locked to the 
orbit (see the discussion below), i.e.,  rigidly rotating. The orbital period then 
provides the spin period of the WR star and implementing these spin periods into 
our stellar models, we can calculate the angular momentum in the star and use this 
angular momentum to obtain the spin period of the collapsed star, assumed to be 
the BH spin. Figure~\ref{fig.tidal} shows the resultant black-hole spin magnitudes 
for our $Z=0.014$ and $Z=0.0004$ MESA models. For most of the models with orbital 
periods in the range $P_{\rm orb}=0.1-1.3$\,d, the resultant BH spin magnitude can 
be fit by an exponential:
\begin{equation}
a_{\rm spin} = e^{-0.1 (P_{\rm orb}/P_0-1)^{1.1}} + 0.125,
\label{eq.tides}
\end{equation}
where $P_{\rm orb}$ is the orbital period in seconds and $P_0=4000$\,s. For 
systems with orbital periods below $0.1$\,d, the resultant BH spin is maximal.  
For systems with orbital periods longer than $\sim 1.3$\,d, tidal locking takes 
longer than the duration of the relevant evolutionary phase and there is no 
significant spin-up (in which case the tidal spin-up is ignored). The fit with 
eq.~\ref{eq.tides} is  shown as the black curve in Figure~\ref{fig.tidal}.

Note that the proposal by \cite{Kushnir2016,Zaldarriaga2017,Kushnir2017,Hotokezaka2017} 
according to which every BH formed from a WR star subject to tidal interactions is 
spun-up to maximum ($a_{\rm spin}=1$) is subject to three caveats. First, for 
systems that are undergoing  tidal synchronization it is assumed that the BH is 
formed with a maximal spin without close scrutiny of the WR star structure and its 
evolution under strong tidal interactions. Second, the scheme ignores the fact that 
WR wind mass--loss (depending on metallicity of the WR star) may widen the 
system pushing it into a regime where synchronization is not maintained. Third, 
the tidal locking of the WR star pumps orbital energy into the WR star spin and
thus causes the orbit to decay. This may lead to a merger of a WR star with its 
companion barring the formation of a BH-BH binary. A detailed analysis of this 
complex interplay of stellar structure, stellar winds, and tidal interactions in 
the context of long Gamma Ray Bursts and BH-BH formation was presented by 
\cite{Detmers2008} and by \cite{Qin2018}. In particular, \cite{Detmers2008} found 
that the majority of close BH-WR systems that are subject to strong tidal interactions 
either evolve to long periods (for high metallicity) or undergo a component merger 
before the BH-BH formation (for low metallicity). On the other hand, \cite{Qin2018} 
found that most close BH-WR systems will increase their periods resulting in a wide 
range of secondary BH spins ($a_{\rm spin}=0-1$). In the light of these detailed 
calculations, our adopted simple model (see eq.~\ref{eq.tides}) resembles the 
\cite{Qin2018} scheme and allows for a broad range of BH spin magnitudes if the WR 
star forming a BH is a subject to strong tidal interactions. We should stress that 
we use this model only as an alternative to our standard assumption according to 
which the effect of tidal spin-up in close binaries on the BH natal spin magnitude 
is ignored (as we use single stellar models to estimate natal BH spins; see 
Sec.~\ref{sec.spins}). 

Recently, ~\cite{Bavera2019} updated some of the \cite{Qin2018} calculations. 
One should note that these studies follow the angular momentum transport in 
stars only during the WR stage that will form the second-born BH 
in BH-BH merger. Additionally, it is assumed that the first BH is born with 
zero spin ($a=0$) and that the WR star that will form the second BH is also 
born spinless. Such simplifying assumptions allow to match LIGO/Virgo BH-BH 
effective spins which are mostly consistent with zero. However, 
it was only tested for initially non-spinning WR stars, and therefore did 
not take into account the fact that for inefficient angular momentum transport 
and high initial stellar rotation, WR stars can be born with high spins. Then 
for high-spinning WR stars and for inefficient angular momentum transport a 
significant fraction of the second-born BHs would have high spins independent 
of tidal interactions. This is demonstrated by our models that employ 
inefficient (Geneva) angular momentum transport, in which BHs (both first-born 
and second-born) may have high spins (see Fig.~\ref{fig.bhspin1} and Fig.
~\ref{fig.xeff1}). 

\subsection{Cosmic Star Formation History and Metallicity Evolution} 
\label{sec.sfr}

Starting from 2016 (e.g., \cite{Belczynski2016b}) we have been using the cosmic 
star formation  density  (SFRD)  determinations from \cite{Madau2014}, which 
are based on a number of deep UV and infrared galaxy surveys. We refer to this 
SFRD as the ``old SFRD" formula.  Here, we adopt two best-fitting comoving
SFRDs from ~\cite{Madau2017},  which update the previous formula by
better reproducing a number of recent $4<z<10$ results:
\begin{equation}
sfr(z)=\mathcal{K}_{\rm IMF} 0.015 {(1.0+z)^{2.6} \over 1.0+((1.0+z)/3.2)^{6.2}} \sfr. 
\label{eq.sfr1}
\end{equation} 
The correction factor $\mathcal{K}_{\rm IMF}$ adjusts the SFRD for the 
assumed IMF to the Salpeter IMF (i.e. $\mathcal{K}_{\rm IMF;Salpeter}=1.0$). 
In this work we adopt a three component broken power-law Kroupa IMF with $\alpha_3=-2.35$ 
(see App.~\ref{sec.Binary} for details), for which the correction factor is  
$\mathcal{K}_{\rm IMF;Kroupa} \approx 0.66$~\citep{Madau2017}. 

\begin{figure}
\hspace*{-0.3cm}
\includegraphics[width=9.2cm]{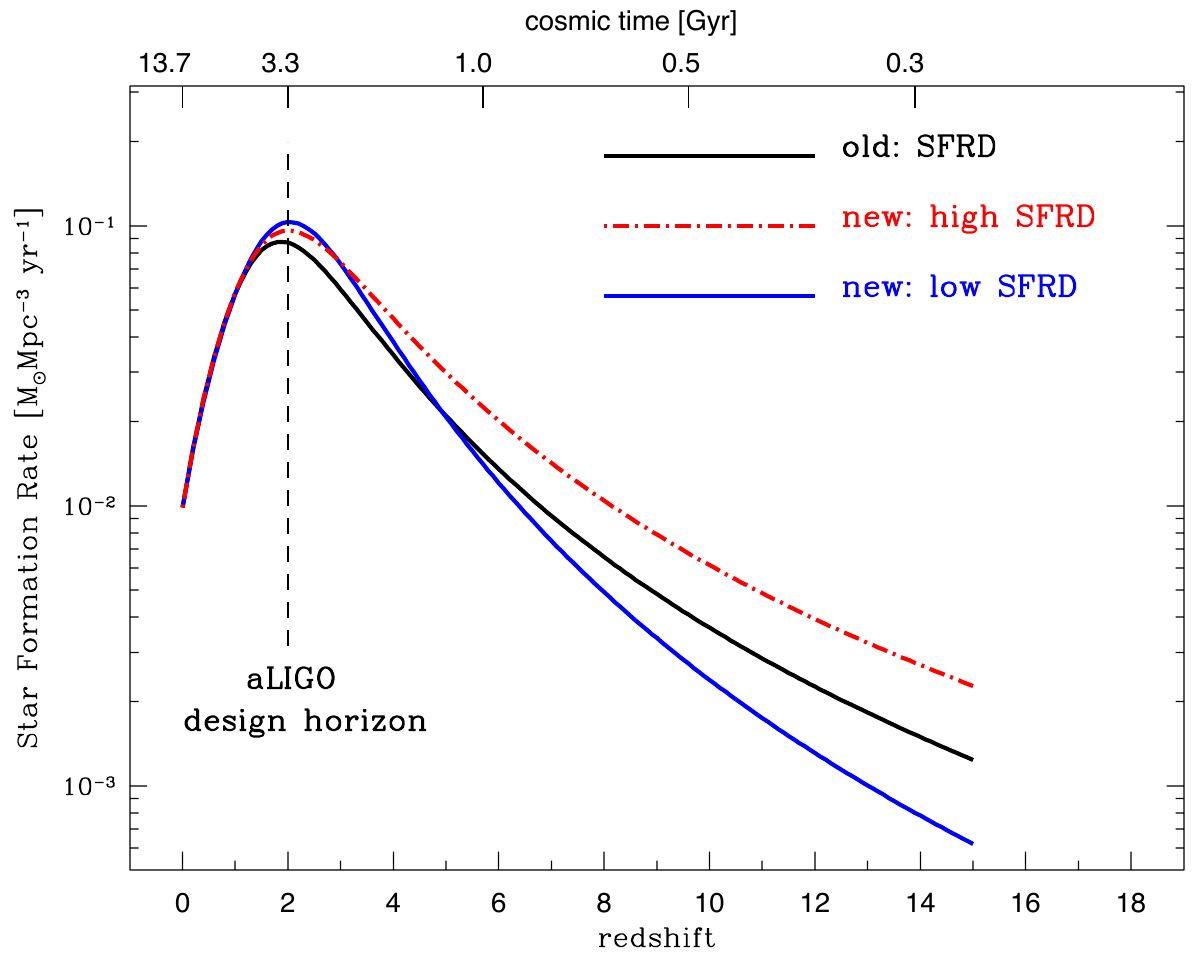}
\caption{The cosmic star formation histories adopted in our modeling (see 
Sec.~\ref{sec.sfr} for details). Note  how the different SFRDs agree at low 
redshifts ($z\lesssim2$), while they disagree by as much as a  factor of three 
at earlier epochs. 
}
\label{fig.sfr}
\end{figure}

We refer to the SFRD given by Eqn.~\ref{eq.sfr1} as to the ``low SFRD",  since it is based 
on a conversion from  luminosity density to SFRD in which all published galaxy luminosity 
functions (LFs) have been conservatively integrated down to the same limiting luminosity 
of $0.03\,L_*$, where $L_*$ is the characteristic luminosity of a Schechter function. 
For completeness, in the following we shall also provide results for a ``high” model in 
which the comoving SFRD is increased at high redshifts  to account for a steep faint end 
of the LF:
\begin{equation}
sfr(z)=\mathcal{K}_{\rm IMF} 0.015 {(1.0+z)^{2.7} \over 1.0+((1.0+z)/3.0)^{5.35}} \sfr.
\label{eq.sfr2}
\end{equation}
We will refer to this formula as the ``high SFRD". These expressions bracket 
uncertainties in the contribution to the early SFRD by galaxies fainter than
$-16$ mag, but agree at $z<2$. Our adopted SFRDs  are shown in Figure~\ref{fig.sfr}.  

Our modeling of the metallicity of the star-forming gas has also been updated 
as follows. In previous calculations, we used the evolving mean  cosmic 
metallicity  $Z(z)$ from \cite{Madau2014}, which is the ratio between the total 
mass density of  heavy elements produced over cosmic time and the cosmic baryon 
density. This quantity had to be increased by $0.5$ dex to better account for 
various supernovae and GRB observations~\citep{Belczynski2016b}. At each redshift 
$z=0-15$ we further assumed that the distribution of $\log(Z/\zsun)$ was a Gaussian 
centered at the average metallicity and with a dispersion of $\sigma=0.5$ dex. 
Stellar populations (Population I/II stars) with various metallicities were 
subsequently evolved, to check whether they produce NS-NS, BH-NS or BH-BH mergers 
detectable by LIGO/Virgo for a given instruments' sensitivity (see Sec.~\ref{sec.GW}). 

This approach does not properly describe the metallicity evolution of the star-forming 
gas within galaxies, as only a small fraction of the baryons in the Universe are 
polluted with heavy elements and take part in the baryon cycle of galaxy evolution 
\citep[e.g.][]{Chruslinska2019}. 
Our improved calculations follow \cite{Madau2017}. The gas-phase oxygen abundance is 
known to correlate strongly with the total stellar mass of star-forming galaxies: this 
``mass-metallicity  relation" (MZR)  has been shown to  extend down to  low-luminosity 
galaxies with stellar masses $\sim 10^6\msun$ \citep{Berg2012} and out to redshift 
$3.5$ \citep{Maiolino2008}. Numerical modeling by \citet{Zahid2014} suggests that the 
MZR originates from a more fundamental, universal relationship between the metallicity and 
the stellar-to-gas mass ratio that is followed by all galaxies as they evolve. We have 
assumed that the \citet{Zahid2014} MZR  holds at all redshifts, and integrated this 
relation over the evolving galaxy stellar mass function at $0<z<7$ to compute a mean 
stellar mass-weighted gas-phase metallicity (see \citealt{Madau2017} for details and 
references) as:
\begin{equation}
\log(Z/\zsun)= 0.153 - 0.074 z^{1.34}.
\label{eq.Z}
\end{equation}
As before, we assume the same distribution of $\log(Z/Z_{\odot})$, but centered around this 
new average. Note that this is a simplification and this distribution may not be symmetric 
\citep{Chruslinska2019}. Recently, \citet{Chruslinska2019b} provided an observation-based 
distribution of the star formation rate density over metallicities and redshifts and 
discussed the uncertainty of this distribution. We note that the metallicity distribution 
used in our paper peaks at similar metallicities as the high-metallicity extreme reported 
by those authors at $z\lesssim 3$ and at noticeably higher metallicities at $z>3$. However, 
at those high redshifts the distribution is poorly constrained by current observations. 

\begin{figure}
\hspace*{-0.3cm}
\includegraphics[width=9.2cm]{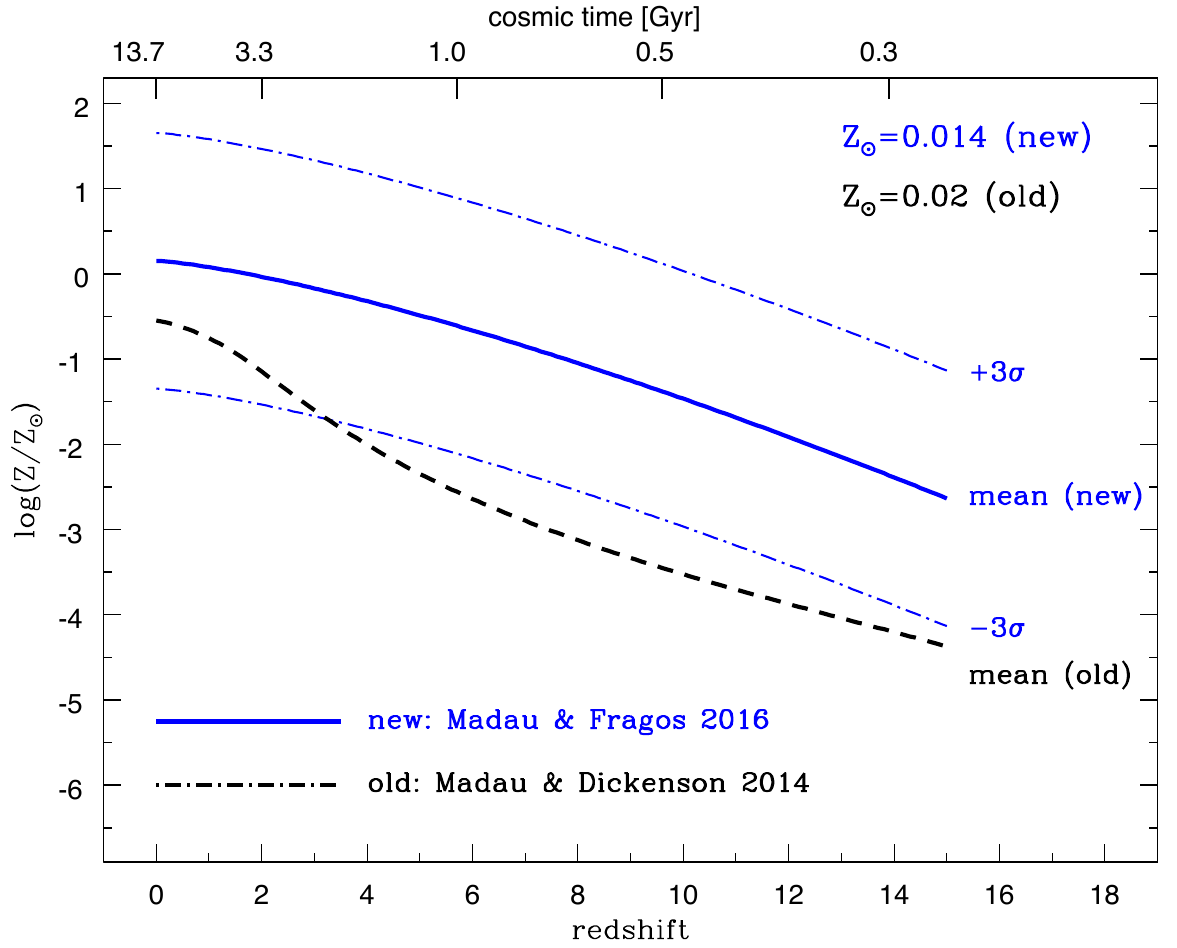}
\caption{Stellar mass-weighted gas-phase metallicity  versus redshift, $Z(z)$.
At every epoch, we adopted a Gaussian distribution of $\log(Z/\zsun)$, centered 
at the mean metallicity and with  dispersion $\sigma=0.5$ dex. Note that the new 
metallicity for the star forming gas  is noticeably higher than the mean cosmic 
metallicity adopted in the past (see Sec.~\ref{sec.sfr}).  
}
\label{fig.Z}
\end{figure}

There is no consensus on the value of solar metallicity,
and no value within range $\zsun=0.012-0.02$ can be rejected at the 
moment \citep{Vagnozzi2019}. In our models we use either $\zsun=0.014$ or $\zsun=0.02$.
A comparison between the adopted old and new mean gas-phase metallicities versus 
redshift is shown in Figure~\ref{fig.Z}. 

\subsection{Calculations}
\label{sec.calcul}

Our binary evolution calculations were performed with the upgraded population
synthesis code {\tt StarTrack}~\citep{Belczynski2002,Belczynski2008a}.
Improvements to the code include updates to the treatment of the common envelope
(CE) evolution, compact object mass calculations including the effect of
pair-instability pulsation supernovae and pair-instability supernovae, and
new BH natal spin prescriptions (Sec.~\ref{sec.spins}), among other upgrades
(see Appendix~\ref{sec.Binary}).

We consider fourteen different realizations (models) of our classical isolated binary 
evolution to test whether it is possible to form LIGO/Virgo BH-BH mergers
with the observed rates, masses and effective spins. The first two models 
correspond to our previous calculations with fallback-decreased, BH mass-dependent 
(M10) and high, mass-independent (M13) BH natal kicks with input physics listed 
in Table~\ref{tab.models} and detailed in ~\citep{Belczynski2016b,Belczynski2016c}. 

The next four models include different input physics on mass transfer, BH accretion 
in the CE phase, and the approximation of effects of stellar rotation on the BH 
mass (see Appendix~\ref{sec.Binary}). These models include natal kicks which are 
fallback-decreased, BH mass-dependent (M20); small, BH mass-independent (M26; 
$\sigma=70\kms$); intermediate, BH mass-independent (M25; $\sigma=130\kms$); and 
high, BH mass-independent (M23; $\sigma=265\kms$).

The above six models employ BH natal spins adopted from the Geneva model with mild
angular momentum transport in massive stars (eq.~\ref{eq.bhspin1}), SFRD and
average metallicity evolution with redshift from \cite{Madau2014}, and high
value of solar metallicity $Z=0.02$. The next eight models represent our current 
update (2019) and tests of input physics. 

M30 employs the rapid supernova engine model for NS/BH mass from \cite{Fryer2012}
supplemented with weak PPSN (eq.~\ref{eq.ppsn3}), $1\%$ neutrino mass loss at the 
BH formation and $10\%$ neutrino mass loss at the NS formation (eq.~\ref{eq.bhmass}), 
fallback-decreased, BH/NS mass-dependent, natal kicks (1D $\sigma=265\kms$), $50\%$ 
non-conservative RLOF, $5\%$ Bondi-Hoyle accretion during CE, inefficient accretion 
onto BH/NS during stable mass transfer and capture from stellar winds (Sec.~\ref{sec.bhacc}), 
BH natal spins from MESA (eq.~\ref{eq.bhspin2}), SFRD and average metallicity evolution 
from \cite{Madau2017} (eqs.~\ref{eq.sfr1} and ~\ref{eq.Z}), and we adopt a low value for 
the solar metallicity $Z=0.014$. In this model we do not take into account additional 
effects of rotation as is done in model M20, we use the standard stellar winds from 
\cite{Vink2001} with addition of LBV winds from \cite{Belczynski2010b}, and we employ 
the initial binary parameters from \cite{Sana2012} as discussed in \cite{deMink2015}. 

Models M33 and M35 are the same as model M30, but mass independent NS/BH
natal kicks (not affected by fallback) are employed: high natal kicks with 1D
$\sigma=265\kms$ (M33) and intermediate natal kicks with 1D $\sigma=130\kms$
(M35).

Model M40 is the same as model M30, but employs a different model for BH
spins: Fuller model (eq.~\ref{eq.bhspin3}). Model M43 is the same as model
M40, but it employs mass independent NS/BH natal kicks with 1D $\sigma=265\kms$. 

Model M50 is the same as model M30, but employs mass loss rates reduced to
$30\%$ for all the stars (see Sec.~\ref{sec.Binary} for a justification). 

Model M60 is the same as model M30, but employs strong PPSN (eq.~\ref{eq.ppsn1}) 
with $10\%$ neutrino mass loss for both BH and NS.

Model M70 is the same as model M30, but employs moderate PPSN
(eq.~\ref{eq.ppsn2}). 

Models that allow CE with HG donors will be marked as submodels MXX.A,
while models that do not allow CE with HG donors will be marked as 
as MXX.B~\citep{Belczynski2007}.  

In all models we assume BH natal kicks to be randomly oriented, thus
generating BH spin misalignments with respect to the orbital angular momentum.
Note that our models are by no means exhaustive in terms of probing the
evolutionary uncertainties. However, they allow us to test the key parameters
that set BH spin magnitudes and misalignments, and thus the  effective spin:
angular momentum transport in massive stars, accretion onto the BH and its 
progenitor, tidal interactions and BH natal kicks. Table~\ref{tab.models} gives 
an overview of the models, and in Appendix~\ref{sec.Binary} we describe 
the details of our binary population synthesis calculations.

The {\tt StarTrack} population synthesis code is used to generate populations of 
BH-BH/BH-NS/NS-NS systems. The star's initial properties (mass and metallicity) 
are used to calculate stellar evolution. Binary interactions (mass gain/loss in 
RLOF and CE events) are taken into account through rejuvenation/de-rejuvenation in 
estimating the final stellar properties (mass and CO core mass) at the time of 
core-collapse, which are then used to obtain the NS/BH natal mass (see Appendix
~\ref{sec.COBH}). In binary calculations we use non-rotating stellar models
~\citep{Hurley2000}. We record whether any of the binary components accreted 
significant amounts of mass ($\gtrsim 10\%$ of its own mass). If accretion occurred 
during main sequence, and if the accreting star was of low metallicity ($Z<0.002$), 
we assume in models M20, M23, M25 and M26 that the star will produce a more 
massive CO core (by $20\%$), and thus a more massive NS/BH, to mimic the effects 
of increased mixing due to rapid rotation induced by accretion (see Appendix
~\ref{sec.Binary}).

At BH formation we use single stellar models to estimate the  BH natal spin
magnitude through the CO-core mass-spin relations proposed in Section~\ref{sec.spins}. 
In these estimates we assume that a star with a given CO core mass (as estimated 
from a binary evolution) forms a  BH with spin given by single stellar models. 
This scheme ignores the effects of mass accretion on the stellar spin of the BH 
progenitor that may increase the BH's natal spin. Obviously this is far from 
perfect, and stellar rotation models will need to be fully integrated into binary 
population synthesis in the future. However, this is impossible at the moment due 
to the lack of stellar models with consistent input physics that would appropriately 
sample mass, metallicity and rotation for massive stars and naked stellar cores (as 
stellar cores are much more often exposed in binary evolution than is single star 
evolution). 

Note also that we use moderately high initial stellar spins that assume $40\%$ of break-up 
velocity in single stellar models. Depending on a mass and stellar structure of a 
model, this gives a range of $250-450\kms$ initial rotation speeds at the equator (see 
Tab.~\ref{tab.spin1} and ~\ref{tab.spin2}). For comparison, the observed spins of 
massive stars show a bimodal distribution~\citep{Ramirez2013}, with one large peak at 
$\sim 100\kms$ and another small peak at $\sim 400\kms$.

A first-born BH can accrete mass from its unevolved companion either during
the RLOF/CE stages or from the companion's  winds. These effects are included in 
our calculations, and the BH spin is increased accordingly. The second-born
BH preserves its natal spin as it does not accrete mass. However, note that we 
allow for the possibility of the tidal spin up of BH progenitors and allow for
significant increase of the BH spin as compared to BH spins that result from single 
stellar models (Sec.~\ref{sec.tides}). 

In all cases, we assume that the stellar spins are initially aligned 
($\Theta_1=\Theta_2=0$) with the orbital angular momentum of the main--sequence star
binary. At BH formation we estimate the spin vector misalignment due to natal kicks. 
We allow for BH spin realignment neither during the  mass transfer phases, nor by 
tidal interaction between the stars in a binary. Note that with this approach we 
may be overestimating the BH spin misalignment, but only for models with high to 
moderate BH natal kicks (see Sec.~\ref{sec.misa} for details). 
\begin{table}
\caption{Binary evolution models}
\centering
\begin{tabular}{c| l}
\hline\hline
Model & Main features\\
\hline
\hline
{\bf M10} &  2016 standard input physics: \\
    & \ \ -- rapid SNa BH masses \citet{Fryer2012} \\
    & \ \ -- with strong PPSN and with PSN\\
    & \ \ -- $10\%$ neutrino mass loss at BH/NS formation\\ 
    & \ \ -- low-to-no BH natal kicks (set by fallback)\\
    & \ \ -- high NS kicks: $\sigma=265\kms$ with fallback\\
    & \ \ -- $50\%$ non-conservative RLOF\\
    & \ \ -- $10\%$ Bondi-Hoyle accretion onto NS/BH in CE\\
    & \ \ -- efficient accretion onto BH in stable MT/winds\\
    & \ \ -- no effects of rotation on stellar evolution\tablefootmark{a}\\
    & \ \ -- initial binary parameters: \citet{Sana2012}\\
    & \ \ -- massive star winds: \citet{Vink2001} + LBV\tablefootmark{b}\\
    & \ \ -- BH spins: Geneva models (eq.~\ref{eq.bhspin1})\\
    & \ \ -- SFRD($z$) and $Z(z)$: \cite{Madau2014}\\
    & \ \ -- solar metallicity: $\zsun=0.02$\\

\hline
{\bf M13} & as in M10, but with:\\
    & \ \ -- high BH/NS natal kicks: $\sigma=265\kms$\\

\hline
\hline
{\bf M20} & modified input physics, as in M10, but with: \\
    & \ \ -- $80\%$ non-conservative RLOF (Sec.~\ref{sec.Binary})\\
    & \ \ -- $5\%$ Bondi-Hoyle accretion onto NS/BH in CE\\
    & \ \ -- rotation increases CO core mass (by $20\%$)\\

\hline
{\bf M26} & as in M20, but with:\\
    & \ \ -- small BH/NS natal kicks: $\sigma=70\kms$\\

\hline
{\bf M25} & as in M20, but with:\\
    & \ \ -- intermediate BH/NS natal kicks: $\sigma=130\kms$\\

\hline
{\bf M23} & as in M20, but with:\\
    & \ \ -- high BH/NS natal kicks: $\sigma=265\kms$\\

\hline
\hline
{\bf M30} &  2019 standard input physics: \\
    & \ \ -- rapid SNa BH masses \citet{Fryer2012} \\
    & \ \ -- with weak PPSN and with PSN\\
    & \ \ -- $1\%$ neutrino mass loss at BH formation\\ 
    & \ \ -- $10\%$ neutrino mass loss at NS formation\\
    & \ \ -- low-to-no BH natal kicks (set by fallback)\\
    & \ \ -- high NS kicks: $\sigma=265\kms$ with fallback\\
    & \ \ -- $50\%$ non-conservative RLOF\\
    & \ \ -- $5\%$ Bondi-Hoyle accretion onto NS/BH in CE\\
    & \ \ -- inefficient accretion onto BH in stable MT/winds\\
    & \ \ -- no effects of rotation on stellar evolution\tablefootmark{a}\\
    & \ \ -- initial binary parameters: \citet{Sana2012}\\
    & \ \ -- massive star winds: \citet{Vink2001} + LBV\tablefootmark{b}\\
    & \ \ -- BH spins: MESA models (eq.~\ref{eq.bhspin2})\\
    & \ \ -- SFRD($z$) and $Z(z)$: \cite{Madau2017}\\
    & \ \ -- solar metallicity: $\zsun=0.014$\\

\hline
{\bf M33} & as in M30, but with:\\
    & \ \ -- high BH/NS natal kicks: $\sigma=265\kms$\\

\hline
{\bf M35} & as in M30, but with:\\
    & \ \ -- intermediate BH/NS natal kicks: $\sigma=130\kms$\\

\hline
\hline
{\bf M40} & as in M30, but with:\\
    & \ \ -- BH spins: Fuller model (eq.~\ref{eq.bhspin3})\\

\hline
{\bf M43} & as in M40, but with:\\
    & \ \ -- high BH/NS natal kicks: $\sigma=265\kms$\\

\hline
\hline
{\bf M50} & as in M30, but with:\\
    & \ \ -- $30\%$ of wind mass loss rates for all stars\\
\hline
\hline
{\bf M60} & as in M30, but with:\\
    & \ \ -- strong PPSN, $10\%$ neutrino mass loss for BH/NS\\

\hline
{\bf M70} & as in M30, but with:\\
    & \ \ -- moderate PPSN \\

\hline
\hline
\end{tabular}
\tablefoot{\\
\tablefootmark{a}{Stellar spins are followed (tides, magnetic braking, 
change of inertia), but rotation does not alter the star properties 
(He/CO core mass).}
\\
\tablefootmark{b}{Luminous Blue Variable winds: $1.5 \times 10^{-4}\mpy$.}
}
\label{tab.models}
\end{table}

\section{Results}

We have estimated the double compact object merger rate densities, merger detection 
rates, merger masses, and BH-BH effective spins using the methods presented in 
\cite{Belczynski2016a,Belczynski2016b} along with updates and revisions presented in 
the present study (see Sec.~\ref{sec.model}). 
In the following sections we will discuss some particular properties of our models. 
In our study, we do not exhaust the information that can be extracted from our models. 
Focusing on BH-BH mergers, we instead show some particular examples of what can be obtained 
with population synthesis modeling in context of the LIGO/Virgo sources. For now, 
we compare our models to the LIGO/Virgo observations showing that some models 
fit the data better than other, but only in terms of the observed allowed ranges 
of  rates, masses and effective spins. Note that we do not yet attempt to match  
particular distributions' shapes (for BH masses and effective spins) or the rate of 
increase of merger rates with redshift. However, anyone interested in such comparisons 
can easily perform them on our models as we make them publicly accessible through our 
website \url{www.syntheticuniverse.org}\footnote{The models will appear on this website
at the moment of the astro-ph submission.}.

\subsection{Binary Evolution of BH-BH Mergers}
\label{sec.evolution}

In this section we present several examples of binary evolution leading to the
formation of BH-BH mergers. In the framework of the Geneva model of angular
momentum transport, it is challenging to explain mergers with very low 
effective spin parameters as lots of BHs are formed with high or moderate 
spin magnitudes (see Fig.~\ref{fig.bhspin1}). Yet, it is not impossible. 
Therefore, we show examples of evolution that can lead to the formation of
merger resembling GW170104, which has one of the lowest measured effective spin
parameters: $-0.24<\chi_{\rm eff}<0.13$ ($90\%$ confidence limits).
In the framework of MESA angular momentum transport, it is challenging to 
explain mergers with moderate and high effective spin parameters as lots of 
BHs are formed with low spin magnitudes (see Fig.~\ref{fig.bhspin2}).
Yet, we show that we can form mergers that are consistent even with
GW170729, which is the merger that has the highest effective spin yet measured: 
$0.11<\chi_{\rm eff}<0.57$ ($90\%$ confidence limits).

\subsubsection{The case of GW170104}
\label{sec.gw170104}

\begin{figure}
\hspace*{-0.3cm}
\includegraphics[width=9.2cm]{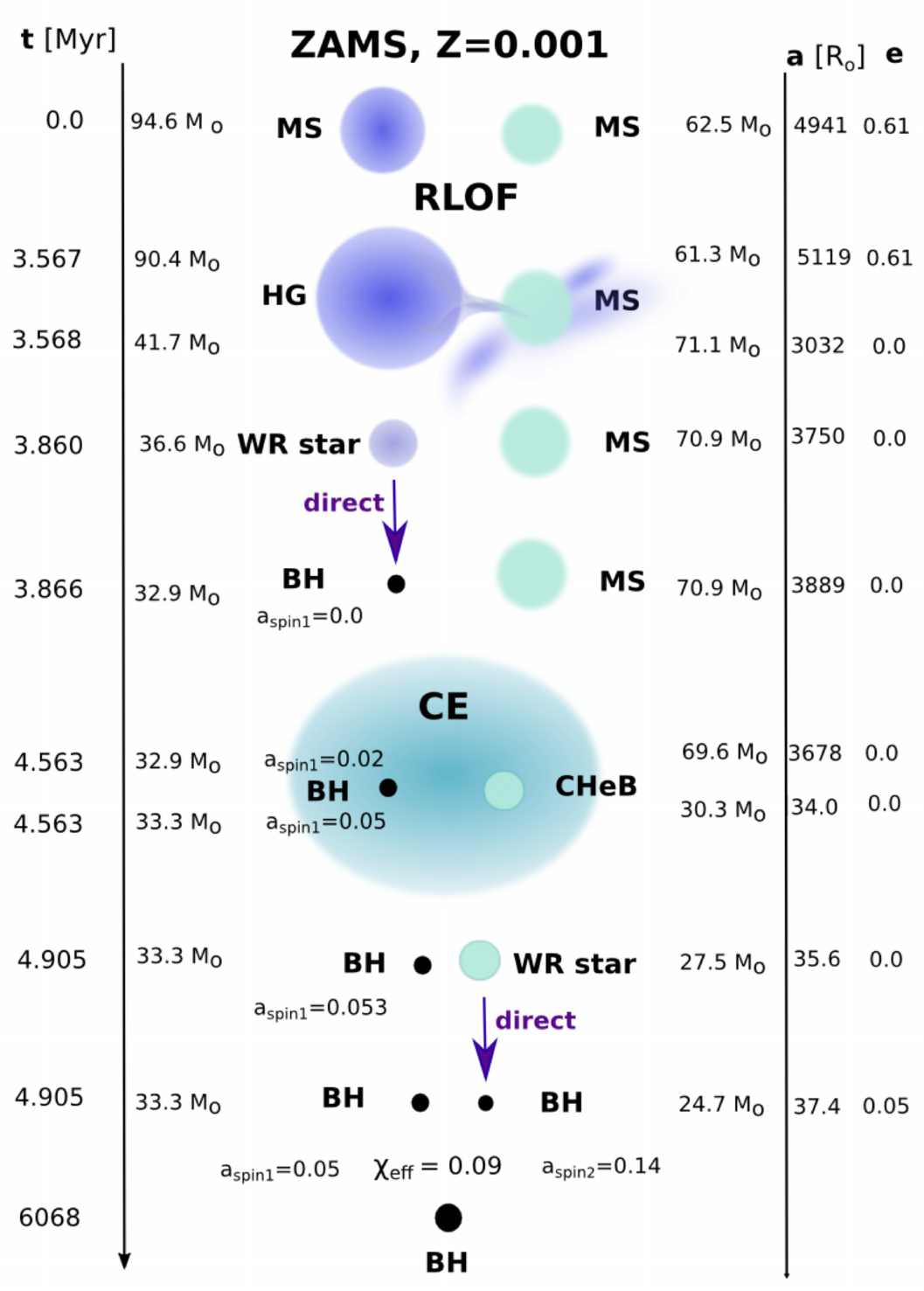}
\caption{Example of a possible route leading to the formation of a
  BH-BH merger similar to GW170104. This example follows the
  classical isolated binary evolution channel. In this model (M20.B) 
  we employ the Geneva BH natal spins, assume that  massive BHs do not receive natal 
  kicks and that their spins are aligned with the binary angular momentum 
  ($\Theta_1=\Theta_2=0^\circ$), producing an upper limit on the 
  effective spin parameter ($\chi_{\rm eff}$). Yet, this system has 
  $\chi_{\rm eff}=0.09$, within LIGO's $90\%$ credible limits for 
  GW170104 $[-0.24$:$0.13]$. Both BH masses are also within the limits: 
  $M_{\rm BH1}=31.0\msun$ $[25.4$:$38.2]$ and $M_{\rm BH2}=20.1\msun$ 
  $[15.6$:$25.0]$.} 
\label{fig.evol1}
\end{figure}

\begin{figure}
\hspace*{-0.3cm}
\includegraphics[width=9.2cm]{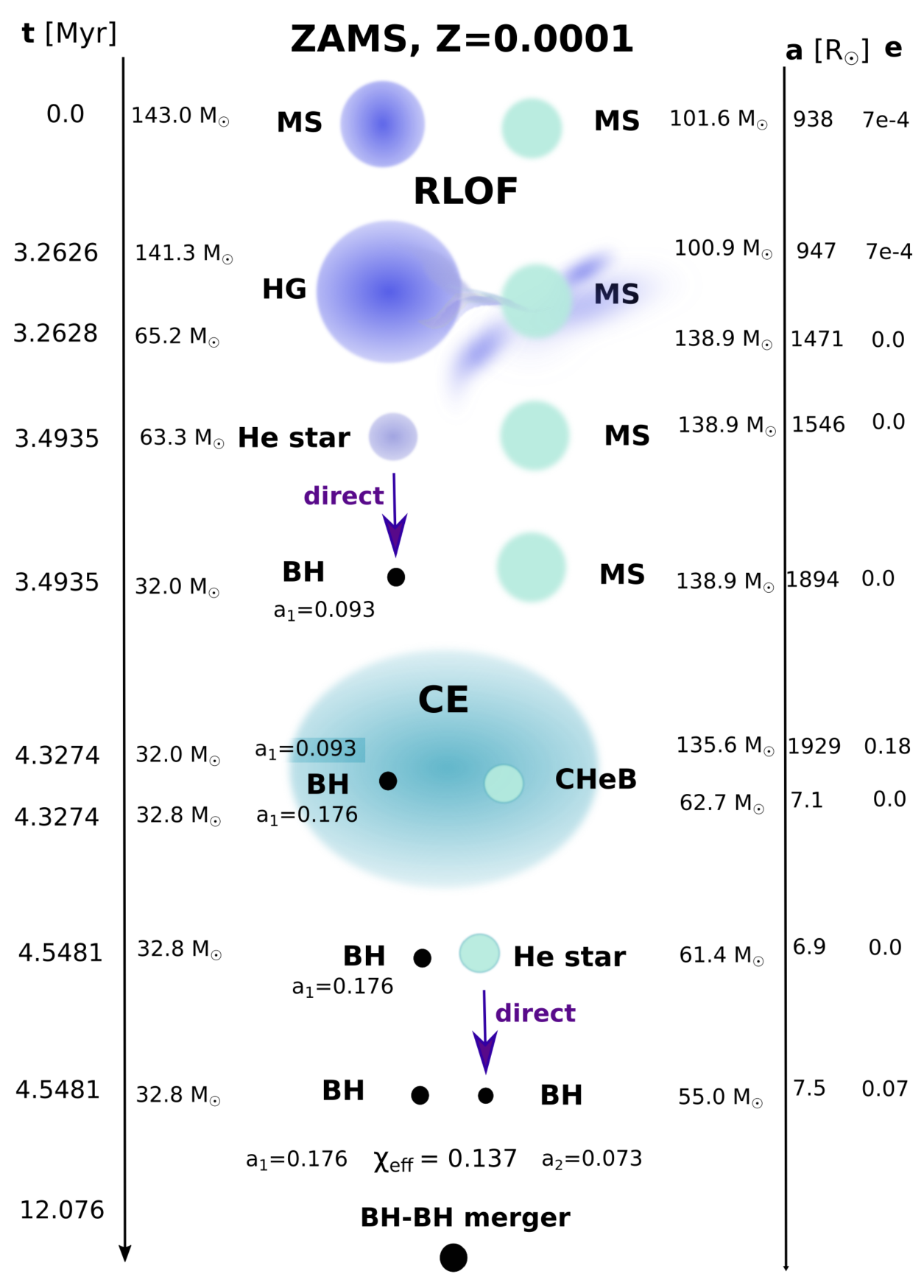}
\caption{Example of a possible route leading to the formation of a
  BH-BH merger similar to GW170729. This example follows  the
  classical isolated binary evolution channel. In this model (M30.B) 
  we employ the MESA BH natal spins, assume that  massive BHs do not receive natal 
  kicks and that their spins are aligned with the binary angular momentum 
  ($\Theta_1=\Theta_2=0^\circ$), producing an upper limit on the 
  effective spin parameter ($\chi_{\rm eff}$). Note that given the small
  binary separation at the BH-WR stage ($a \approx 7\rsun$), the WR star 
  is most likely going to become  tidally synchronized (see Sec.~\ref{sec.tides}). 
  If we assume that the  rapidly rotating WR star collapses into rapidly spinning BH 
  ($a_{\rm spin2} = 1$) then the effective spin of the presented system would 
  increase from $\chi_{\rm eff} = 0.137$ (no tidal spin-up; presented on the
  Figure) to $\chi_{\rm eff} = 0.484$ (full tidal spin-up). Both values are 
  within the LIGO/Virgo $90\%$ credible limits for GW170729 $[0.11$:$0.57]$. 
  Both BH masses are also within the limits: $M_{\rm BH1}=50.6\msun$ $[40.4$:$67.2]$ 
  and $M_{\rm BH2}=34.3\msun$ $[24.2$:$43.4]$.} 
\label{fig.evol2}
\end{figure}

Here we present a proof-of-principle scenario demonstrating that isolated 
binary evolution with the Geneva model of angular momentum transport can form 
a BH-BH merger with BH masses and effective spin compatible with LIGO's 
observation of GW170104, in particular, its very low effective spin.

Within our models with the Geneva angular momentum transport (M10, M13, M20,
M23, M25, M26) we search for systems with BH masses and effective spins 
within LIGO's $90\%$ credible limits: $25.4<M_{\rm BH1}<38.2\msun$, 
$15.6<M_{\rm BH2}<25.0\msun$, $-0.24<\chi_{\rm eff}<0.13$. The upper bound 
on $\chi_{\rm eff}$ may actually be as high as $\chi_{\rm eff} \approx 0.2$
(Appendix~\ref{sec.Xeff}). For example, within model M20 it is indeed 
possible to form a BH-BH merger resembling GW170104: $M_{\rm BH1}=33.3\msun$, 
$M_{\rm BH2}=24.7\msun$, $\chi_{\rm eff}=0.09$. The evolutionary history of 
such a merger is presented in Figure~\ref{fig.evol1}. 
Note that model M20 is rather conservative regarding assumptions on natal 
kicks, which are strongly suppressed by the fallback material. Massive BH 
spins are mostly aligned with the binary angular momentum 
($\cos \Theta_1=\cos \Theta_2=1$), thus maximizing the value of the 
effective spin. For all the other models (with the exception of M10) the 
BH spins will tend to be misaligned, decreasing the value of the effective 
spin and making it even easier to produce systems with low effective spin 
values (as observed in GW170104).

Although the BH spin can be modified by accretion from a binary companion,
the amount of matter accreted in our calculations is very modest, and the
accretion-induced spin-up of BHs is not significant. In the example shown
in Figure~\ref{fig.evol1}, the first-born massive BH forms with no spin
($a_{\rm spin1}=0$), then it accretes very little mass from its MS companion
wind increasing its spin only to $a_{\rm spin1}=0.02$. Most of the accretion 
occurs during the CE phase ($0.4\msun$) and the BH increases its spin to
$a_{\rm spin1}=0.05$. Finally, the BH accretes very little mass from its
Wolf-Rayet star companion wind, increasing its spin to its final value
of $a_{\rm spin1}=0.053$. The second-born BH forms with a natal spin of
$a_{\rm spin1}=0.14$ and it is not spun up, as it does not accrete any
material. In other words we predict that LIGO/Virgo observations of BH-BH
mergers will probe the natal BH spin distribution, up to evolutionary effects 
of order $0.05$ in dimensionless spin.

Note also that this particular system is not subject to a potential WR-star tidal 
spin-up in a BH-WR binary (the last evolutionary stage before the BH-BH formation). 
The system separation at this stage ($a \approx 35\rsun$) is too large for the 
tides to effectively spin-up WR star ($P_{\rm orb} > 1.3$d, see Sec.~\ref{sec.tides}).
The full details of this evolutionary example are given in Appendix~\ref{sec.GW170104}.

\subsubsection{The case of GW170729}
\label{sec.gw170729}

In a subset of our binary evolution models M30.A/B, M33.A/B, M35.A/B, M50.A/B, 
M60.A/B, and M70.A/B the natal BH spins are obtained from stellar models calculated 
with the MESA code under the assumption of efficient angular momentum transport 
in the stellar interiors (see Sect.~\ref{sec.spins} and Fig.~\ref{fig.bhspin2}). 
In this framework, the initial BH spins are always small ($a_{\rm spin}\lesssim0.15$), 
mostly independent of the progenitor mass and metallicity. Small natal BH spin 
values can in principle be increased during further evolution as result of 
mass accretion. However, in the isolated binary evolution channel for BH-BH 
mergers, the first formed BH can only accrete mass during the CE inspiral 
and through accretion of the wind from its stellar companion. We find that, 
in our simulations,
neither of those two processes leads to a significant increase of the BH spin 
 which is primarily the consequence of: {\em (i)} inefficient 
($5$-$10\%$) Bondi-Hoyle accretion rate onto BH in the short-lived CE phase 
\citep[e.g.,][]{MacLeod2017} and {\em (ii)} small wind mass loss rates from 
low metallicity stars~\citep{Vink2001}. This means that the small natal BH 
spins in the framework of efficient angular momentum transport in stellar 
interiors do result in small effective spin values of BH-BH mergers (typically 
$\chi_{\rm eff}\lesssim0.25$). 

Here we present a proof-of-principle scenario demonstrating that even the 
BH-BH merger event with the largest effective spin value reported to date 
(GW170729, $\chi_{\rm eff} = 0.36^{+0.21}_{-0.25}$) can be reconstructed in 
the isolated--binary evolution channel with small natal BH spin values given by
eq.~\ref{eq.bhspin2}. Within our models with efficient (MESA) angular momentum 
transport we search for systems with BH masses and effective spin within LIGO's 
$90\%$ credible limits for GW170729: $40.4<M_{\rm BH1}<67.2\msun$, 
$24.2<M_{\rm BH2}<43.4\msun$, $0.11<\chi_{\rm eff}<0.57$. For example, within 
the model M30.B it is indeed possible to form a BH-BH merger resembling GW170729: 
$M_{\rm BH1}=55.0\msun$, $M_{\rm BH2}=32.8\msun$, $\chi_{\rm eff}=0.137$. The 
evolutionary history of such a merger is presented in Figure~\ref{fig.evol2}. 
Both the BHs where formed in direct collapse with small but non-zero natal 
spins ($a_{\rm spin1} = 0.093$ and $a_{\rm spin2} = 0.073$). For reasons 
discussed above, the BH formed first accreted only a very modest amount of 
mass during binary evolution ($<1.0\msun$), which led to an increase of its 
spin to $a_{\rm spin1} = 0.176$. The final effective spin of the merging BH-BH 
system is $\chi_{\rm eff}=0.137$, and thus it is in (marginal) agreement with 
the $90\%$ credible limits for GW170729. 

It should be noted that the spin value for the secondary BH presented in the
above example (Fig.~\ref{fig.evol2}; $a_{\rm spin2} = 0.073$) is most likely 
underestimated because this example neglects the potential tidal spin-up of the
secondary WR star (direct progenitor of the second BH) during the BH-WR stage 
(see Sec.~\ref{sec.tides}). Since the orbital separation during the BH-WR 
phase is very small ($a \approx 7\rsun$), tidal interactions are expected to 
be efficient in spinning-up the WR star. In fact, in the case of this system 
the timescale of WR tidal synchronization as approximated by \citet[][see our 
Eq.~\ref{eq.tides}]{Zaldarriaga2017} is only $t_{\rm sync} \approx 11$ yr,
which is a few orders of magnitude shorter than the duration of BH-WR stage 
($\sim 2 \times 10^5$ yr). Thus, the secondary WR star has most likely become 
fully synchronized with the orbit by the time it collapsed to form the second 
BH. If we assume that such a rapidly rotating WR star collapses into an also 
rapidly spinning BH ($a_{\rm spin2}=1.0$) then the effective spin of the BH-BH 
merger in Fig.~\ref{fig.evol2} would become $\chi_{\rm eff}=0.484$, still well 
within the $90\%$ credible limits for GW170729. The large range of $\chi_{\rm eff}$ 
values that spans from $0.137$ to $0.484$ encloses all the possible results of  
the uncertain process of tidal spin-up in BH-WR binaries and its effect on the 
secondary BH spin.

We caution that the simple fact that a particular BH-BH system can be reproduced 
in a given binary evolution channel does not in itself guarantee consistency 
between population synthesis results and observations. Such consistency can only 
be tested by comparing the entire populations. We discuss the distributions of 
BH-BH mergers parameters obtained in our simulation in the following sections.

\begin{figure}
\hspace*{-0.3cm}
\includegraphics[width=9.2cm]{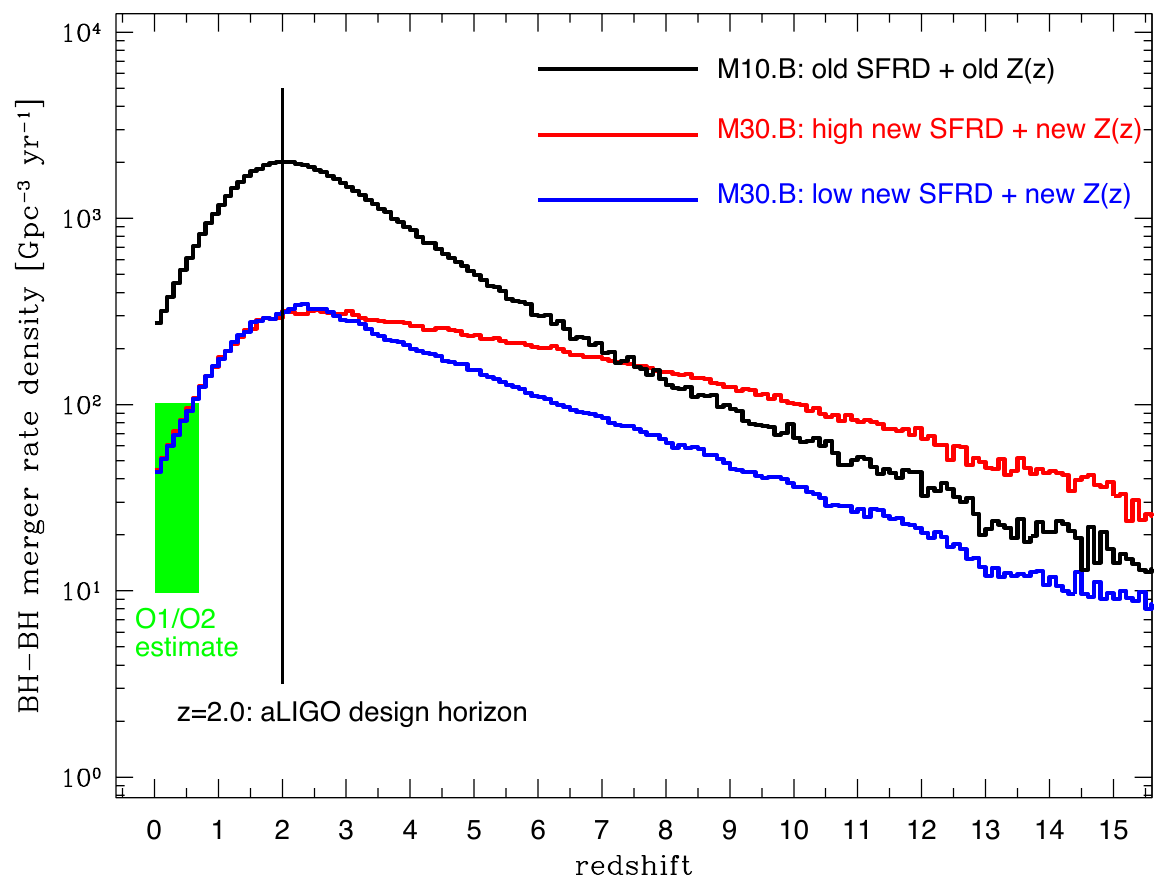}
\caption{Merger rate density of BH-BH mergers from Population I/II stars. 
We employ models M10.B and M30.B to illustrate the effects of our assumptions 
on the star formation rate and cosmic metallicity evolution on the BH-BH mergers. 
LIGO/Virgo O1/O2 constraint on the BH-BH merger rate in local Universe is also 
shown. Note that the decrease of the BH-BH merger rate density at low-redshift 
(from old to new models) is due to the average metallicity of stars at any given 
redshift in old models, being much lower than the average metallicity of stars in 
new models (see Sec.~\ref{sec.rates} for details).
}
\label{fig.sfrrates}
\end{figure}

\begin{figure}
\hspace*{-0.3cm}
\includegraphics[width=9.2cm]{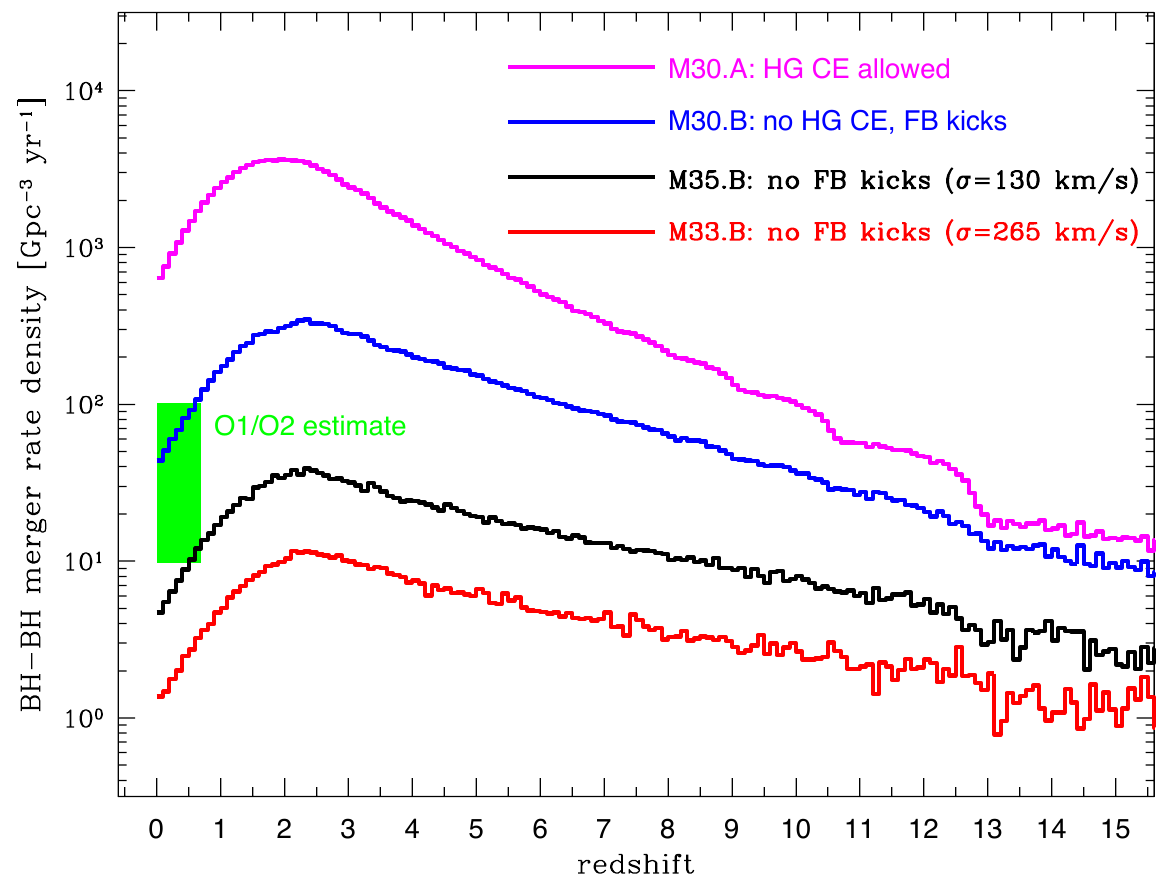}
\caption{Merger rate density of BH-BH mergers for various assumptions about the
common envelope and BH natal kicks. The possibility of development and survival 
of the CE phase with a Hertzsprung gap donor, increases the low-redshift BH-BH 
merger rate density by $\sim 1$ order of magnitude: compare models M30.A and M30.B 
(a similar effect is found for other models). Natal kicks tend to decrease the 
BH-BH merger rate density by $\sim 1.5$ order of magnitude: from fall-back 
attenuated natal kicks (almost no BH natal kicks: model M30.B) to full scale BH 
natal kicks (as high as observed for Galactic single pulsars: model M33.B). Note 
the degeneracy between the tested input physics; by applying high natal kicks we
can bring model M30.A down to agree with the LIGO/Virgo estimate, or by allowing
for HG CE we can bring models M35.B and M33.B up to also match the LIGO/Virgo
constraint.  
}
\label{fig.kickrates}
\end{figure}

\begin{figure}
\hspace*{-0.3cm}
\includegraphics[width=9.2cm]{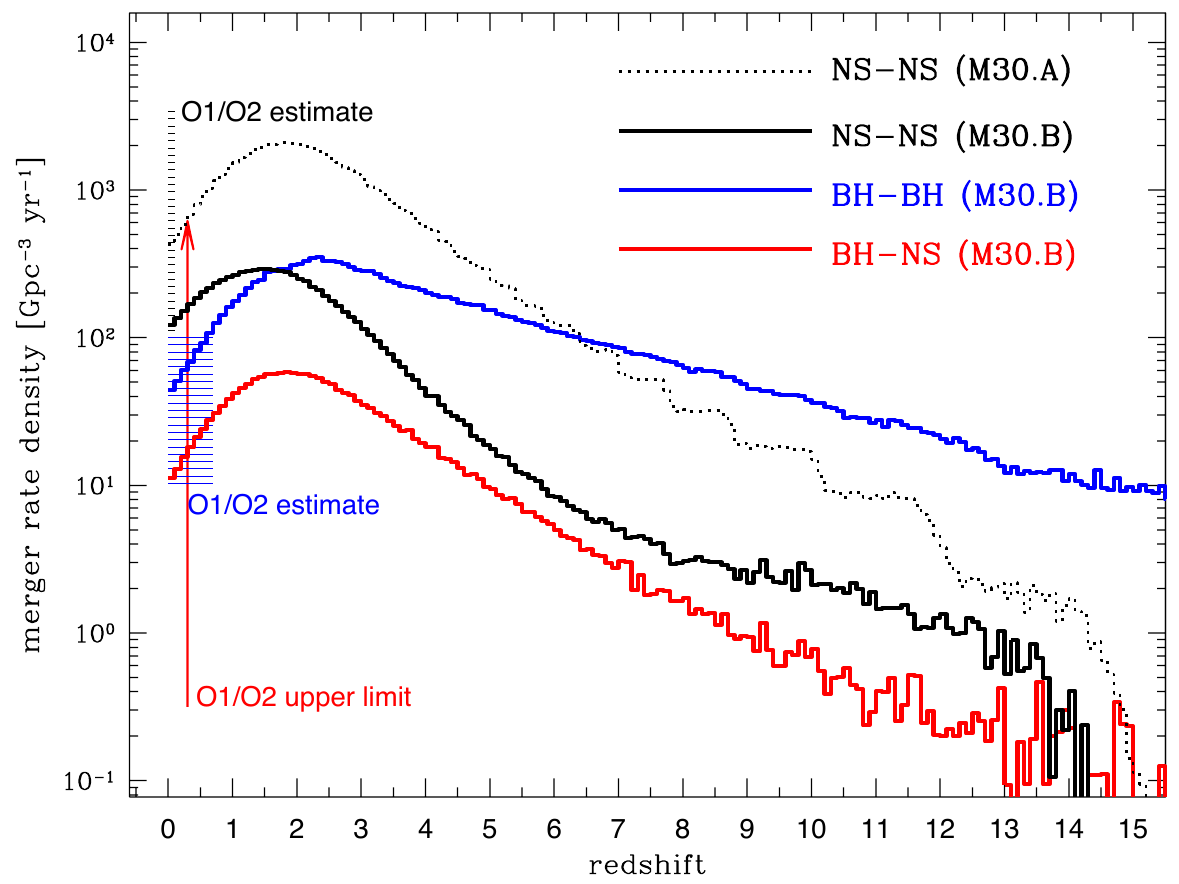}
\caption{Merger rate density of NS-NS, BH-NS and BH-BH mergers for model M30. 
For comparison, we show LIGO/Virgo O1/O2 constraints on merger rate densities. 
While actual estimates are available for the BH-BH and NS-NS mergers, only an 
upper limit is available for the BH-NS mergers from O1/O2 data (but see Sec.
~\ref{sec.app_BHNS_limits}). Note that the presented merger rate densities are 
consistent with the LIGO/Virgo estimates. 
}
\label{fig.dcorates}
\end{figure}

\begin{table}
\caption{Local merger rate densities and LIGO/Virgo detection rates.} 

\centering
\begin{tabular}{c| c| c c c}
\hline\hline
Model & Merger & Rate density\tablefootmark{a} & detection rate\tablefootmark{b} & $z_{\rm hor}$\tablefootmark{c}\\
      &        & [Gpc$^{-3}$ yr$^{-1}$] & [yr$^{-1}$] &   \\
\hline
\hline
{\bf M10.B}  & NS-NS         &  65.2      &   0.289 & 0.081 \\
             & BH-NS         &  28.8      &   2.419 & 0.404 \\
             & {\bf BH-BH}   & {\bf 274}  &   536.2 & 0.961 \\
\hline
{\bf M13.B}  & NS-NS         &  49.3      &   0.226 & 0.082 \\
             & BH-NS         &  2.43      &   0.246 & 0.397 \\
             & {\bf BH-BH}   & {\bf 8.63} &   16.60 & 1.031 \\
\hline
\hline
{\bf M20.B}  & NS-NS         &  84.3      &   0.358 & 0.081  \\
             & BH-NS         &  96.6      &   6.525 & 0.405  \\
             & {\bf BH-BH}   & {\bf 469}  &   783.7 & 0.955  \\
\hline
{\bf M26.B}  & NS-NS         &  107       &   0.453 & 0.082  \\
             & BH-NS         &  42.4      &   2.880 & 0.401  \\
             & {\bf BH-BH}   & {\bf 132}  &   192.8 & 0.958  \\
\hline
{\bf M25.B}  & NS-NS         &  105       &   0.447 & 0.081  \\
             & BH-NS         &  17.2      &   1.486 & 0.426  \\
             & {\bf BH-BH}   & {\bf 62.8} &   105.4 & 0.953  \\
\hline
{\bf M23.B}  & NS-NS         &  53.0      &   0.225 & 0.080  \\
             & BH-NS         &  4.17      &   0.337 & 0.397  \\
             & {\bf BH-BH}   & {\bf 16.6} &   26.29 & 1.024  \\
\hline
\hline
{\bf M30.B}  & NS-NS         &  122       &   0.514 & 0.081  \\
             & BH-NS         &  11.1      &   0.800 & 0.433  \\
             & {\bf BH-BH}   & {\bf 43.7} &   50.95 & 1.152  \\
\hline
{\bf M35.B}  & NS-NS         &  143       &   0.608 & 0.081  \\
             & BH-NS         &  3.10      &   0.206 & 0.444  \\
             & {\bf BH-BH}   & {\bf 4.69} &   6.627 & 1.152  \\
\hline
{\bf M33.B}  & NS-NS         &  79.6      &   0.333 & 0.080  \\
             & BH-NS         &  0.48      &   0.031 & 0.392  \\
             & {\bf BH-BH}   & {\bf 1.37} &   1.714 & 1.147  \\
\hline
\hline
{\bf M40.B}  & NS-NS         &  122       &   0.508 & 0.081  \\
             & BH-NS         &  10.2      &   0.752 & 0.391  \\
             & {\bf BH-BH}   & {\bf 39.2} &   42.98 & 1.152  \\
\hline
{\bf M43.B}  & NS-NS         &  76.7      &   0.319 &  0.080 \\
             & BH-NS         &  0.61      &   0.044 &  0.381 \\
             & {\bf BH-BH}   & {\bf 1.24} &   1.578 &  1.150 \\
\hline
\hline
{\bf M50.B}  & NS-NS         &  93.2      &   0.395 & 0.081 \\
             & BH-NS         &  8.54      &   0.702 & 0.489 \\
             & {\bf BH-BH}   & {\bf 64.2} &   133.4 & 1.153 \\
\hline
{\bf M60.B}  & NS-NS         &  118       &   0.499 & 0.080 \\
             & BH-NS         &  12.1      &   0.817 & 0.363 \\
             & {\bf BH-BH}   & {\bf 46.2} &   41.88 & 0.949 \\
\hline
{\bf M70.B}  & NS-NS         &  120       &   0.505 & 0.081  \\
             & BH-NS         &  10.7      &   0.765 & 0.396  \\
             & {\bf BH-BH}   & {\bf 43.9} &   51.37 & 1.107  \\
\hline

\end{tabular}
\tablefoot{\\
\tablefootmark{a}{Local merger rate density at redshift $z=0$ (note rate
increase with $z$).}\\
\tablefootmark{b}{Detection rate for LIGO/Virgo mid-high range (O3 proxy).}\\
\tablefootmark{c}{Redshift for the most distant source detectable (best
located/oriented).\\}
}
\label{tab.ratesB}
\end{table}

\begin{table}
\caption{Local merger rate densities and LIGO/Virgo detection rates.} 

\centering
\begin{tabular}{c| c| c c c}
\hline\hline
Model & Merger & Rate density\tablefootmark{a} & detection rate\tablefootmark{b} & $z_{\rm hor}$\tablefootmark{c}\\
      &        & [Gpc$^{-3}$ yr$^{-1}$] & [yr$^{-1}$] &   \\
\hline
\hline
{\bf M10.A}  & NS-NS         & 230         &   1.022 & 0.083  \\
             & BH-NS         & 99.6        &   6.289 & 0.404  \\
             & {\bf BH-BH}   & {\bf 1368}  &    3464 & 0.973  \\
\hline
{\bf M13.A}  & NS-NS         & 193         &   0.858 & 0.083  \\
             & BH-NS         & 23.3        &   1.179 & 0.396  \\
             & {\bf BH-BH}   & {\bf 52.6}  &   77.19 & 1.031  \\
\hline
\hline
{\bf M20.A}  & NS-NS         & 260         &   1.094 & 0.082  \\
             & BH-NS         & 272         &   15.85 & 0.405  \\
             & {\bf BH-BH}   & {\bf 1285}  &    2765 & 0.973 \\
\hline
{\bf M26.A}  & NS-NS         & 374         &   1.604 & 0.082  \\
             & BH-NS         & 297         &   15.78 & 0.402  \\
             & {\bf BH-BH}   & {\bf 531}   &   706.6 & 0.958  \\
\hline
{\bf M25.A}  & NS-NS         & 310         &   1.333 & 0.082  \\
             & BH-NS         & 152         &   8.272 & 0.426  \\
             & {\bf BH-BH}   & {\bf 226}   &   292.0 & 0.953  \\
\hline
{\bf M23.A}  & NS-NS         & 204         &   0.858 & 0.082  \\
             & BH-NS         & 45.4        &   2.280 & 0.397  \\
             & {\bf BH-BH}   & {\bf 50.8}  &   63.02 & 1.024  \\
\hline
\hline
{\bf M30.A}  & NS-NS         &  426        &   1.789 & 0.083  \\
             & BH-NS         &  113        &   6.451 & 0.433  \\
             & {\bf BH-BH}   & {\bf 641}   &    1023 & 1.152  \\
\hline
{\bf M35.A}  & NS-NS         &  524        &   2.237 & 0.082  \\
             & BH-NS         &  73.3       &   3.526 & 0.444  \\
             & {\bf BH-BH}   & {\bf 109}   &   108.2 & 1.152  \\
\hline
{\bf M33.A}  & NS-NS         &  321        &   1.329 & 0.082  \\
             & BH-NS         &  21.1       &   0.935 & 0.392  \\
             & {\bf BH-BH}   & {\bf 19.5}  &   19.41 & 1.149  \\
\hline
\hline
{\bf M40.A}  & NS-NS         &  434        &   1.822 & 0.082  \\
             & BH-NS         &  114        &   6.531 & 0.422  \\
             & {\bf BH-BH}   & {\bf 627}   &   947.3 & 1.152  \\
\hline
{\bf M43.A}  & NS-NS         &  323        &   1.346 &  0.082 \\
             & BH-NS         &  21.5       &   0.944 &  0.381 \\
             & {\bf BH-BH}   & {\bf 19.5}  &   20.07 &  1.151 \\
\hline
\hline
{\bf M50.A}  & NS-NS         &  408        &   1.691 & 0.082 \\
             & BH-NS         &  101        &   5.595 & 0.489 \\
             & {\bf BH-BH}   & {\bf 1114}  &    2843 & 1.153 \\
\hline
{\bf M60.A}  & NS-NS         &  429        &   1.805 & 0.082 \\
             & BH-NS         &  108        &   5.727 & 0.363 \\
             & {\bf BH-BH}   & {\bf 640}   &   837.0 & 0.949 \\
\hline
{\bf M70.A}  & NS-NS         &  428        &   1.805 & 0.082  \\
             & BH-NS         &  108        &   6.107 & 0.408  \\
             & {\bf BH-BH}   & {\bf 637}   &    1017 & 1.107  \\
\hline

\end{tabular}
\tablefoot{\\
\tablefootmark{a}{Local merger rate density at redshift $z=0$ (note rate
increase with $z$).}\\
\tablefootmark{b}{Detection rate for LIGO/Virgo mid-high range (O3 proxy).}\\
\tablefootmark{c}{Redshift for the most distant source detectable (best
located/oriented).\\}
}
\label{tab.ratesA}
\end{table}

\subsection{Merger Rate Density and Detection Rate} 
\label{sec.rates}

Tables~\ref{tab.ratesB} and ~\ref{tab.ratesA} summarize our predictions for the 
local (redshift $z \sim 0$) merger rate density, the detection rate for LIGO/Virgo's 
mid-high sensitivity curve (that may be taken as an approximation of O3  observing 
run) along with the maximum horizon redshift for the best located/oriented source in 
a given merger type category: NS-NS, BH-NS, BH-BH.  

The predicted BH-BH merger rate densities vary between $1.24$ and $1368\gpy$ across 
our models. A number of models (M13.A, M23.A/B, M25.B, M26.B, M30.B, M33.A, M40.B,
M43.A, M50.B, M60.B, M70.B) produce rates within the allowable range determined by the 
first 10 LIGO/Virgo detections ($9.7$--$101\gpy$: \cite{LIGO2019b}).
The predicted NS-NS merger rate densities vary between $49.3$ and $524\gpy$ for the 
tested models. A number of models (M10.A,M13.A, M20.A, M23.A, M25.A, M26.A, M30.A/B, 
M33.A, M35.A/B, M40.A/B, M43.A, M50.A, M60.A/B, M70.A/B) produce rates within 
allowable range determined by the first LIGO/Virgo detection ($110$--$3840\gpy$: 
\cite{LIGO2019b}).
The predicted BH-NS merger rate densities vary between $0.48$ and $297\gpy$ for the 
tested models. So far all the models produce rates within an upper limit determined 
by the non-detection in the O1/O2 LIGO/Virgo runs ($<610\gpy$: \cite{LIGO2019b}).

Figure~\ref{fig.sfrrates} shows the intrinsic BH-BH merger rate density (not weighted 
by LIGO/Virgo detection probability) evolution with redshift for models M10.B and 
M30.B that employ different star formation rate and cosmic evolution of metallicity. 
We note that model M10.B which employs $Z(z)$ measured from stars generates higher 
(by factor of $\sim 5$ at low redshifts $z<2$) rates than the model M30.B that employs 
$Z(z)$ as measured by the metallicity of star forming gas at any given redshift.  The 
choice of $Z(z)$ is one of the most important factors affecting local (low $z$) merger 
rate densities for BH-BH mergers. This is because with decreasing metallicity 
we expect {\em (i)} an  increase of the BH mass~\citep{Belczynski2010b} and {\em (ii)} 
an increase of the BH-BH merger formation efficiency per unit mass~\citep{Belczynski2010a}. 
Therefore, models with lower $Z(z)$ result in higher BH-BH merger rates. 

Note that the difference in SFRD($z$) between the two models does not affect significantly 
the rates as the differences in star formation are relatively small (see Fig.
~\ref{fig.sfr}). Note also that for the two updated (new) models of the SFRD($z$), there 
is virtually no difference in the BH-BH merger rates at low redshifts ($z<2$), while 
differences ($\sim$ factor of a few) begin to appear only at higher redshifts. For our 
updated models we use a SFRD($z$) with low star formation at high redshifts. This 
particular choice does not affect any of our conclusions for LIGO/Virgo as these 
instruments are not expected to probe sources at high redshifts ($z>2$).

There are different PPSN models employed in models M10.B (strong PPSN; low maximum 
BH mass) and M30.B (weak PPSN; high maximum BH mass). However, this does not affect 
significantly the intrinsic merger rate density in a given volume that is presented in
Figure~\ref{fig.sfrrates}. This is because the number of BH-BH binaries is about
the same in each of the PPSN models. However, the differences in BH masses affect notably the
detection rates. For example, compare the detection rate of BH-BH mergers in models M30.A/B 
and M60.A/B that differ only by PPSN input physics (see Tables~\ref{tab.ratesB} and 
~\ref{tab.ratesA}). 

Figure~\ref{fig.kickrates} demonstrates the effects of the CE phase and natal kicks on 
the intrinsic BH-BH merger rate density. We use a sequence of four models with exactly 
the same input physics that differ only in various assumptions about the CE phase and natal 
kicks. With the optimistic approach to the CE and natal kicks (that allows for survival of 
the CE phase with HG donors and very low or zero BH natal kicks; model M30.A) the local 
($z=0$) merger rate density is relatively high: $R_{\rm BHBH}=641\gpy$. When we apply a 
more restrictive approach to the CE phase, but keep the same natal kicks as above (CE 
survival not allowed for HG donors; model M30.B) then we note a significant decrease in 
the rate: $R_{\rm BHBH}=43.7\gpy$. Adding to the above moderate natal kicks and keeping the
restrictive CE-phase approach (model M35.B) decreases the rate further: $R_{\rm BHBH}=4.69\gpy$. 
An additional increase of natal kicks (model M33.B) with the same restrictive CE-phase 
approach, brings the rate to a very low value: $R_{\rm BHBH}=1.37\gpy$. Comparison with 
the LIGO/Virgo rate estimate shows that models M30.A, M35.B, M33.B are excluded. It means, 
that within our limited sample of evolutionary models, some combinations of CE-phase 
approach and natal kicks can be excluded as not likely. For example, moderate to high natal 
kicks and a restrictive CE-phase treatment are an unlikely combination. Note, however, that 
we cannot draw conclusions about the rates based solely on the natal kicks or solely on the 
CE-phase approach, because the results are degenerate with respect to these two major 
factors. For example, if we apply the  optimistic CE-phase approach to models with moderate 
to  high natal kicks then both these models fit within (or very close to) the LIGO/Virgo 
rate estimate. In particular, we find $R_{\rm BHBH}=109\gpy$ for model M35.A (moderate 
natal kicks and optimistic CE=phase) and $R_{\rm BHBH}=19.5\gpy$ for model M33.A (high 
natal kicks and optimistic CE-phase). 

Figure~\ref{fig.dcorates} shows the intrinsic merger rate density for all types of mergers: 
NS-NS, BH-NS and BH-BH. As an example we use model M30.A/B. As discussed above, the 
local BH-BH intrinsic merger rate density for model M30.B ($R_{\rm BHBH}=43.7\gpy$) is 
within the LIGO/Virgo estimate. The predicted local BH-NS intrinsic merger rate density 
for model M30.B ($R_{\rm BHNS}=11.1\gpy$) is within the LIGO/Virgo upper limit.
The local intrinsic merger rate density for NS-NS systems is only just above LIGO/Virgo 
$90\%$ level lower limit for model M30.B ($R_{\rm NSNS}=122\gpy$), while it is
well within the LIGO/Virgo estimate for model M30.A ($R_{\rm NSNS}=426\gpy$). 

Detection rates (Tables~\ref{tab.ratesB} and ~\ref{tab.ratesA}) are based on each 
source merger redshift, its mass (we use mass dependent waveforms for each merger) 
and take into account the LIGO detector antenna (peanut-shaped) pattern. For this particular 
estimate we employ LIGO sensitivity labeled "mid-high" that approximately corresponds 
to the current (O3) LIGO sensitivity (for details see Appendix~\ref{sec.GW}). The detection 
rate is then a convolution of the merger rate density (and its change with redshift within the
LIGO horizon for a given source type) and the mass distribution of given source type (NS-NS, 
BH-NS, BH-BH). For clarity, we list the horizon redshift for the best located/oriented 
(directly overhead source with an orbital plane perpendicular to the line of sight) source 
within each merger type. This naturally corresponds to the furthest detectable redshift 
(distance) of the most massive source within each merger type found in our simulations.

\subsection{BH Masses}
\label{sec.bhmass}

In this section we discuss the masses of BHs in BH-BH mergers, focusing on either the 
individual component masses, $M_{\rm BH1}$ (more massive component) and $M_{\rm BH2}$ 
(less massive component), or the total merger mass, $M_{\rm tot}=M_{\rm BH1}+M_{\rm BH2}$.

\begin{figure}
\hspace*{-0.3cm}
\includegraphics[width=9.2cm]{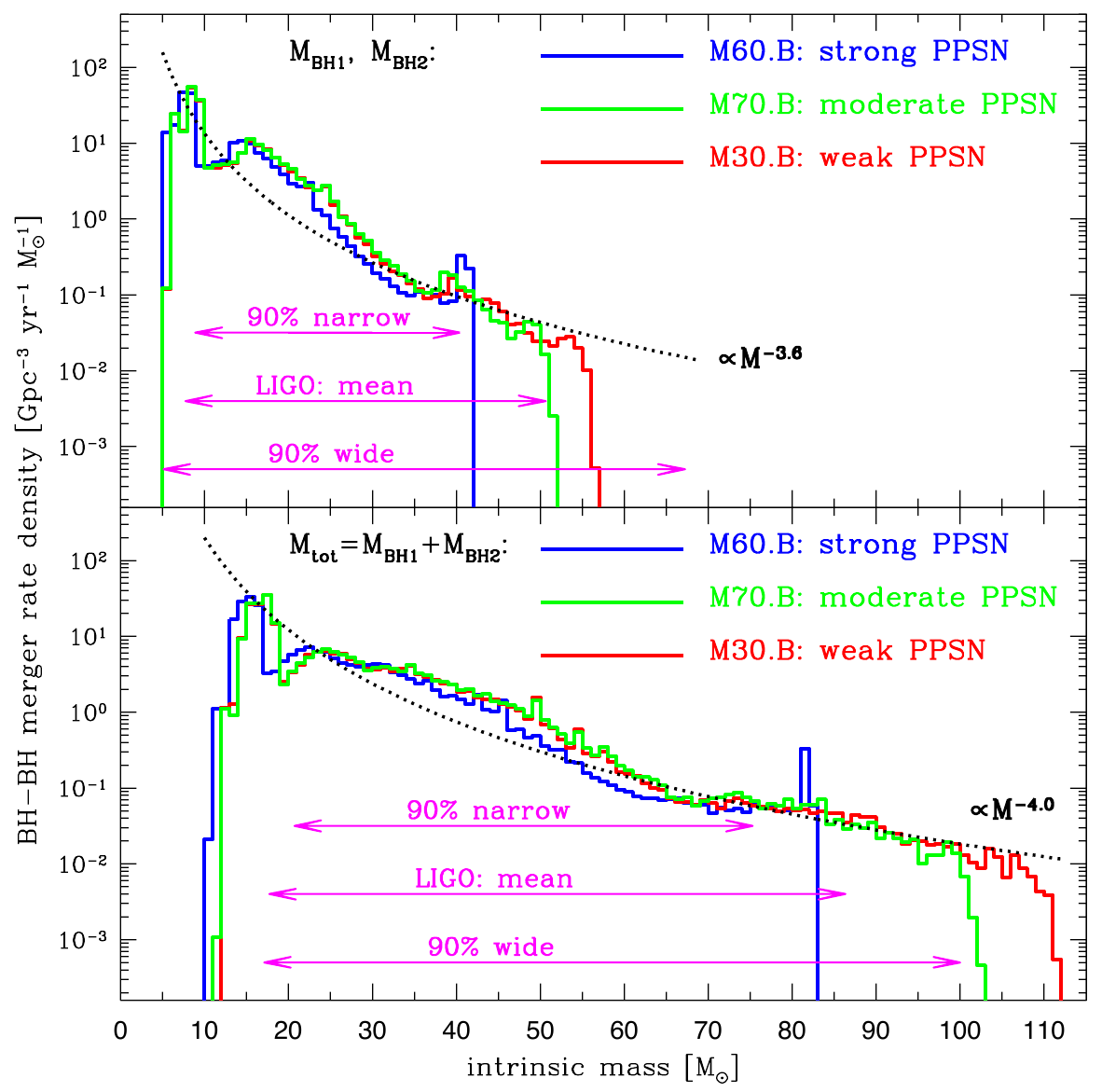}
\caption{BH masses in BH-BH mergers within the design advanced--LIGO horizon  
($z<2$) for models M60.B (strong pair-instability pulsation supernovae), 
M70.B (moderate PPSN), and M30.B (weak PPSN). The top panel shows the distribution 
of individual BH masses, while the bottom panel shows the  total BH-BH system mass.
All distributions are intrinsic (neither redshifted nor weighted by detection
probability). For comparison, we also indicate the range of LIGO/Virgo O1/O2 mean 
mass estimates and their most narrow and the widest allowed range within 
$90\%$ confidence limits. 
Individual BH masses may reach $\sim 40\msun$ (M60.B) and $\sim 55\msun$ 
(M30.B), while the total BH-BH mass may reach $\sim 80\msun$ (M60.B) and 
$\sim 110\msun$ (M30.B). Note that these distributions only very
approximately resemble power-laws: $\sim M^{-3.6}$ (for individual BH masses) and 
$\sim M^{-4.0}$ (for total BH-BH mass). Power-law fits (dashed black lines) 
were performed for model M30.B in the log-log space. 
}
\label{fig.mass1a}
\end{figure}

Figure~\ref{fig.mass1a} shows the intrinsic (source frame) distributions of the individual 
BH masses ($M_{\rm BH1}$ and $M_{\rm BH2}$ shown together in one distribution) and the total 
BH-BH merger masses for our three models of PPSN: M30.B (weak PPSN), M70.B (moderate PPSN),
and M60.B (strong PPSN). These distributions are weighted by the intrinsic merger rate
density for all BH-BH mergers (independent of mass) within a given redshift ($z<2$).
As clearly seen, the different PPSN treatments in the models shown here impact the maximum 
BH mass generated. In our calculations, the individual BH masses in merging binaries extend to 
$M_{\rm BH,max} \sim 40\msun$ in M60.B, $M_{\rm BH,max} \sim 50\msun$ in M70.B, and
$M_{\rm BH,max} \sim 55\msun$ in M30.B. For comparison, among all the candidates reported 
by LIGO/Virgo (see Tab.~\ref{tab.ligodata}), the largest mean mass for any individual BH is 
$50.6\msun$ (for GW170729); even most optimistically (within the $90\%$ credible limits), the 
largest BH  reported in 
a binary is only  $67.2\msun$ (again for GW170729). The total BH-BH merger mass reaches 
$M_{\rm BH,max} \sim 80\msun$ in M60.B, $M_{\rm BH,max} \sim 100\msun$ in M70.B, and
$M_{\rm BH,max} \sim 110\msun$ in M30.B. For comparison, LIGO/Virgo total mean BH-BH mass 
estimates reach $86.3\msun$, while the $90\%$ confidence level on the most massive BH-BH merger 
is as high as $100\msun$.
The distributions approximately resemble steep power-laws: $\sim M^{-3.6}$ for individual 
BH mass, and $\sim M^{-4.0}$ for total BH-BH mass. [Unless otherwise noted, all exponents 
refer to a power law fit to all merging binaries, covering the full mass range.] The power 
law index depends weakly on the PPSN model. 

\begin{figure}
\hspace*{-0.3cm}
\includegraphics[width=9.2cm]{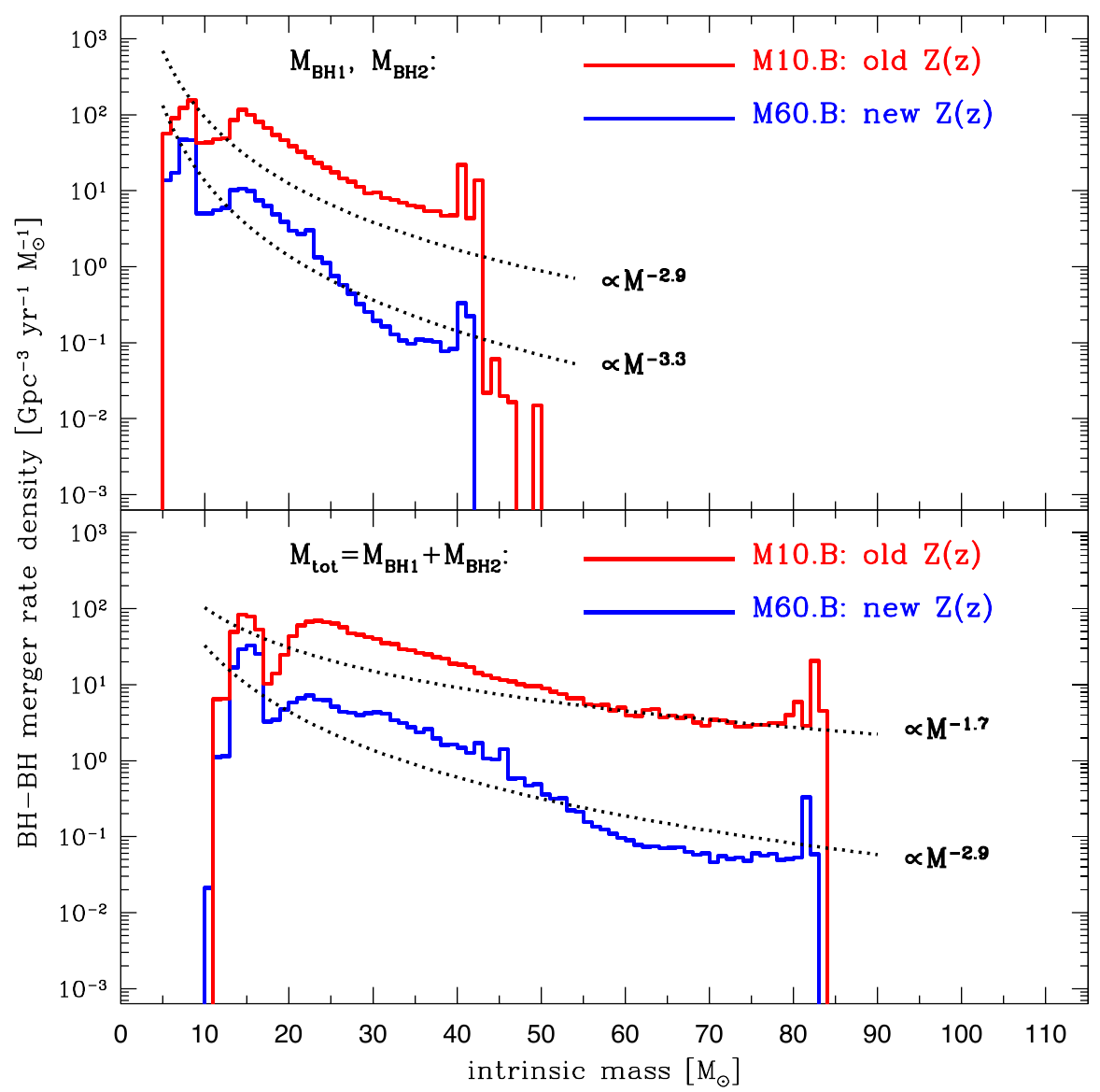}
\caption{BH masses in BH-BH mergers within design advanced--LIGO sensitivity
($z<2$) for models M10.B (that employs old average metallicity cosmic
evolution) and M60.B (new average metallicity evolution). BH masses are
calculated with the same formulas in both models. Top panel shows distribution 
of individual BH masses, while the bottom panel shows the total BH-BH system mass.
All distributions are intrinsic (not redshifted nor weighted by detection
probability). Note that application of new metallicity evolution with redshift 
(less low-Z stars: less high mass BHs) results in somewhat steeper BH mass 
distributions as contrasted with application of old metallicity evolution 
(more low-Z stars: more high mass BHs). Power-law fits (dashed black lines) 
were performed for both models in the log-log space.
}
\label{fig.mass1b}
\end{figure}

Figure~\ref{fig.mass1b} illustrates the effect of different  $Z(z)$ assumptions on 
source frame BH mass distributions. We choose models M10.B and M60.B that utilize the 
same strong PPSN model, but employ slow $Z(z)$ evolution based on an older estimate 
and faster $Z(z)$ evolution based on a more recent estimate, respectively. For both 
models the range of BH masses (whether individual BH masses or total BH-BH merger 
masses) is very similar. The same PPSN model leads to a similar cutoff BH mass at the 
high mass end.  However, we note the change in steepness of the power-laws which 
approximate these distributions. For individual BH masses the slope changes from 
$\sim M^{-2.9}$ for model M10.B to $\sim M^{-3.3}$ for model M60.B. For total BH-BH 
masses the slope changes from $\sim M^{-1.7}$ for model M10.B to $\sim M^{-2.9}$ for 
model M60.B. This comes from the fact that in models with high numbers of low 
metallicity stars (e.g., M10.B) BH masses are (on average) higher and therefore the 
distributions are flatter than in models with small number of low metallicity stars 
(e.g., M60.B).

\begin{figure}
\hspace*{-0.3cm}
\includegraphics[width=9.2cm]{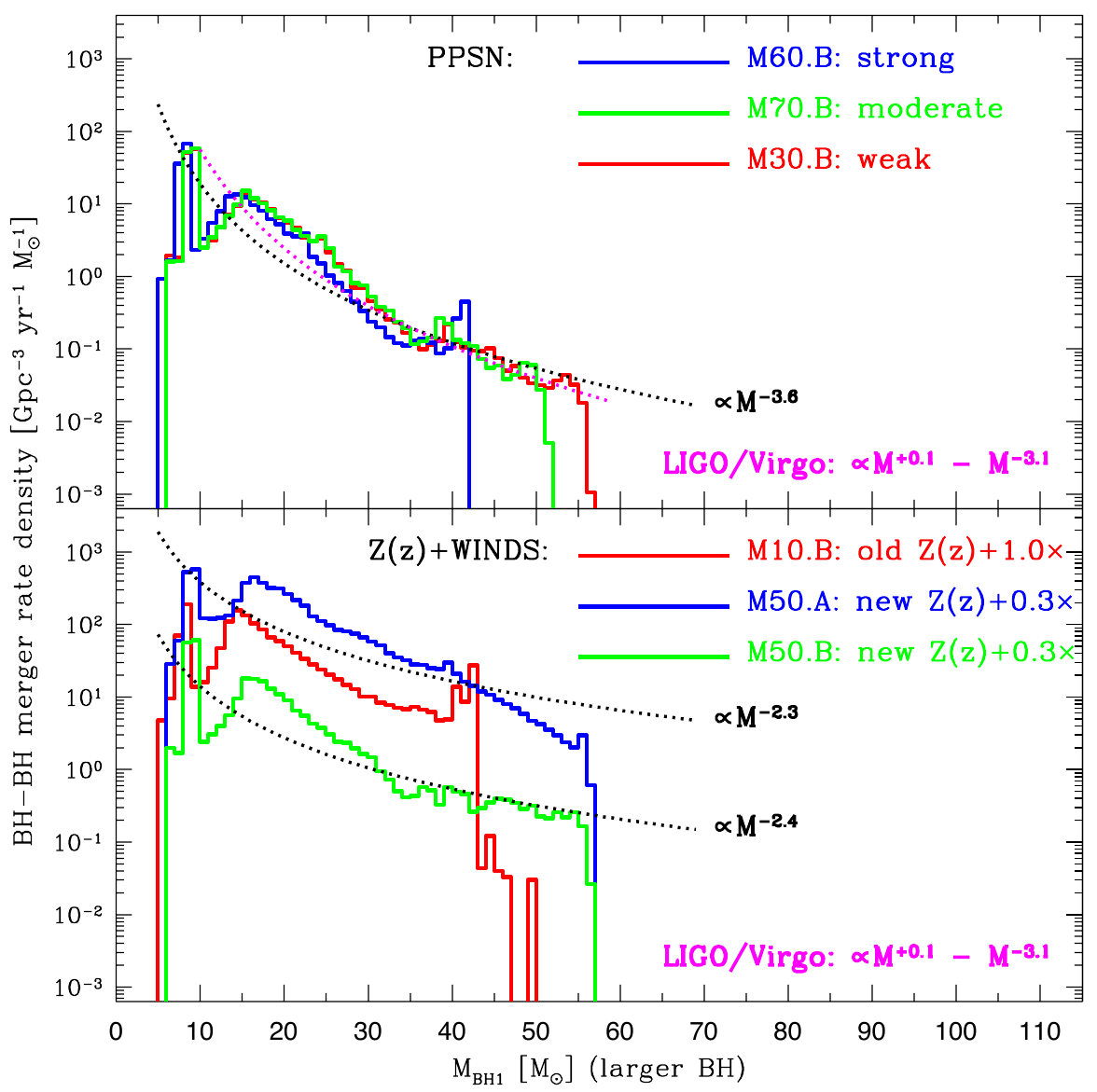}
\caption{The larger (more massive) BH mass distributions in BH-BH mergers within design 
advanced--LIGO sensitivity ($z<2$) for models M60.B, M70.B, M30.B, M10.B, M50.A, M50.B. 
Power-law fits (dashed black lines) were performed for model M30.B in the top panel 
($\propto M^{-3.6}$), and for models M50.A ($\propto M^{-2.3}$) and M50.B 
($\propto M^{-2.4}$) in the bottom panel in the log-log space. These distributions can be 
compared with other theoretical predictions (e.g., Fig.~1 of \cite{Mapelli2019}, or Fig.~4 
of \cite{Stevenson2019} that seem to be close to $\propto M^{-2} - M^{-3}$; note that 
power-laws shown in the latter work are not correct) or with the LIGO/Virgo observational 
estimate of $\propto M^{-1.6}$ with large uncertainty on the power-law index: 
$\propto M^{+0.1}-M^{-3.1}$ (Fig.~1 in \cite{LIGO2019a}). Some of our models (e.g., 
top panel) show somewhat steeper relations, while other models (bottom panel) are in 
agreement with LIGO/Virgo estimate. 
}
\label{fig.mass1c}
\end{figure}

In Figure~\ref{fig.mass1c} we show the intrinsic (source frame) distributions of the more 
massive BH mass ($M_{\rm BH1}$) weighted by the intrinsic merger rate density for all 
BH-BH mergers (independent of mass) within a given redshift ($z<2$). We present two 
groups of models. In one, we alter the amount of PPSN mass-loss: M60.B (strong 
mass-loss), M70.B (moderate mass-loss) and M30.B (weak mass-loss). All these models 
tend to have similar and rather steep power-law distribution of larger BH mass 
($\propto M^{-3.6}$). In the second group we show models that tend to produce 
shallower power-law distributions ($\propto M^{-2.4}$): M10.B (old $Z(z)$ relation
with standard stellar winds and restrictive CE treatment), M50.A (new $Z(z)$ relation 
with $30\%$ reduced stellar winds and optimistic CE), and M50.B (new $Z(z)$ relation 
with $30\%$ reduced stellar winds and restrictive CE). Clearly models that tend to 
produce more low metallicity stars (e.g., M10) and models with reduced stellar winds 
(e.g., M50) generate more massive BHs and produce shallower distributions. 
However, our mass distributions are not perfect power laws, particularly at low mass. 
For example, a power law approximation to our simulation data restricted to the mass 
range of reported LIGO/Virgo observations will recover a steeper power law exponent. 
For comparison, LIGO/Virgo infers a phenomenological pure power law distribution of 
$\propto M^{+0.1}-M^{-3.1}$ (\cite{LIGO2019a} based on observations spanning $M_{\rm BH1}$; 
see their Fig.1). When comparing our calculations to LIGO/Virgo observations in this 
way, the estimated power law exponent depends sensitively on the choice of  the low-mass 
cutoff $M_{\rm BH1}$, where  our models predict a substantial structure in the mass 
distribution. 

\begin{figure}
\hspace*{-0.3cm}
\includegraphics[width=9.2cm]{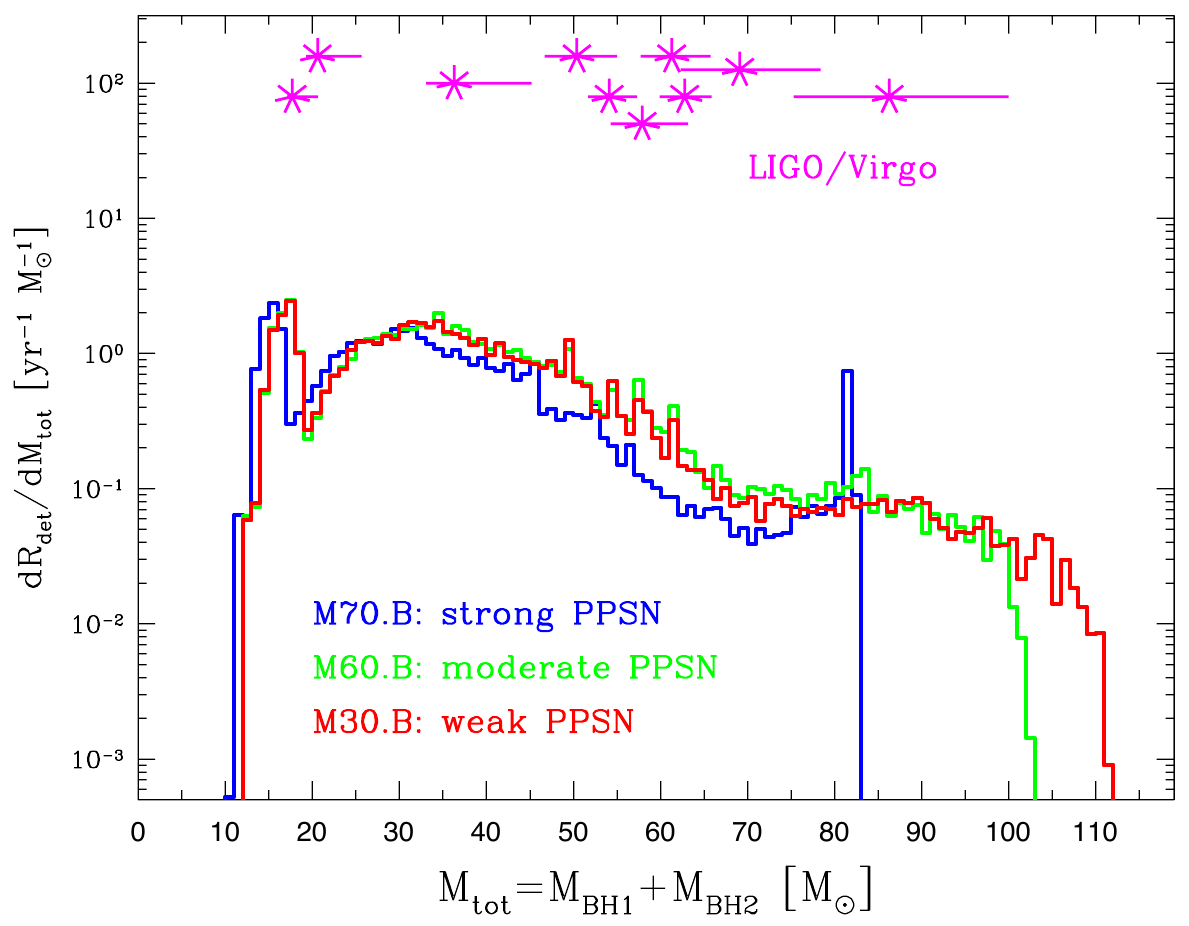}
\caption{Detection weighted BH-BH merger total intrinsic (not redshifted) mass for 
models with  different assumptions on the pair-instability pulsation supernovae: M70.B 
(strong PPSN), M60.B (moderate PPSN), M30.B (weak PPSN). For comparison, we also 
show LIGO/Virgo O1/O2 mean total mass estimates and their $90\%$ confidence limits. 
}
\label{fig.mass2}
\end{figure}

Figure~\ref{fig.mass2} shows the total BH-BH merger mass distributions weighted by the LIGO
detection probability (dependent on mass). We show models with different assumptions
on PPSN (M30.B, M60.B, M70.B) and contrast them with LIGO/Virgo observations.
These distributions are flatter than the intrinsic mass distributions as more massive
mergers are detected in larger volumes (and thus in larger numbers) than lower mass
mergers. Naturally, the maximum BH-BH mass cut-offs are the same as for the intrinsic
distributions. Note that within observational uncertainties, all these three models
can reproduce the range of the detected BH-BH total masses. However, a closer inspection
of the distribution of observational points shows an overabundance of points near
$M_{\rm tot} \sim 50$--$70\msun$ with respect to the models. At this point we do not
attempt to match this apparent peak in the observed distribution (this can wait for 
LIGO/Virgo to publish another set of detections), but we have just tested one potential 
change to the input physics (see below) that may affect the shape of the total mass 
distribution for massive BH-BH mergers.

\begin{figure}
\hspace*{-0.3cm}
\includegraphics[width=9.2cm]{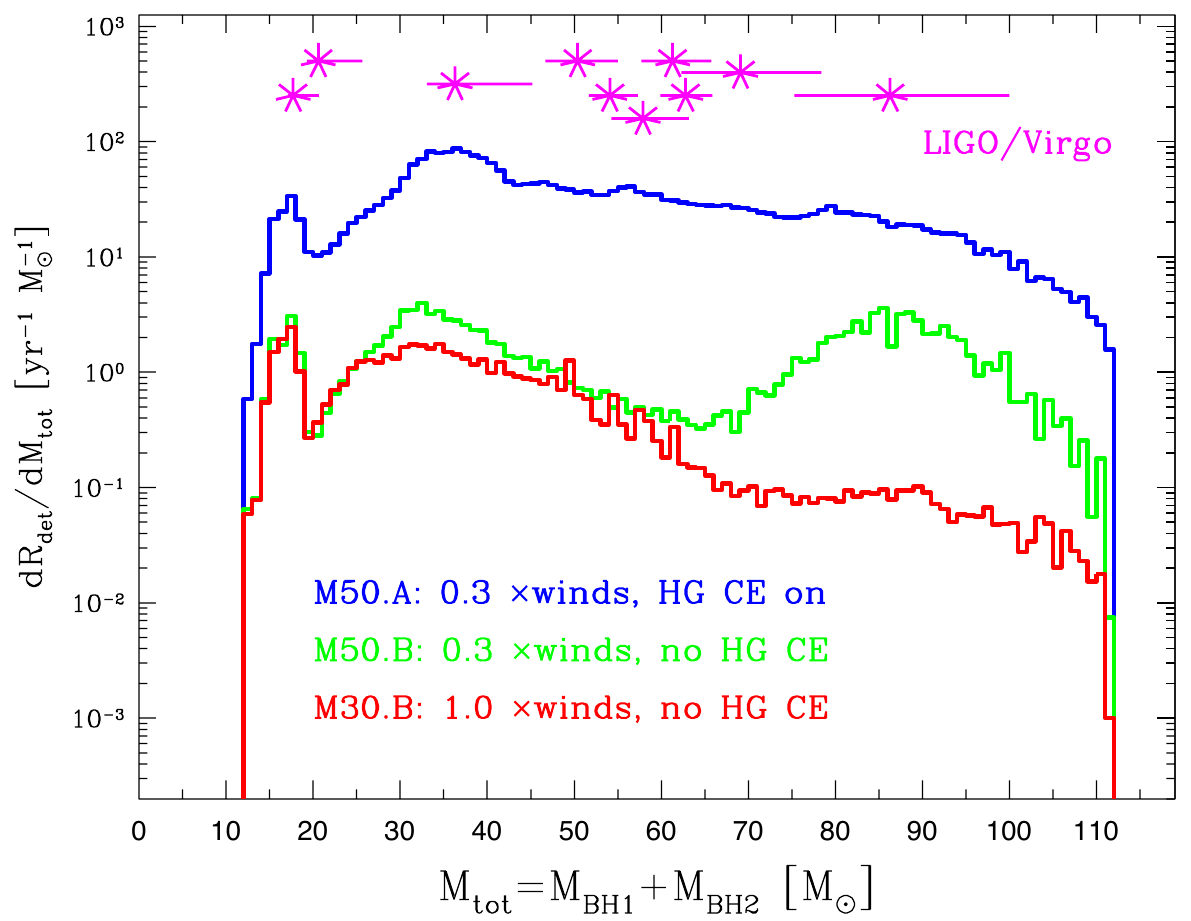}
\caption{Detection weighted BH-BH merger total intrinsic (not redshifted) mass 
for models with  different assumptions on wind mass loss and common envelope development: 
M50.B (weak stellar winds and optimistic CE), M50.B (weak stellar winds and 
pessimistic CE), M30.B (strong stellar winds and pessimistic CE). For comparison, 
we also show LIGO/Virgo O1/O2 mean total mass estimates and their $90\%$ confidence 
limits. Note that although no model seems to reproduce the observed
LIGO/Virgo BH-BH total mass distribution, one can expect that some interplay of
several parameters (winds, CE, PPSN, Z(z)) may possibly in future reproduce
the shape of the observed distribution. For example, stronger PPSN and
higher Z(z) (not shown here) will tend to shift the highest BH masses to lower 
values. This calls for further parameter study calculations.  
}
\label{fig.mass3}
\end{figure}

In Figure~\ref{fig.mass3} we re--plot the model M30.B (with weak PPSN and restrictive 
CE-phase treatment) employing the standard theoretical estimates of wind mass-loss rates 
(rather high wind mass-loss rates; \cite{Vink2001}). Additionally, we show the distributions 
for models M50.A and M50.B for which we reduce all stellar wind mass loss to $30\%$ of the
values used in all other models, but the rest of the input physics is exactly the same as 
in model M30.A/B. We note the appearance of a very prominent peak in the total BH-BH mass 
distribution at $M_{\rm tot} \sim 80$--$100\msun$ in model M50.B (weak PPSN and restrictive 
CE-phase treatment) as compared to the model M30.B distribution. That peak disappears in 
model M50.A (weak PPSN and optimistic CE-phase treatment), but the distribution becomes 
much flatter than for model M30.B. These models seem to better reproduce the distribution 
of observational points. In Sec.~\ref{sec.dismasses} we present a brief discussion 
comparing synthetic {\tt StarTrack} data with LIGO/Virgo data on BH masses.

\subsection{BH-BH Effective Spins}             
\label{sec.xeff}

In this section we present our predictions for the effective spins of BH-BH mergers. 
Having shown that we can reproduce the effective spins (whether low or high) with 
some of our adopted BH spin models (Geneva or MESA), we wish to see whether we 
can also reproduce the effective spin distribution (or rather range of observed 
values) of \textsl{all} the reported LIGO/Virgo BH-BH mergers. 

\begin{figure}
    \includegraphics[width=\columnwidth]{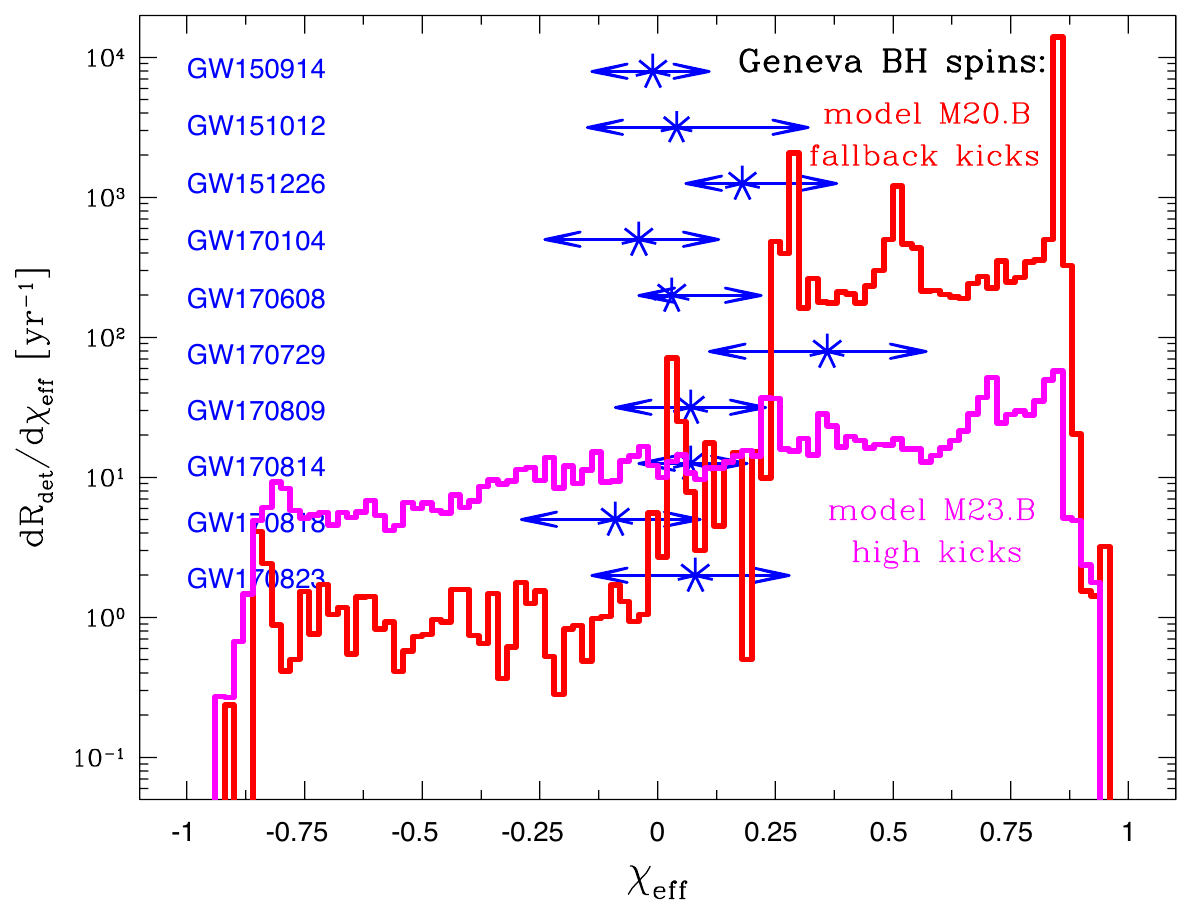}
\caption{
Detection weighted distribution of effective spin parameter of BH-BH mergers for model 
M20.B (fallback decreased BH kicks; no BH kicks for massive BHs) and M23.B (high BH natal 
kicks with $1D\ \sigma=265 \kms$ for all BHs) with Geneva mildly efficient angular 
momentum transport.
Note that both distributions peak at high values ($\chi_{\rm eff} \sim 0.9$).
Natal kicks decrease the effective spin parameter (average $\chi_{\rm eff}=0.3$; M23.B)
as compared to model with almost no BH kicks (average $\chi_{\rm eff}=0.7$; M20.B).
For comparison we show the $90\%$ credible limits (blue arrows) and the most likely values
(blue stars) of the effective spin parameter for ten LIGO/Virgo BH-BH mergers.
Although we can recover all the reported values, the predicted peak of $\chi_{\rm eff}$
distribution is not coincident with the current LIGO/Virgo data. It indicates that BHs
have typically lower spins than resulting from the Geneva model for BH natal spins, or that
the detected BH-BH mergers are not formed in the classical isolated binary evolution.
}
\label{fig.xeff1}
\end{figure}

In Figure~\ref{fig.xeff1} we show the measurements of the effective spins in the LIGO/Virgo 
BH-BH merger observations, superimposed with  the $\chi_{\rm eff}$ predictions 
calculated using the BH natal spins from the Geneva evolutionary calculations. All of the 
LIGO/Virgo observations cluster around zero ($\chi_{\rm eff}\sim0$). For the 
comparison we use two models: model M20.B, with fallback-decreased BH natal kicks 
(effectively no kicks for massive BHs, and small kicks for low mass BHs) and the model 
M23.B, with high BH kicks that are independent of the BH mass (with 1-dimensional 
$\sigma=265\kms$). Our model distributions peak at high values 
($\chi_{\rm eff} \sim 0.9$) and then quickly fall off toward small values (note 
logarithmic scale on Fig.~\ref{fig.xeff1}).

Mergers with high values of the effective spin host BHs with high spins. In the
Geneva model, these BHs can form with either low or high masses, depending on 
the chemical composition of the progenitor stars (Fig.~\ref{fig.bhspin1}). 
Mergers with low effective spins originate predominantly from systems with 
high-mass BHs. Strong natal kicks (as in model M23.B) produce higher misalignment 
of the BH spins with respect to the binary's angular momentum vector during the BH 
formation, and this decreases the value of the effective spin. Note that although 
both distributions are similar and peak at high values, natal kicks decrease 
the effective spin parameter (average $\chi_{\rm eff}=0.3$; M23.B) as compared to 
model with almost no BH kicks ($\chi_{\rm eff}=0.7$; M20.B). The fraction of BH-BH 
mergers with negative $\chi_{\rm eff}$ is sizeable in model M23.B ($27.4\%$), while 
it is negligible in model M20.B ($\sim 0.3\%$).

We can recover all the reported values within both models (more easily with model 
M23.B than with M20.B), but our predicted $\chi_{\rm eff}$ distributions using 
Geneva spins as inputs are not consistent with the current LIGO/Virgo data. 
For example, for each of our models and for $N$ observations, we can 
ask whether we would expect at least one measurement with a maximum value 
of $\chi_{\rm eff}$ above what has been observed thus far; i.e., 
$1-P(<\chi_{\rm eff})^N$, where $P(<\chi_{\rm eff})$ is the cumulative 
distribution implied by the detection-weighted $\chi_{\rm eff}$ distribution 
reported in Figure~\ref{fig.xeff1}. Even considering only subsets of events, 
such as the high-mass or low-mass events, we find a small ($1.1 \times 10^{-4}$ 
for M23.B) to infinitesimal ($6.7 \times 10^{-10}$ for M20.B) probability that 
the predicted spin distribution is compatible with observations reported to 
date.

Note that to reach this conclusion we have used two extreme natal kick models. 
In particular, we have tested a model with high BH natal kicks, that tends to 
maximize the BH spin-orbit misalignment and therefore decrease our predicted 
effective spins. Although BH natal kicks as high as those adopted in model M23.B 
(with average 3-dimensional velocity of $\sim 400 \kms$) cannot yet be observationally 
excluded~\citep{Belczynski2016a}, it is unlikely for BHs to receive larger natal 
kicks than adopted in this model~\citep{Mandel2016b,Repetto2017}. Note that in these 
models, we allow neither for any processes that could realign the BH spins, nor for 
an effective tidal spin-up of stars in the binary progenitors of BH-BH mergers. These 
processes might increase the effective spin parameter for some BH-BH progenitors, 
thus shifting our results for Geneva model of BH natal spins even further away from 
the LIGO/Virgo observations.

\begin{figure}
    \includegraphics[width=\columnwidth]{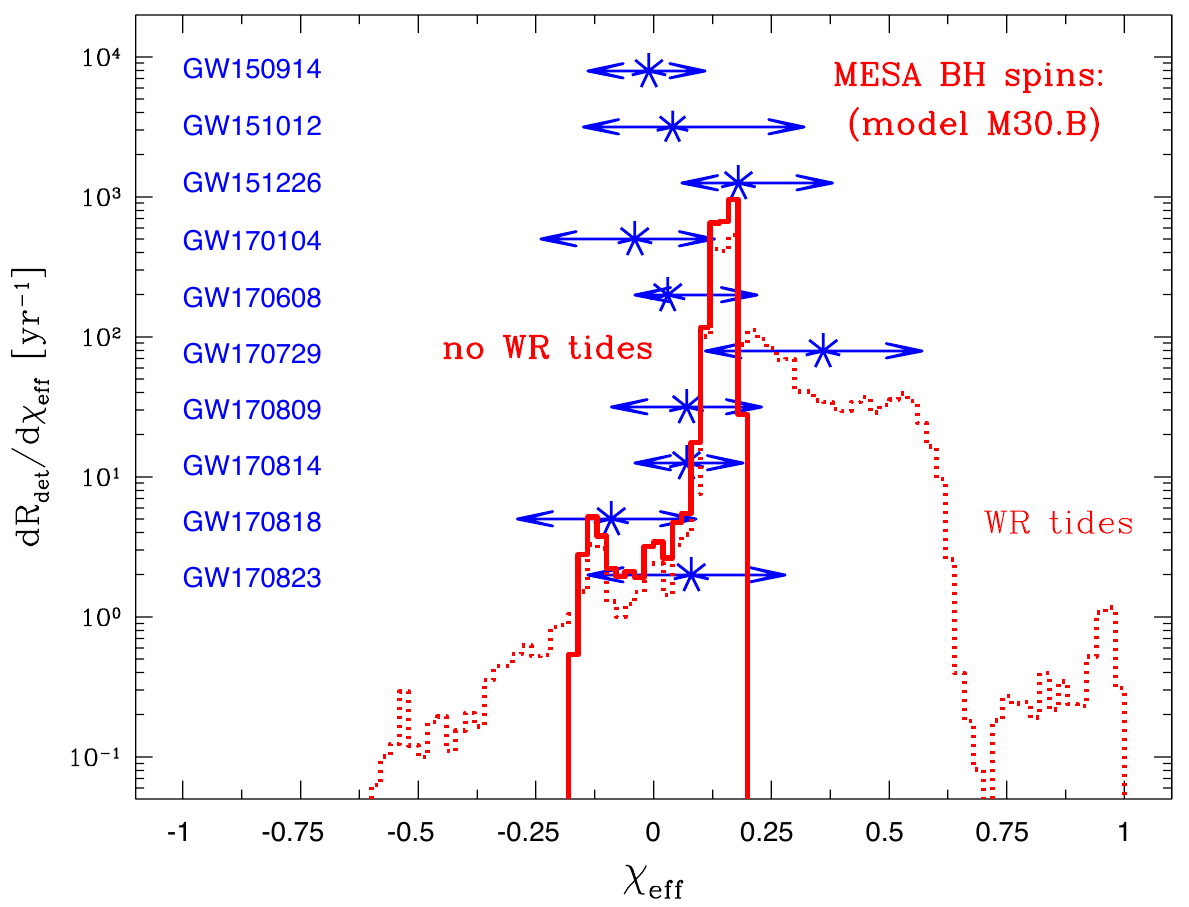}
\caption{
Detection weighted distribution of the effective spin parameter of BH-BH mergers for 
model M30.B with MESA efficient angular momentum transport and fallback decreased BH 
kicks (no BH kicks for massive BHs). We either do not allow for efficient tidal 
spin-up of WR stars that are the most common immediate progenitors of BHs in our 
models (natal BH spin is calculated directly from MESA stellar models), or we take it 
into account (natal BH spin is then calculated as described in Sec.~\ref{sec.tides} 
if the WR star progenitor was subject to an efficient tidal spin-up). 
For the "no WR tides" approach we find a rather narrow distribution of effective spins 
($-0.2\lesssim \chi_{eff} \lesssim 0.2$) that is peaked at positive values (average 
$\chi_{\rm eff}=0.15$). For efficient "WR tides" the distribution is broad 
($-0.5\lesssim \chi_{eff} \lesssim 1.0$) with a peak at $\chi_{eff} \sim 0.15$ 
($\sim 73\%$) and a tail with $\chi_{eff} \gtrsim 0.25$ ($27\%$). 
For comparison we show the $90\%$ credible limits (blue arrows) and the most likely 
values (blue stars) of the effective spin parameter for ten LIGO/Virgo BH-BH mergers.
}
\label{fig.xeff2}
\end{figure}

In Figure~\ref{fig.xeff2} we present the effective spin distribution for the model 
M30.B that employs the MESA BH natal spins. We show two versions of this model: 
one in which we do not allow and one that allows for efficient tidal interactions 
in close binaries (that host WR stars; see Sec.~\ref{sec.tides}). The version 
with no efficient tides shows a narrow distribution of effective spins: 
$-0.2\lesssim \chi_{eff} \lesssim 0.2$ that is peaked at positive values: average 
$\chi_{\rm eff}=0.15$). This limited range of effective spins follows directly 
from the underlying BH natal spin model that allows only for narrow range of BH 
spin magnitudes $a_{\rm spin}\sim0.05$-$0.15$ (see Fig.~\ref{fig.bhspin2}). The BH 
spins are, at most, moderately increased by accretion from the binary companion (see 
Figs.~\ref{fig.evol1} and ~\ref{fig.evol2}) and this tends to slightly extend the 
distribution to higher positive values. The distribution is only moderately affected 
by small-to-no BH natal kicks (fall-back decreased kicks are employed in this model) 
producing a small population of BH-BH mergers with negative effective spins. 
On the other hand the variation with the efficient WR tides generates a rather broad 
effective spin distribution: $-0.5\lesssim \chi_{eff} \lesssim 1.0$ with a peak at: 
$\chi_{eff} \sim 0.15$ ($\sim 80\%$) and a tail with $\chi_{eff} \gtrsim 0.25$ 
($27\%$). This is because in this variation about $1/3$ of systems are 
subject to significant tidal spin-up of at least one binary component and they 
produce large BH spins. The tail shows a significant drop beyond 
$\chi_{eff} \gtrsim 0.6$. Systems with $0.25\lesssim \chi_{eff} \lesssim 0.6$ are 
these in which only one binary component was subject to tidal spin-up, while systems 
with $\chi_{eff} \gtrsim 0.6$ are those with both binary components being subject
to tidal spin-up in close WR-WR binaries.

\begin{figure}
    \includegraphics[width=\columnwidth]{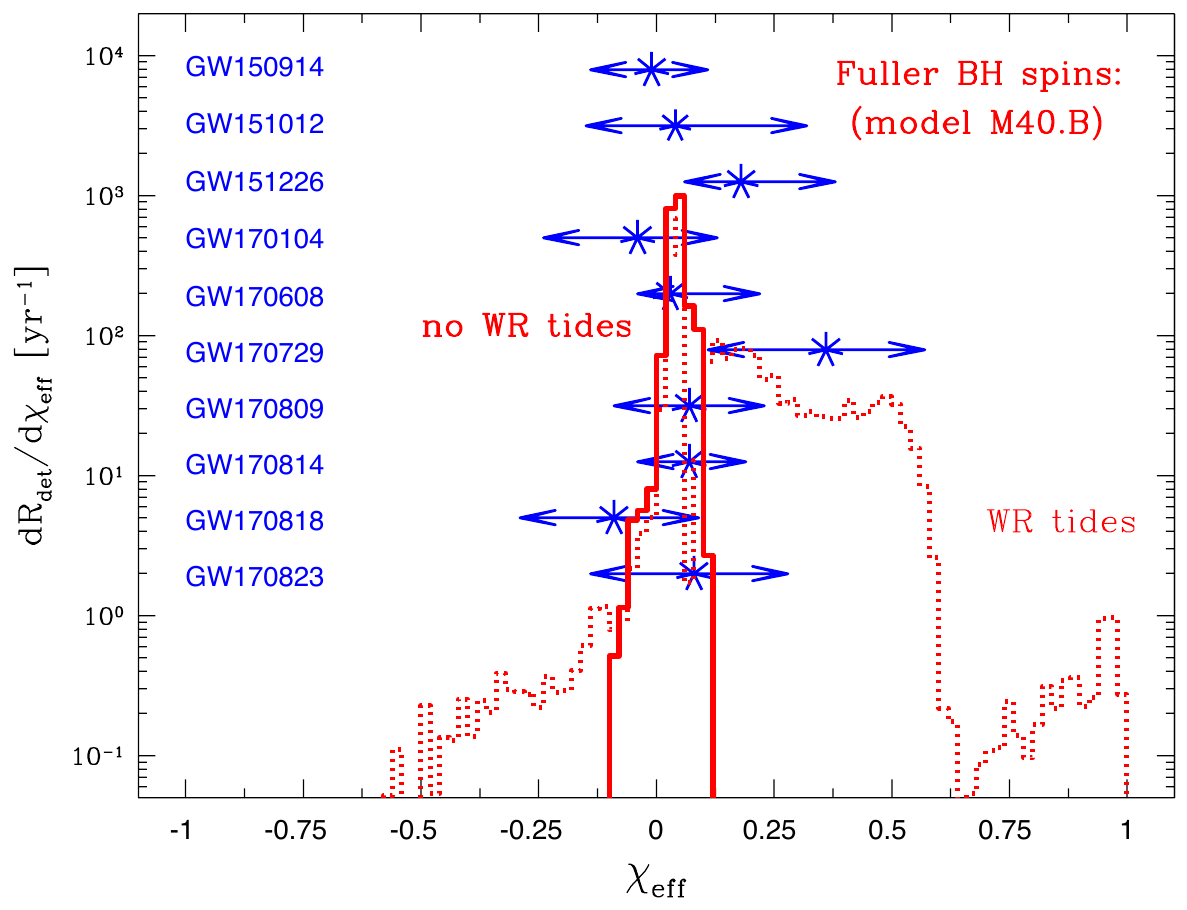}
\caption{
Detection weighted distribution of the effective spin parameter of BH-BH mergers for model 
M40.B with Fuller super-efficient angular momentum transport and fallback decreased BH kicks 
(no BH kicks for massive BHs). We either do not allow for the efficient tidal spin-up 
of WR stars that are the most common immediate progenitors of BHs in our models (natal 
BH spin is calculated directly from MESA stellar models), or we take it into account 
(natal BH spin is then calculated as described in Sec.~\ref{sec.tides} if the WR star 
progenitor was subject to an efficient tidal spin-up).
For the "no WR tides" approach we find very narrow distribution of effective spins 
($-0.1\lesssim \chi_{eff} \lesssim 0.1$) that is peaked at positive values (average 
$\chi_{\rm eff}=0.05$). For efficient "WR tides" the distribution is broad 
($-0.5\lesssim \chi_{eff} \lesssim 1.0$) with a peak at $\chi_{eff} \sim 0.05$ 
($\sim 78\%$) and a tail with $\chi_{eff} \gtrsim 0.25$ ($22\%$). 
For comparison we show $90\%$ credible limits (blue arrows) and the most likely values
(blue stars) of effective spin parameter for ten LIGO/Virgo BH-BH mergers.
}
\label{fig.xeff3}
\end{figure}

In Figure~\ref{fig.xeff3} we present the effective spin distribution for model M40.B 
that employs Fuller BH natal spins. We also show two versions of this model: one
without efficient WR tides and one with the tides. The distribution is very narrow: 
$-0.1\lesssim \chi_{eff} \lesssim 0.1$ and is peaked at positive values: average
$\chi_{\rm eff}=0.05$. This comes directly from the assumption of very low natal BH 
spins: $a_{\rm spin}=0.01$ (see eq.~\ref{eq.bhspin3}). For such a low value of 
natal BH spins, the effective spin of BH-BH mergers is $\chi_{\rm eff} \sim 0$.
Binary accretion onto the second-born BH spreads effective spins to 
$\chi_{\rm eff} \sim 0.1$, while small natal kicks applied to some (low-mass) BHs in 
this model create a tail extending to low negative values $\chi_{\rm eff}\sim-0.1$.
For efficient "WR tides" the distribution is broad: $-0.5\lesssim \chi_{eff} \lesssim 1.0$ 
with a peak at $\chi_{eff} \sim 0.05$ ($\sim 78\%$) and a tail with 
$\chi_{eff} \gtrsim 0.25$ ($22\%$).

\begin{figure}
    \includegraphics[width=\columnwidth]{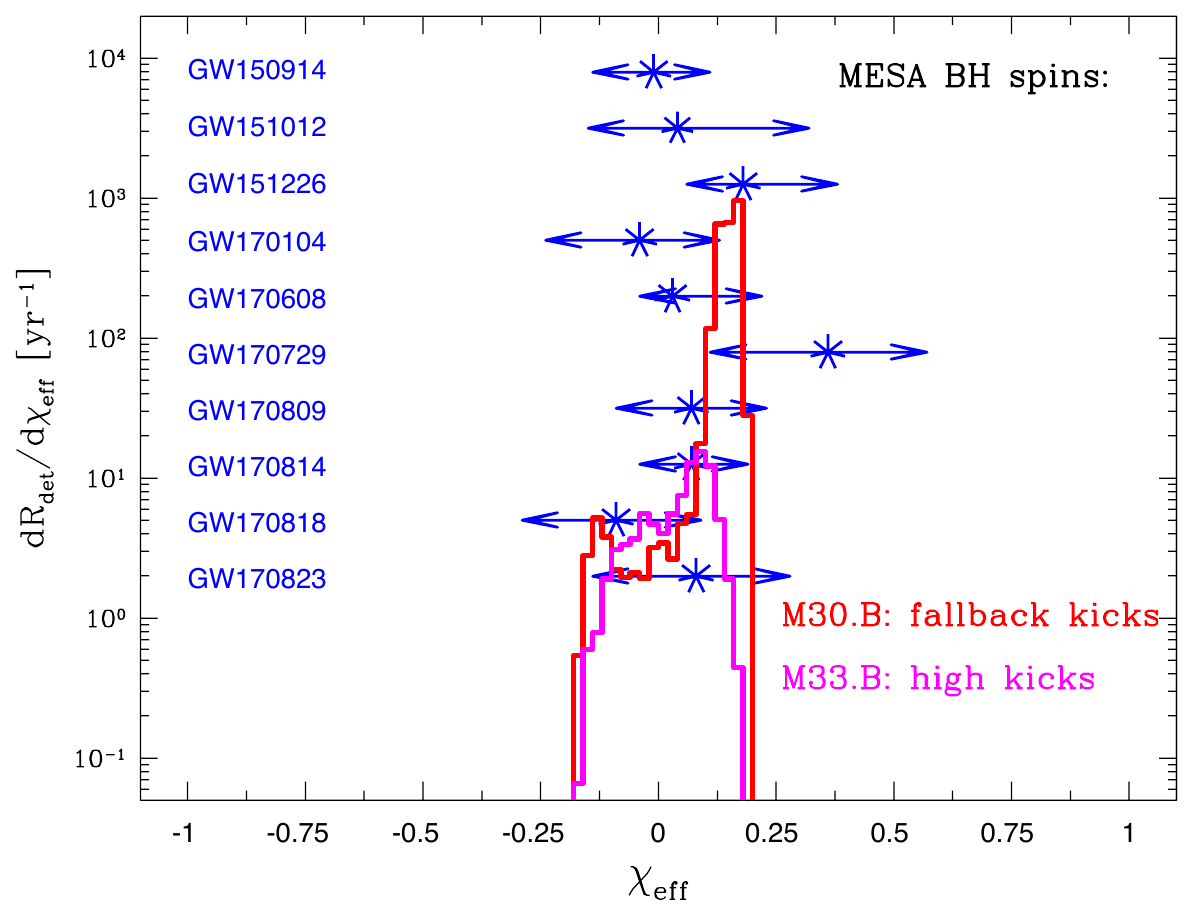}
\caption{
Detection weighted distribution of the effective spin parameter of BH-BH mergers for 
model M30.B with fallback decreased natal kicks (low natal BH kicks for low-mass
BHs, and no natal kicks for high-mass BHs) and for model M33.B with high natal kicks 
(all BHs, independent of mass, are subject to natal kicks drawn from a 1D Maxwellian 
distribution with $\sigma=265\kms$). 
Note that the average effective spin decreases from model M30.B ($\chi_{\rm eff}=0.15$) 
to model M33.B ($\chi_{\rm eff}=0.04$) as an effect of natal kicks that tend
to misalign BH spins and lower the effective spin (see eq.~\ref{eq.xeff}).
For comparison we show the $90\%$ credible limits (blue arrows) and the most likely values
(blue stars) of the effective spin parameter for ten LIGO/Virgo BH-BH mergers.
}
\label{fig.xeff4}
\end{figure}

\begin{figure}
    \includegraphics[width=\columnwidth]{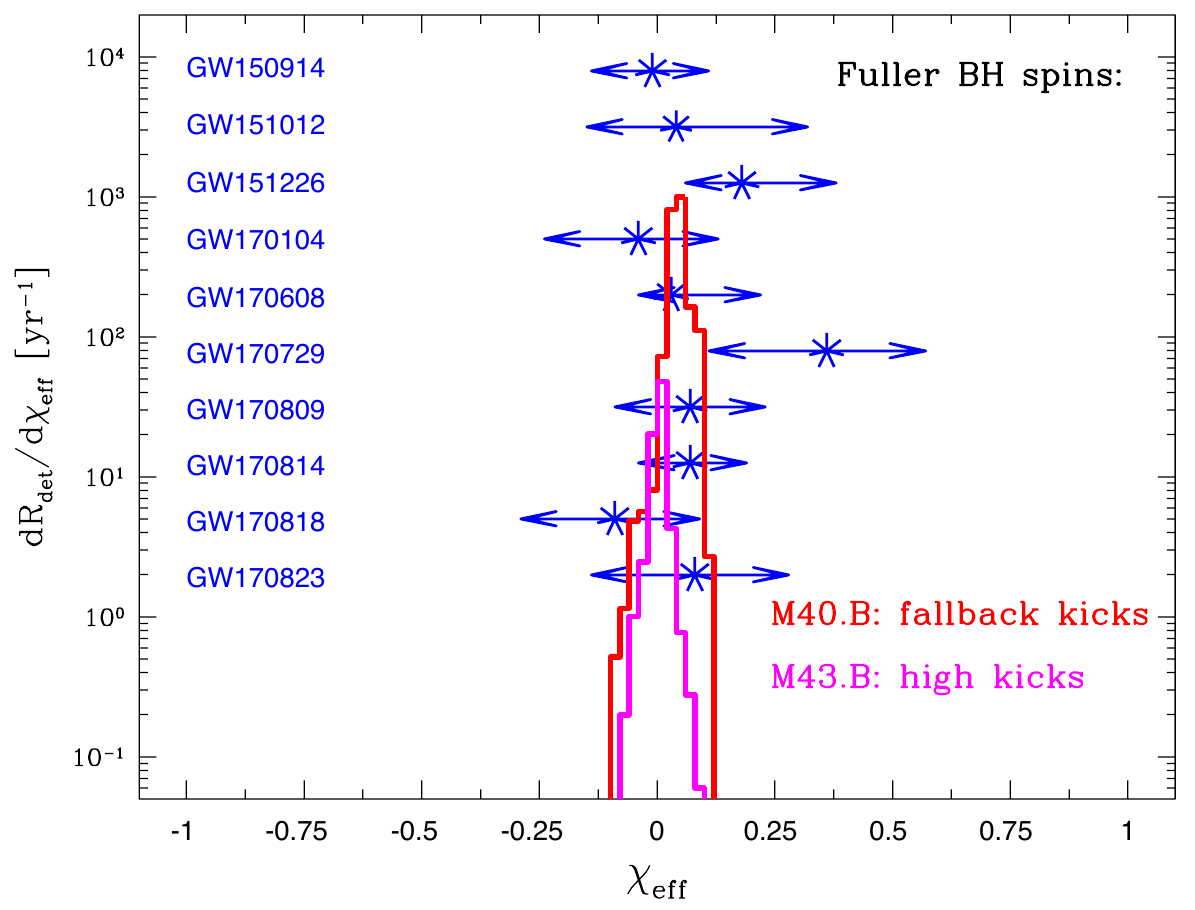}
\caption{
Detection weighted distribution of the effective spin parameter of BH-BH mergers for 
model M40.B with fallback decreased natal kicks (low natal BH kicks for low-mass
BHs, and no natal kicks for high-mass BHs) and for model M43.B with high natal kicks 
(all BHs, independent of mass, are subject to natal kicks drawn from a 1D Maxwellian 
distribution with $\sigma=265\kms$). 
Note that the average effective spin decreases from model M40.B ($\chi_{\rm eff}=0.05$) 
to model M43.B ($\chi_{\rm eff}=0.004$) as an effect of the natal kicks that tend
to misalign BH spins and lower the effective spin (see eq.~\ref{eq.xeff}).
For comparison we show the $90\%$ credible limits (blue arrows) and the most likely values
(blue stars) of the effective spin parameter for ten LIGO/Virgo BH-BH mergers.
}
\label{fig.xeff5}
\end{figure}

In Figures~\ref{fig.xeff4} and ~\ref{fig.xeff5} we present the effect of high BH 
natal kicks on the effective spin distributions in models that employ MESA and Fuller
BH natal spins. For MESA BH natal kicks we use two models: M30.B (fallback-decreased 
natal kicks: low-to-no BH kicks), and M33.B (high natal kicks, independent of BH mass, 
drawn from a 1D Maxwellian distribution with $\sigma=265\kms$). Note that the average 
effective spin decreases from model M30.B: $\chi_{\rm eff}=0.15$ to model M33.B: 
$\chi_{\rm eff}=0.04$. This is an effect of the natal kicks that tend to misalign BH 
spins and lower the effective spin. For Fuller BH natal kicks we use two models:
M40.B (fallback-decreased natal kicks), and M43.B (high natal kicks). We also note
that the average effective spin decreases with increasing natal kicks: from model M40.B: 
$\chi_{\rm eff}=0.05$ to model M43.B: $\chi_{\rm eff}=0.004$.

\section{Discussion}
\label{sec.talk}

\subsection{Angular Momentum Transport in Massive Stars}
\label{sec.ang_mom_transport}

In this section we discuss the dependence of the final angular momentum of the 
star at core collapse on angular momentum transport prescriptions and initial 
(ZAMS) rotation rate.  

\begin{figure}
    \includegraphics[width=\columnwidth]{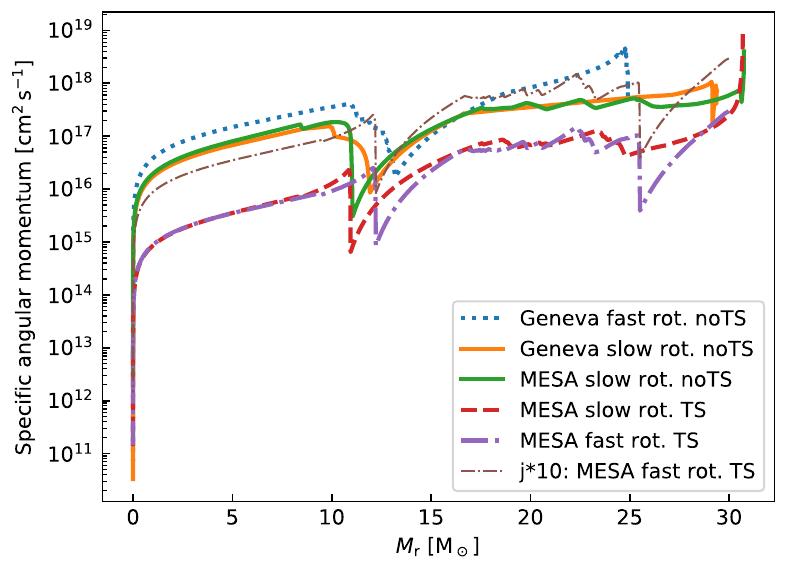}
\caption{
Specific angular momentum profile at the end of core He-burning for the  
$32\msun$ models at $Z=0.002$ calculated with a range of physical ingredients 
and initial rotation rates: the Geneva fast rotation model (model listed in 
Table~\ref{tab.spin1} with $V_{\rm ini}/V_{\rm crit}=40\%, V_{\rm ini}=341.9\kms$, 
dotted blue line), the Geneva slow rotation model ($V_{\rm ini}=100\kms$, solid orange  
line). Both these Geneva models do not include the Tayler-Spruit magnetic dynamo
(``noTS''). 
We also show: the MESA slow rotation non-magnetic (``noTS'') model ($V_{\rm ini}=100\kms$, 
solid green line), the MESA slow rotation magnetic (``TS'') model ($V_{\rm ini}=100\kms$, 
dashed red line), the MESA fast rotation magnetic (``TS'') model (model listed 
in Table~\ref{tab.spin2}, dot-dashed purple line). The brown thin dot-dashed line
corresponds to the specific angular momentum of the MESA fast rotation model 
multiplied by a factor of 10 to facilitate comparisons.
}
\label{fig.jrot_endHe}
\end{figure}

In order to evaluate the dependence of the final angular momentum on angular
momentum transport prescriptions and initial rotation rate, we ran three
additional $32\msun$ models at a metallicity of $Z=0.002$: a slow initial 
rotation ($V_{\rm ini}=100\kms$) non-magnetic (``noTS'', where TS stands for 
the Tayler-Spruit dynamo) model with both the Geneva and MESA code as well as 
a slow initial rotation ($V_{\rm ini}=100\kms$) magnetic (``TS'') model with 
the MESA code. The specific angular momentum profile of these models at the 
end of core He-burning are shown in Figure~\ref{fig.jrot_endHe}. The helium 
core, which will form the bulk of the black hole, extends from the center to 
about $12\msun$ in all the models plotted. For comparison, we also show the
specific angular momentum profiles for our two fast initial rotation models 
($V_{\rm ini}/V_{\rm crit}=40\%: V_{\rm ini}\approx340\kms$) for $32\msun$ 
star with $Z=0.002$ calculated with MESA-TS and Geneva no-TS assumptions. 
Note that these two fast models are used to calculate BH spin magnitudes 
employed in our population synthesis calculations (and that are referred to 
as MESA BH spins and Geneva BH spins).

Comparing various models, we see the following dependencies:
\begin{enumerate}

\item Comparing the magnetic slow and fast rotation MESA models (dashed red 
and purple lines, respectively), we see that they end with a very similar 
final angular momentum ($a_{\rm spin} =0.084$ and $a_{\rm spin}=0.087$ for 
the slow and fast rotation models, respectively\footnote{The values of dimensionless
spun parameter, correspond to a specific angular momentum of Fig.~\ref{fig.jrot_endHe}
$\sim 4.54 \times 10^{15}$ cm$^2$/s; see Eq. (\ref{eq.bhmag}).}). 
There is thus very little 
dependence on the initial velocity in magnetic models. This is due to a weaker 
magnetic instability (and thus less efficient angular momentum transport) in 
slower rotation models. We would thus expect slower rotation magnetic models 
to have a very similar final angular momentum content as the fast models 
listed in Table~\ref{tab.spin2}.

\item Comparing the magnetic (MESA slow and fast rotation; dashed red and purple 
lines) and non-magnetic models (MESA and GENEVA no-TS slow rotation and GENEVA fast 
rotation no-TS models; blue, green and orange lines), we see that the non-magnetic 
models have final angular momenta that are more than a factor of $10$ higher than 
the magnetic models (the angular momentum profile of the MESA fast rotation TS 
model multiplied by a factor of $10$ is shown to facilitate the comparison; brown 
line). This means that for the ``lower'' end of the massive star range, non-magnetic 
models predict BHs rotating near or above critical rotation 
($a_{\rm spin} \gtrsim 0.9$). 

\item Comparing the slow rotation, {\it non-magnetic}, Geneva and MESA
models (solid orange and green lines, respectively), the final angular
momentum is very similar. The small differences in angular momentum profiles
arise from small differences of the core sizes between the two codes (a result 
of the different treatment of the convective boundary mixing). We thus see that 
the two codes give consistent results when using a similar treatment of the angular 
momentum transport (without magnetic fields).

\end{enumerate}

The correlations of the angular momentum content listed above apply directly to 
black hole spins. We thus expect (slow and fast) rotating magnetic models to end up
in low BH spin ($a_{\rm spin} \sim 0.1$) and non-magnetic (slow and fast) rotating 
models to result in BH with large spins ($a_{\rm spin} \gtrsim 0.8$). The change of 
the initial stellar rotation from fast ($\sim 340\kms$; $40\%$ critical) to slow 
($\sim 100\kms$) does not significantly affect our estimates of the BH natal spin. 
In Sec.~\ref{sec.MESA} we present more detailed information and show that even when we 
increase rotation to $\sim 680\kms$ ($80\%$ critical) our conclusions remain unchanged 
(see also Fig.~\ref{fig.jrot_endHe_MESA}).
Note that the observed distribution of massive star spins is rather broad
($0-600\kms$), with two distinctive peaks: one (dominant) at $\sim 100\kms$ and 
one (small) at $\sim 400\kms$~\citep{Ramirez2013}.
These correlations were studied for a single mass (32\,M$_\odot$: the lower end of 
the massive star range) at one metallicity ($Z=0.002$) and with two models
of angular momentum transport (the Tayler-Spruit theory for the magnetic dynamo 
and meridional currents which dominate angular momentum transport while TS
dynamo is switched off). More in-depth studies, especially for higher masses are 
needed to confirm these correlations. 

Our models also make predictions for the birth spin periods of neutron
stars and it is worthwhile comparing these predictions with the
estimates of these spin periods from observations.  Stars with
zero-age main sequence masses lying between $\sim
8-20$\,M$_\odot$~\citep{fryer99b,fryer12} and more massive stars whose
mass loss is sufficiently extreme to dramatically alter the CO core
mass will collapse to form neutron stars (through normal supernovae).  If
the CO cores are rotating and the magnetic fields are sufficiently strong,
these neutron stars will emit as pulsars. By studying the distribution 
of spins in these pulsars, we can place constraints on the rotation 
period of the stars prior to collapse. This spin distribution also could
provide some indication of the role angular momentum can play in 
normal ``core-collapse'' supernovae. However, in section \ref{sec:neuspin} in
 the Appendix we study this problem and arrive to the conclusion that
 NS spins cannot be directly used to test angular momentum transport 
in massive stars as various (highly uncertain) mechanisms may spin down or spin 
up NS after its formation.

\subsection{Using Single Stellar Models in Binary Simulations}
\label{sec.sinbin}

Although to estimate the BH natal spins we use single stellar models (see Sec.
~\ref{sec.spins}), we use binary evolution models to estimate the BH-BH merger 
effective spins (see Sec.~\ref{sec.sfr}). This is not fully consistent as our 
approach does not take into account an important effect in binary evolution that 
may influence the natal BH spin, that is the spin-up of stars (progenitors of BHs) 
during the RLOF phases. This process typically affects the secondary star of the
BH-BH progenitors (Figs.~\ref{fig.evol1} and ~\ref{fig.evol2}: see the second 
evolutionary stage in both cases). Depending on the donor mass and our assumption 
on how much mass is accreted during the RLOF (in most models we allow for $50\%$ 
of mass lost by the donor to be accreted by its companion; but see models 
M20.A/B--M26.A/B) the main sequence secondary accretes anywhere from several to 
several tens of solar masses. This is most likely enough to significantly spin-up 
even a massive star~\citep{Packet1981}. However, there is a good chance that at 
least some of this extra angular momentum is later removed from the secondary star.

When the accretion and spin-up end, the secondary star is still on the main 
sequence. It will be therefore subject to angular momentum transport and losses 
(through winds) during a significant fraction of its subsequent lifetime. Obviously, 
there is less time in our binary evolution sequences for secondaries to lose extra 
angular momentum as compared with single stellar models initiated at ZAMS. On the 
other hand, there are binary interactions that help the secondary stars to get rid 
of extra angular momentum after the RLOF/accretion phase. 

This happens when the secondary star expands significantly after the RLOF and leaving 
the main sequence and is subject to tidal spin-down as this occurs on a very wide orbit 
(Figs.~\ref{fig.evol1} and ~\ref{fig.evol2}: see the third and fourth evolutionary stage 
in both cases). Then the expansion leads to a CE phase, and the entire secondary's H-rich 
envelope (with all its angular momentum content) is removed from the binary (Figs.
~\ref{fig.evol1} and ~\ref{fig.evol2}: see the fifth evolutionary stage in both cases). 
Any extra angular momentum from accretion during the main sequence that is left in the 
He-core of the secondary can contribute to an extra spin of the BH with respect to 
single stellar models but this would not change our conclusions significantly. 
If we increased the spin magnitudes of the secondary BHs, then 
the distributions of BH-BH effective spin parameters would become somewhat wider, and 
the effective spins would mostly shift to larger positive values. This would be 
qualitatively similar to the effect of efficient WR tides on secondary stars (see 
Figs.~\ref{fig.xeff2} and ~\ref{fig.xeff3}) for models with the effective angular 
momentum transport (MESA and Fuller models). For the model with inefficient angular 
momentum transport (Geneva model) the BH spins (on average) are already high so 
increasing stellar spin may only shift this model (in terms of effective spin 
parameter) even further away from LIGO/Virgo observations.  

Also the scenario in which the star is stripped of its envelope does not generally affect 
the trend in rotation rates discussed in Sec.~\ref{sec.ang_mom_transport}. Indeed, 
the effective angular momentum transport through a magnetic dynamo only operates 
when there is a shear allowing the development of a magnetohydrodynamic instability. Hence, 
different rotation rates between the core and the envelope are crucial since they create
the shear. In models including magnetic fields, the transport of most of the angular 
momentum occurs during the short phase between core hydrogen depletion and core helium 
ignition (during this short phase, the core contracts and spins up whereas the 
envelope expands and slows down, thus, creating a strong shear). Therefore, any 
stripping through binary interaction during core helium burning or afterwards does 
not affect significantly the slowdown of the cores. Stripping during the MS, however, 
is more complicated. There are two opposite effects. The first is that mass loss 
removes also angular momentum. The second is that if the entire hydrogen-rich layer 
is removed, then there is no envelope left to expand at the end of the MS and  the spin-down 
of the core by internal angular momentum transport (core-envelope coupling) is reduced. 
In non-magnetic models, spin-down of the core is weak anyway and models still end up with 
relatively large spin values, $a_{\rm spin}: 0.25$ to maximum spin (see Tab.~\ref{tab.spin1}).
In magnetic models, removal of the hydrogen-rich outer layer significantly reduces the 
spin-down of the core in the contraction phase after the MS and the core may keep its 
fast rotation rate (see also \cite{Fuller2019c}).

The majority of binaries that produce NS-NS/BH-NS/BH-BH mergers in our models 
have large initial orbital separations ($\gtrsim 1000 \rsun$; ~\cite{deMink2015}). 
Therefore, binary interactions (CE or stable RLOF; see Fig.~\ref{fig.evol1} and 
~\ref{fig.evol2}) that remove H-rich stellar envelopes happen either at the end of 
HG or during CHeB so that values of the natal compact body spin are not significantly
affected.

Therefore one may safely conclude that neither accretion spin-up of stars nor 
increasing the initial stellar rotation rate (see Sec.~\ref{sec.ang_mom_transport}) 
can significantly modify our conclusions about the natal spins of merging binary 
compact objects resulting from isolated evolution. However,we do not include 
quasi-homogeneous evolution scenarios discussed by~\cite{Chrimes2020}.

Note that other processes that can affect BH spin in binary evolution are
taken into account in our calculations. Besides tidal interactions that we probe 
with recent models presented in literature (see Sec.~\ref{sec.tides}), we also 
account for accretion onto BH that may lead to BH spin-up. The spin-up is 
calculated using the formalism presented by ~\cite{Belczynski2008b}. There are two 
potential phases of accretion onto the first-formed BH in our major evolutionary 
scenario: during the CE phase from the  H-rich envelope of the secondary and after the CE 
phase from the He-rich wind of the WR secondary (Figs.~\ref{fig.evol1} and ~\ref{fig.evol2}: 
see the fifth and sixth evolutionary stage in both cases). Since accretion leads
to (usually very small or only modest) increase of the first BH spin magnitude, 
it tends to slightly broaden the effective spin parameter distribution.

\subsection{BH-BH/BH-NS/NS-NS Merger Rate Densities} 
\label{sec.disrates}

The predicted merger rate density of coalescing double compact object binaries 
depends on the particular formation scenario which is assumed. In the following 
two sections we discuss our merger rates obtained for isolated binary evolution
and contrast them with BH-BH merger rates obtained from dynamical evolution in 
globular clusters. In summary, the typical globular cluster rates ($5\gpy$) can 
explain about $10\%$ of the LIGO/Virgo BH-BH mergers (the empirical rate peaks at 
$\sim 50\gpy$) while the remaining $90\%$ of the observed rate can be easily 
explained by isolated binary BH-BH formation (see Tab.~\ref{tab.ratesB} and 
~\ref{tab.ratesA}).

\subsubsection{Isolated Binary Evolution}
\label{sec.isolated}

\begin{figure}
\includegraphics[width=\columnwidth]{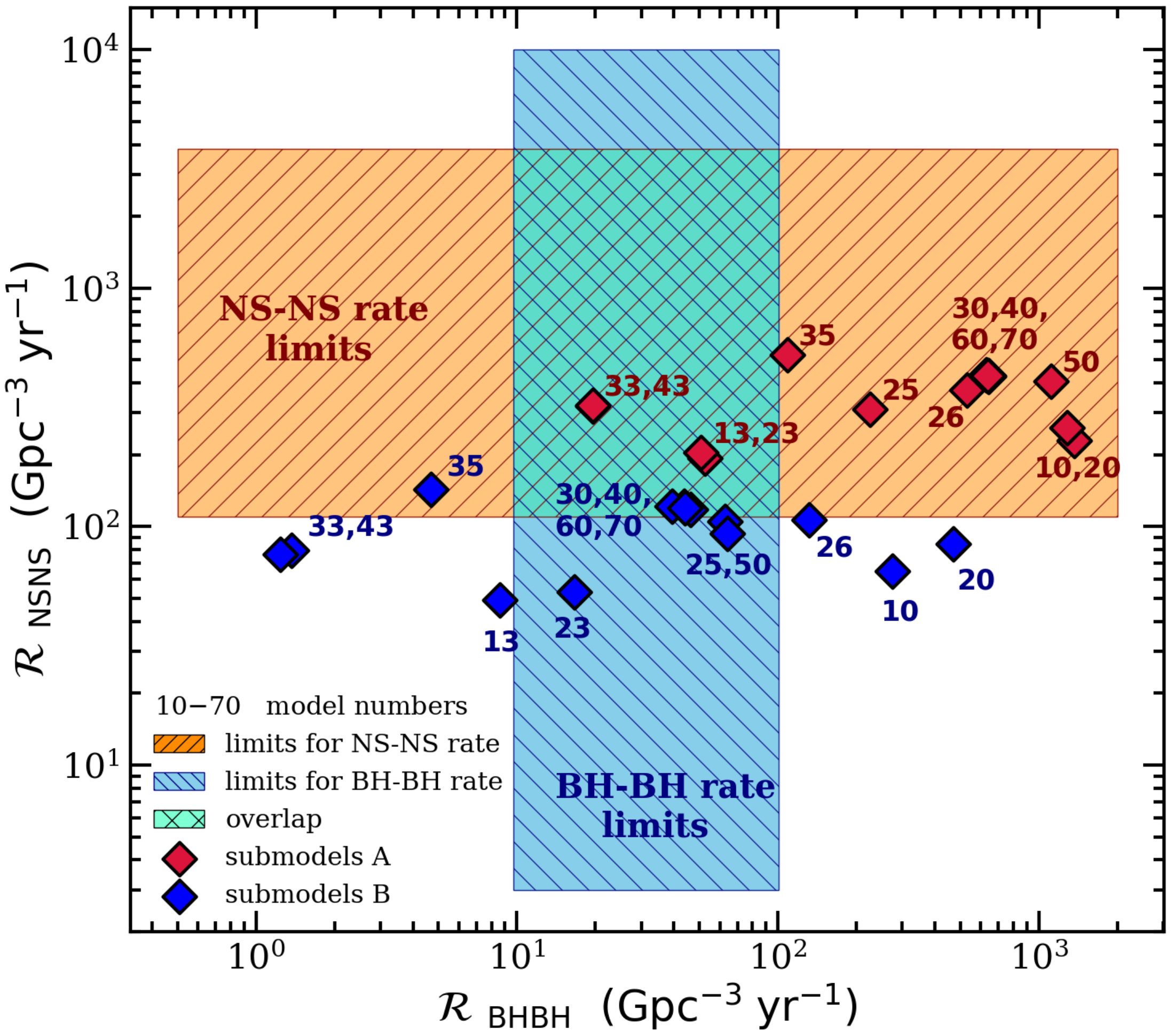}
\caption{
Comparison of the local merger rate densities of BH-BH and NS-NS mergers from all 
our models (see Tables~\ref{tab.ratesB} and ~\ref{tab.ratesA}) with the current 
limits inferred from the O1/O2 LIGO/Virgo observational runs: $9.7$--$101\gpy$ 
for the BH-BH mergers and $110$--$3840\gpy$ for the NS-NS events \citep{LIGO2019b}. 
Models consistent with the observational limits are: M13.A, M23.A, M33.A, M43.A,  
M30.B, M40.B, M60.B, M70.B. Note that while some of the models are consistent with 
the observational constraints and others are not, it is not straightforward to draw 
conclusions about the physical ingredients of the models at this stage due to 
degeneracies in the impact of various assumptions on the theoretical merger rates 
(see Sect.~\ref{sec.isolated}).
}
\label{fig.rates1}
\end{figure}

We have presented a range of models and their rate densities. The rate density 
in each case is the result of several assumptions: starting from the cosmic star 
formation, metallicity evolution, going on through the initial binary parameters 
and the binary evolution model to the implied delay time (between the birth of a 
binary and the final merger of two compact objects) distribution. For each model we 
have calculated the local merger rate density ($z=0$) as well as the predicted 
LIGO/Virgo detection rate in O3, see Tables~\ref{tab.ratesB} and ~\ref{tab.ratesA}. 
The local rate densities of BH-BH and NS-NS mergers that we have obtained are 
directly comparable with the rate limits inferred by LIGO/Virgo at the 
conclusion of the O1/O2 observing runs, see Figure~\ref{fig.rates1}. 
Additionally, based on the recent LIGO/Virgo candidate (S190814bv) for a very 
likely ($\sim 99\%$ in the reported mass-based source classification) BH-NS 
merger from August 14 \citep{GCN_324,GCN_333}, we estimate the BH-NS merger rate 
density of ${\cal R} = 1.6-60\gpy$ (see Appendix~\ref{sec.app_BHNS_limits} 
for explanation), which can also be confronted with our models.

\begin{figure}
\includegraphics[width=\columnwidth]{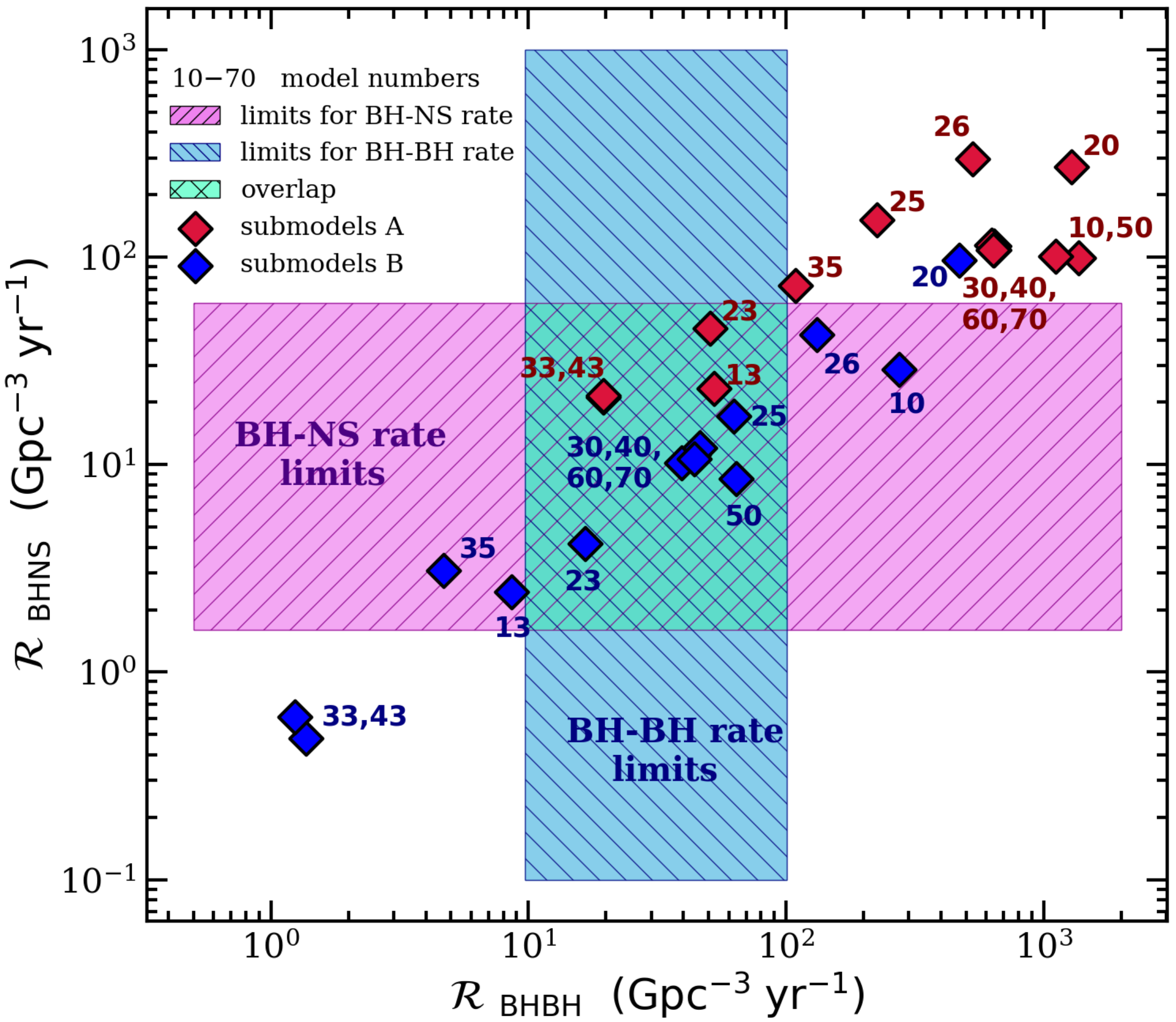}
\caption{
Comparison of the local merger rate densities of BH-BH and BH-NS mergers from all 
our models (see Tables~\ref{tab.ratesB} and ~\ref{tab.ratesA}) with the current 
limits: $9.7$--$101\gpy$ for the BH-BH mergers (LIGO/Virgo O1/O2) and $1.6$--$60\gpy$ 
for the BH-NS events (based on the LIGO/Virgo O3 candidate of the first BH-NS 
system, S190814bv \citep{GCN_324,GCN_333}; see Sec.~\ref{sec.app_BHNS_limits} for 
details). Models consistent with the observational limits are: M13.A, M23.A, M33.A, 
M43.A, M23.B, M25.B, M30.B, M40.B, M50.B, M60.B, M70.B. 
Note also that the BH-NS merger rates from our models span the range between $0.48$ 
and $297\gpy$, which is consistent with the upper limit determined by their 
non-detection in O1/O2 LIGO/Virgo runs ($<610\gpy$, \cite{LIGO2019b}).
}
\label{fig.rates2}
\end{figure}

In addition to the overall rate of events, the relative rates of the three generic 
categories of events (BH-BH, BH-NS, NS-NS) provide particularly valuable constraints. 
With this in mind, in Figures~\ref{fig.rates1} and ~\ref{fig.rates2} we show the 
rates of all three source categories. The horizontal and vertical bands 
represent the allowed rates consistent with current LIGO/Virgo data. Superimposed 
on the LIGO/Virgo bands are the rate predictions from a range of our models, 
with the red diamonds from submodel A (the optimistic approach to common
envelope, in which CE events with Hertzsprung gap donors are allowed to 
survive), and the blue diamonds from submodel B (the pessimistic approach to
common envelope, in which CE events with Hertzsprung gap donors are not 
allowed to survive). We note that in both figures there are many models that 
fall in the central sweet spot of the figure.

Looking at Figure~\ref{fig.rates1}, we see that the existing O1/O2 run bounds on the 
NS-NS and BH-BH merger rate densities already allow excluding some of the models. In 
the standard model group with the 2016 assumptions the only model consistent with 
the data is M13.A, which requires high BH/NS natal kicks. If we allow some 
modified physics we can also add model M23.A. From the group of models with updated 
physics of 2019, we can select two: M30.B ad M33.A, that are consistent with 
observations. Allowing for models with some modifications we can also include 
the models M40.B and M43.A as well as M60.B and M70.B. Thus we can see that the 
2016 standard model is consistent with the merger rate density O1/O2 runs data 
provided that the natal kicks are large. This may be in contradiction to some 
observations of BH in binaries that indicate small BH natal kicks, however, such 
a conclusion requires more detailed studies. The updated model of 2019 is consistent 
with the data. It also allows for certain modifications such as probing the origin of
BH spins or the inclusion of various models of PPSNs.

The inclusion of the BH-NS merger rate estimate (based on the O1/O2 runs and 
ongoing O3 run\footnote{Under assumption that the reported gravitational-wave 
signal from S190814bv is in fact BH-NS merger and not BH-BH merger, both options 
being allowed.} (see Appendix~\ref{sec.app_BHNS_limits}) in Figure~\ref{fig.rates2} 
offers additional insights. Of particular interest are models which fall in the 
central overlap region of {\em both}\/ Figures ~\ref{fig.rates1} and 
~\ref{fig.rates2}. For example, M13.A, M23.A, M33.A, M43.A, M30.B, M40.B, M60.B 
and M70.B fall within the observed range for all three rates. A number of trends 
are evident from looking at both figures together. For example, we note that the 
B class of submodels, without the inclusion of HG donors in the CE evolution, 
brings many of the models into the center of the BH-BH/BH-NS plot. However, this 
same change brings some of the models into tension or only into marginal agreement 
with the data on the BH-BH/NS-NS plot. In other words, submodel A is almost 
entirely excluded in Figure~\ref{fig.rates2}, while submodel B is almost entirely 
excluded in Figure~\ref{fig.rates1}. This might be a preliminary indication that 
the NS-NS systems do allow for CE with HG donors, while BH-NS systems do not. 
However, there are other factors that are relevant for the relative merger rates 
of different source types. For example, improving the metallicity evolution 
(which shifts the net star formation across the cosmic history towards higher 
metallicities) appears to shift down the BH-BH merger rates while leaving the 
NS-NS merger rates relatively unchanged (see also \cite{Chruslinska2019}).

In all cases that we consider the merger rate density increases with 
increasing redshift, peaks at about $z\approx 2$ and then decreases. A 
rough approximation of the scaling of the merger rate density with redshift in 
the range $z\lesssim2$ can be approximated as $\propto (1+z)^{1.5}$, however, 
the exponent is not well constrained. It is quite interesting to note that 
the assumed shape of the star formation rate history within the allowed 
bounds does not strongly influence the shape of the dependence of the merger 
rate density on redshift within $z\lesssim 2$, see Figure~\ref{fig.sfrrates}.

We expect that most of the remaining ambiguities should be resolved with 
the release of analysis of the O3 data. Assuming that these observations will 
yield about a hundred detections, we expect the bounds on the rates to 
narrow by a factor of three. This should allow us to strongly constrain 
the models presented in this paper based on the event rates alone.
It should be stressed, however, that even then it will not be straightforward 
to constrain the underlying physics. Variations in different ingredients 
of a model can have a degenerate effect on the resulting merger rates. For 
instance, the inclusion of HG donors in the CE evolution (going from submodels 
B to A) leads to an increase in the merger rates, but so does the lowering of 
BH/NS natal kicks (see Sect.~\ref{sec.rates} and Fig.~\ref{fig.kickrates} for 
details). The CE treatment and the natal kicks are only two of the many ingredients 
of our models, each of those ingredients being to some extent uncertain. This 
introduces a degeneracy that cannot easily be resolved at the moment. In the 
future, we can hope that the higher number of detected mergers allow us to combine 
all the inferred observables (e.g., rates, masses, spins, redshifts, host 
galaxies) in order to break the degeneracies and put better constraints on the 
model's input physics.

This being said, submodels A: M13.A, M33.A, M43.A appear to be consistent with 
all rate estimates (although on the high end of the BH-NS limits). This shows 
the effect of the high natal kicks that reduce high BH-BH merger rates typically
found in the submodels A. At the same time these high natal kicks do no affect 
the NS-NS merger rates significantly, as in our standard approach (fallback
decreased kicks) NS natal kicks are already high. 
Submodels B: M30.B, M40.B, M60.B and M70.B, that appear consistent with all 
rate estimates, indicate that high average cosmic metallicity in unison with 
exclusion of CE events with HG donors may reduce high BH-BH merger rates in 
comparison with older models and with submodels A. Note that these models 
include our current standard approach to input physics, with the two efficient 
angular momentum transport mechanism (MESA and Fuller models), and with the 
three prescriptions for PPSN (weak, moderate and strong) mass loss.   
Note that some models are very close to be consistent with all rate constraints. 
Model M50.B is one such example. In this model we have reduced stellar winds to 
$30\%$ of their standard values. Typically, BHs will form with higher mass in 
this model. Additionally, with such low wind mass-loss rates for massive 
stars it is possible to form $\sim 60\msun$ BHs in a solar metallicity environment 
($Z=0.02$) while still avoiding PPSN mass loss (see~\cite{Belczynski2020a} for 
discussion). At such mass-loss rates single stars (or stars in wide non-interacting 
binaries) may form $\sim 40\msun$ He cores with $\sim 20-30\msun$ H-rich envelopes 
leading to massive BH formation even at high metallicity. Such massive BHs would be 
found either as single BHs through microlensing surveys (e.g., \cite{Wyrzykowski2019}) 
or in wide binaries through radial velocity surveys (motion of the companion star). 
Note that such massive BHs would not necessarily add to either LIGO/Virgo or X-ray 
binary populations that form (at least in classical isolated binary evolution) from 
interacting binaries in which stars lose their H-rich envelopes (e.g., this study, 
~\cite{Wiktorowicz2014}). There is, of course, a complicated interdependence of the 
various aspects of these models, and this subset of models does not fully cover the 
parameter space. Nonetheless, these general trends may provide important clues about 
the underlying stellar and binary physics.

\subsubsection{Dynamical Evolution in Globular Clusters}
\label{sec.dynamics}

In dense stellar environments, even binaries that are initially too wide
to inspiral via gravitational radiation in a Hubble time can be hardened
and induced to merge by binary-single and binary-binary interactions.
Moreover, single BHs can segregate to the cluster center due to dynamical
friction and form close binary systems that can merge due to gravitational
radiation. Thus dense stellar systems such as globular clusters can be highly
efficient per stellar mass at producing BH-BH mergers.  However, only a
small fraction of stars are in globular clusters or similarly dense
systems ($\sim 10^{-4}-10^{-3}$, depending on the type of galaxy).  This
turns out to mean that the plausible rate density of mergers in dense
stellar systems is low.  A strong upper bound can be obtained (see
\citet{Mandel2018}) by noting that there is roughly one
globular cluster per Mpc$^3$ in the local Universe.  If each globular has
$\sim 10^6$ stars (the actual average is a few times lower than this),
$\sim 10^{-3}$ of stars become black holes, and all black holes pair up
and merge in a Hubble time of $\sim 10^{10}$ years, then at most the rate
will be $\sim 10^6\times 10^{-3}/(10^{10}~{\rm yr}\times 1~{\rm Mpc}^3)$,
or $\sim 100~{\rm Gpc}^{-3}~{\rm yr}^{-1}$. In reality, the expected rate 
from clusters is 10 to 100 times lower, for a discussion see~\cite{Mandel2018}.

It does seem likely that globular cluster masses at formation were a few
times larger than they are today.  For example, it has been estimated that
$\sim 60\%-70\%$ of cluster mass is lost in a Hubble time due to dynamical
evolution and interactions with the host galaxy~\citep{Chatterjee2010,Giersz2013,
Webb2015}. However, recent realistic estimates of BH-BH merger rate densities 
from globular clusters are typically ${\rm few}~{\rm Gpc}^{-3}~{\rm yr}^{-1}$
~\citep{Rodriguez2016b,Askar2017,Park2017,Hong2018,Choski2019}. Recoil
during binary-single and binary-binary interactions can be sufficient to
eject a binary from its host globular, with the result that the majority
of such mergers are expected to occur outside the globular (\cite{Portegies2000} 
and many subsequent papers).

\subsection{BH Masses}
\label{sec.dismasses}

An important test of population synthesis models is the direct comparison of their 
predictions of the mass distribution of binary black holes with the observed 
distribution from gravitational-wave detections. The masses of black holes are 
among the best measured quantities of the LIGO/Virgo sources. In particular, 
the chirp mass is often measured to high accuracy ($O(5\%)$ or better) for lower--mass 
mergers with a significant number of inspiral cycles detected in the LIGO/Virgo
sensitivity band. For the higher mass systems, the total mass can be measured with 
reasonable accuracy. The individual component masses are often more poorly 
measured, and equivalently, the mass ratio is often relatively poorly constrained. 
For this discussion we focus on the shape of the mass distribution. 

As LIGO/Virgo continues to add to the sample of binary black holes, the
inferred underlying mass distributions become increasingly well constrained. 
Some preliminary limits are already presented in~\citet{LIGO2019a}. In particular, 
that paper employs a number of different fits to the mass distribution of the 
more massive components of the detected binaries. For the purpose of comparison 
we use the Model B from this paper, which consists of a power-law in $M_{\rm BH1}$ 
with upper and lower mass cutoffs and a power-law in the binary mass ratio. 
The mass cutoff accounts for the dearth of high-mass black holes
~\citep{Fishbach2017b}, as would be expected from PSN and PPSN~\citep{Belczynski2016c}. 
LIGO/Virgo finds a power law index in range $\alpha=-3.1$ to $\alpha=0.1$
with peak probability at $\alpha=-1.6$ \footnote{Note that the LIGO/Virgo power 
law index is defined with a minus sign in front (see eq.~2 of~\citet{LIGO2019a}.}, 
and with an upper mass cutoff between $36.3\msun$ and $57\msun$. 

\begin{figure}
\includegraphics[width=\columnwidth]{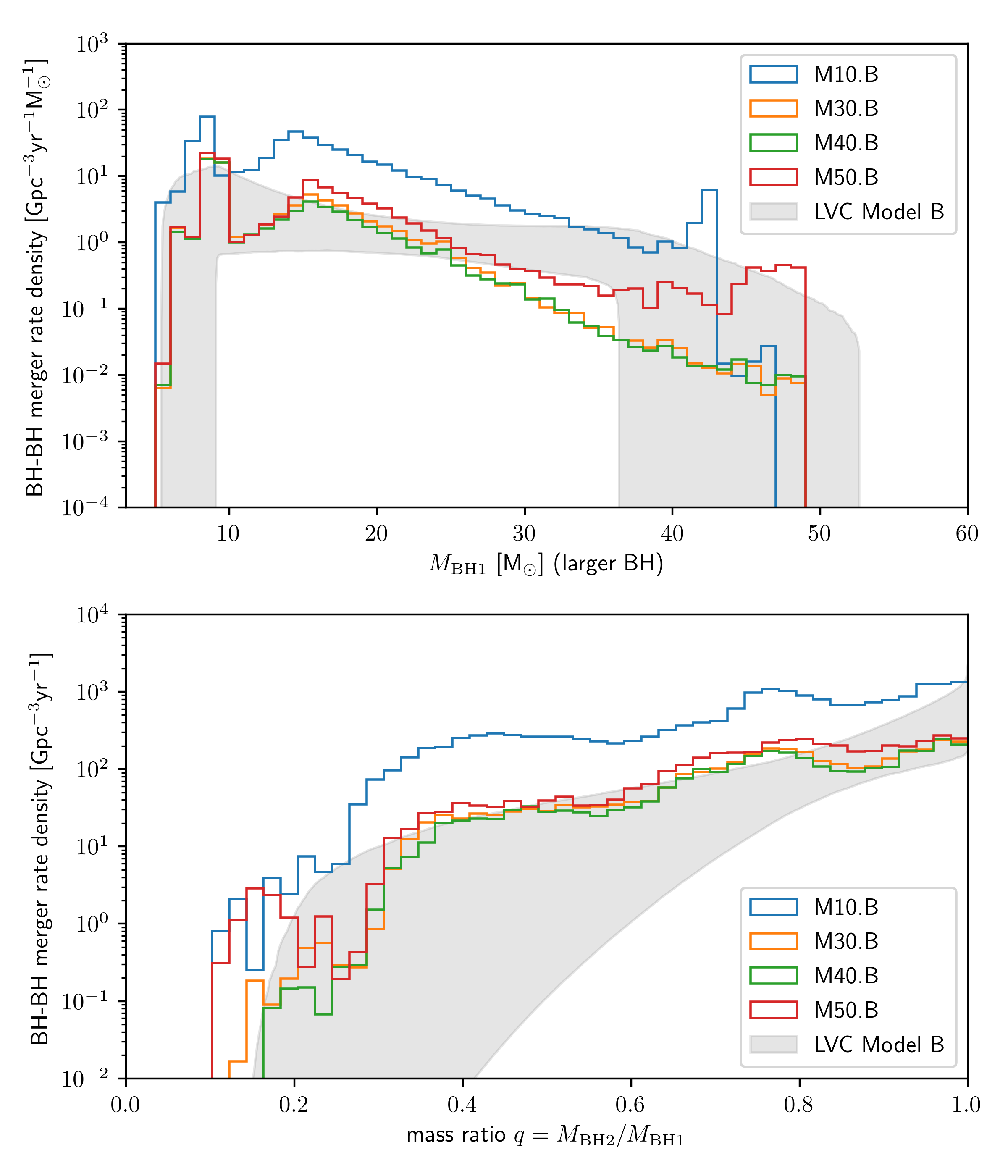} 
\caption{The intrinsic merger rate distribution as a function of primary black 
hole mass $M_{\rm BH1}$ ({\it top}) and binary mass ratio $M_{\rm BH2}/M_{\rm BH1}$ 
of BH-BH mergers ({\it bottom}) for different {\tt StarTrack} model choices. The gray 
bands show the $90\%$ confidence limits on the populations inferred in~\citet{LIGO2019a} 
using the phenomenological Model B. 
}
\label{fig.mass.pdf}
\end{figure}  

In Fig.~\ref{fig.mass.pdf}, we show the intrinsic rate density of mergers 
averaged out to $z=0.5$ versus the mass of the more massive BHs in BH-BH mergers 
($M_{\rm BH1}$; upper figure) and the mass ratio $q=M_{\rm BH2}/M_{\rm BH1}$ (lower 
figure) for several of our models and compare them with the fits to Model B from
~\citet{LIGO2019a}. Although Model B does not have the freedom to capture some of 
the features of the population synthesis distributions, the overall rate densities 
of the mass distribution found in the LIGO/Virgo collaboration analysis (shown in 
gray) are quite similar to those obtained, for example, in models M30.B, M40.B, and 
M50.B. We emphasize that the shapes of the primary mass distributions we obtain in 
these {\tt StarTrack} models are better fit to exponentials than to power-laws and 
that there are some interesting features in the mass distributions at low BH mass. 
Nevertheless, compared to our theoretical models, the LIGO/Virgo Model B fits to the 
$10$ O1/O2 BH-BHs exhibit a similar falloff of the primary mass distribution towards 
higher masses, lower and upper mass cutoffs, and a falloff of the mass ratio 
distribution at more unequal masses.

At this point we do not attempt any more elaborate fits to the O1/O2 data than the 
comparison to the LIGO/Virgo phenomenological models, as these can wait till more 
LIGO/Virgo data points become available. For anyone interested in such fits, all our 
data is available online. With $O(100)$ BH-BH mergers expected for the entirety of 
O3, we will be able to test the existence of some of the finer features seen in our 
mass distributions as well as the goodness-of-fit of exponentials versus power-laws 
to the primary masses.

\subsection{BH-BH Effective Spins}

\begin{figure}
\includegraphics[width=1.0\columnwidth]{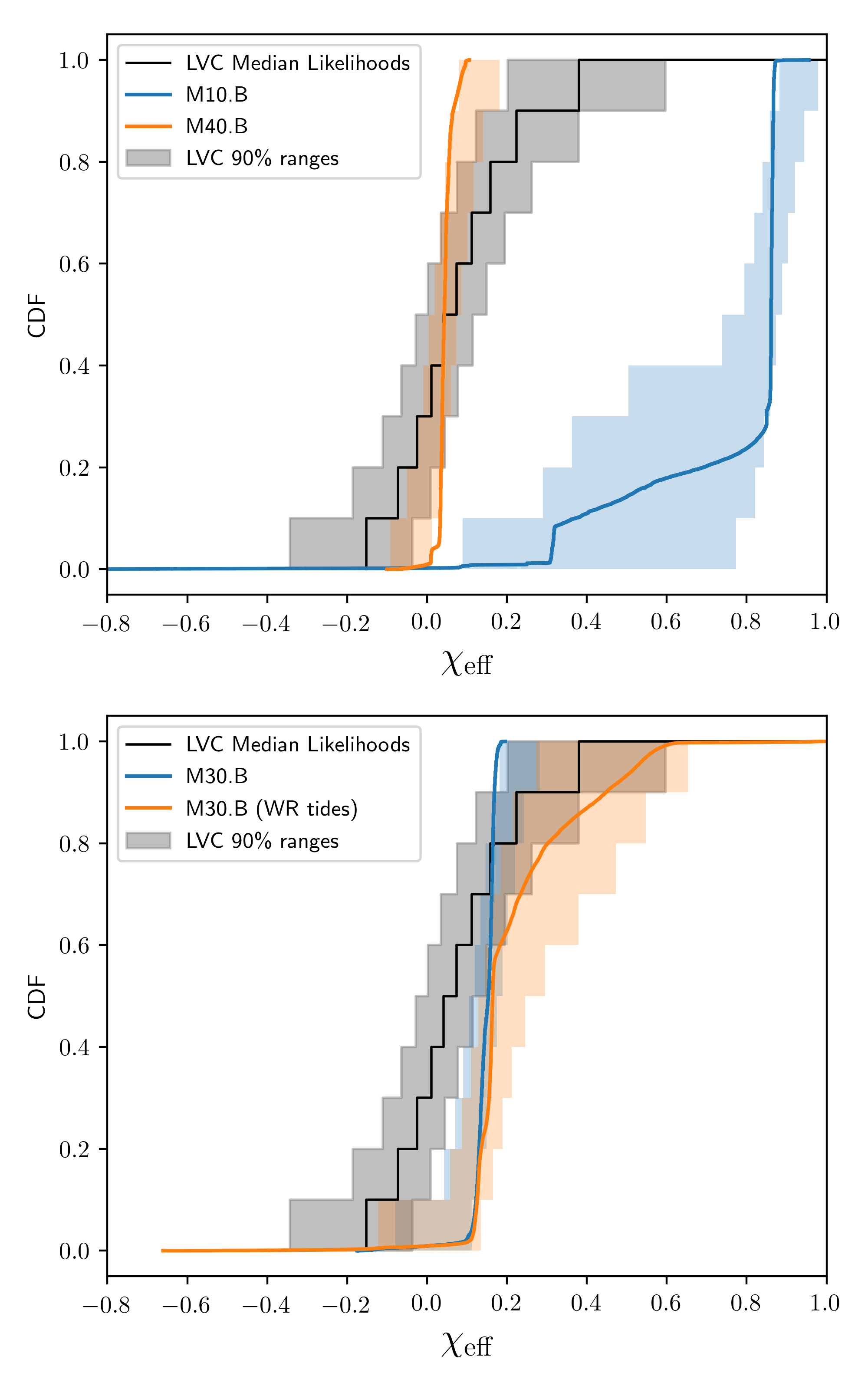}
\caption{
The cumulative density functions (CDFs) of four {\tt StarTrack} models
versus effective spins $\chi_{\rm eff}$. Solid color lines correspond to
$\chi_{\rm eff}$ CDF of detectable events for each model. The black line 
shows the $\chi_{\rm eff }$ CDF of median likelihoods of the O1/O2 LVC 
events, and the gray region bounds CDFs of the 5th and 95th percentiles of 
the likelihoods from LVC parameter estimation. In the corresponding colored
shaded regions, we show the $90\%$ range of 5000 mock observed CDFs under
each model. To generate a mock observed CDF, we draw 10 $\chi_{\rm eff}$
values from the detected population under a given model and add random
Gaussian noise with $\sigma=0.05$, which is approximately the uncertainty 
in $\chi_{\rm eff}$ likelihoods from the events of GWTC-1 (we resample any 
samples with $|\chi_{\rm eff}|>1$ after adding noise). 
}
\label{fig.chieff.cdf}
\end{figure}

The rate of BH-BH mergers with different effective spins provides several 
reliable features that we can use to corroborate and constrain our models 
with present and future GW observations. 

First and foremost, as has been discussed in our and other previous work, our 
binary evolution models only rarely if ever produce binaries with $\chi_{\rm eff}$ 
significantly below zero (see Figs.~\ref{fig.xeff1}, \ref{fig.xeff2}, 
\ref{fig.xeff3}, \ref{fig.xeff4}). This void provides an opportunity to 
identify the unique contribution from alternative formation channels, as our 
models cannot produce a preponderance of events with negative effective spins. 

Second, our predicted effective spin distributions have sharp cutoff-like 
features in each of the one (or sometimes two) subpopulations that dominate 
the detection rate versus effective spin. For example, the black hole spin 
distribution drops sharply for $\chi_{\rm eff} \gtrsim 0.9$ in the inefficient 
Geneva angular momentum transport mechanism (Fig.~\ref{fig.xeff1}), above $0.2$ 
for the binary black holes not spun up by tides in the MESA models; above $0.6$ 
for the subpopulation of BH-BH with a tidally-spun-up WR progenitor in the MESA
models (Fig.~\ref{fig.xeff2}); and near $0.1$ for the Fuller models (Fig.
~\ref{fig.xeff3}). Sharp cutoffs like these are rapidly identified empirically, 
allowing future gravitational-wave observations to constrain the BH natal spin 
distribution, and the physics of tidal spinup.

Third, for models with WR tides (e.g., M30.B and M40.B shown in Figs~\ref{fig.xeff2} 
and \ref{fig.xeff3}), our effective spin distribution has two well-separated 
subpopulations (a peak and a tail), associated with the principal channel and 
the WR-spinup channel. In other words, the gravitational-wave population may 
allow us to reliably associate specific binary physics to specific binary black 
hole mergers, and thus to measure the relative proportions (and properties) of 
BH-BH systems forming through each channel. 

While our binary evolution models have many parameters, because of the limited 
role of accretion on the BH spin, very few parameters have more impact 
on the $\chi_{\rm eff}$ distribution than the physics we have described above.
For example, we have shown with Figure~\ref{fig.xeff5} that BH kicks have a 
relatively modest effect on the shape of the $\chi_{\rm eff}$ distribution. 
Strong BH kicks can for example make the $\chi_{\rm eff}$ distribution more 
symmetric and isotropic, effectively by disrupting binaries which would otherwise 
dominate the sharp cutoffs at the largest values of $\chi_{\rm eff}$. In other 
words, these kicks make the $\chi_{\rm eff}$ merger rate smaller and with a less 
prominent (and therefore more difficult to identify) cutoff. Though this poorly
constrained physics of natal kicks can reduce the rate at which $\chi_{\rm eff}$ 
measurements can inform about the binary evolution and diminish the contrast of the 
features we have described above, they cannot erase them, particularly features 
such as the second population of WR tidally spun-up stars or the low probability of 
mergers with large negative $\chi_{\rm eff}$.

In Fig.~\ref{fig.chieff.cdf}, we show the cumulative density functions
(CDFs) of the effective spin $\chi_{\rm eff}$ for a select set of our 
models and compare them to the $\chi_{\rm eff}$ measured for the $10$ 
O1/O2 BH-BH mergers. For each model we show the $\chi_{\rm eff}$ CDF of 
detectable events as a color solid line. In the corresponding colored 
shaded regions, we show the $90\%$ range of $5000$ mock observed CDFs 
under each model. To generate a mock observed CDF, we draw 10 
$\chi_{\rm eff}$ values from the detected population under a given model    
and add random Gaussian noise with $\sigma=0.05$, which is approximately    
the uncertainty in $\chi_{\rm eff}$ likelihoods from the events of GWTC-1 
(we resample any samples with $|\chi_{\rm eff}|>1$ after adding noise). 
The black line shows the $\chi_{\rm eff}$ CDF of median likelihoods of 
the O1/O2 LVC events, and the gray region bounds CDFs of the 5th and 95th 
percentiles of the likelihoods from LVC parameter estimation. Notably, 
these effective spins are clustered around zero effective spin, indicating 
that model M10.B's predicted spins are disfavored by the data. Models M30.B 
(no WR tides), M30.B (WR tides), and M40.B are much more consistent with 
these observations, and with the LIGO/Virgo collaboration analysis in
~\citet{LIGO2019a}, which finds a preference for most effective spins to
be very close to zero (see Fig.~12 in~\citet{LIGO2019a}). That analysis 
finds that only a few tens of percent of the observed BH-BH mergers have 
$|\chi_{\rm eff}|>0.05$, and of those with $|\chi_{\rm eff}|>0.05$, almost 
all are likely to have positive $\chi_{\rm eff}$. For the moment the 
observed $\chi_{\rm eff}$ distribution does not allow to test the importance 
of WR star tides during the evolution towards a BH even if it seems to 
slightly favor the no-tide path. Overall, the limited BH-BH sample from 
O1/O2 does not strongly constrain the BH spins other than to the conclusion 
that the effective spins tend to be near zero. However, it suggests that O3 
should provide data that will allow selecting evolutionary path that lead to 
BH-BH mergers.

\subsubsection{Dichotomy of LIGO/Virgo BHs and HMXB BHs}

LIGO/Virgo BHs are mostly massive (see Tab.~\ref{tab.ligodata}) and it seems that 
the spins of these BHs are small. Unfortunately, the spin magnitudes of massive 
BHs with $M_{\rm BH}>15\msun$ are not constrained by other observational means. 
For example, the three most massive BHs in wind-fed high mass X-ray binaries 
(HMXB) for which we have spin estimates are at roughly $1/3$ or $1/2$ the mass of 
the larger-mass LIGO/Virgo BHs. The estimated masses and spins of these BHs can 
be found online at \url{https://universeathome.pl/universe/blackholes.php}: 
LMC X-1, $M_{\rm BH}=10.9\pm1.6\msun$ ($a_{\rm spin}=0.92$); 
Cyg X-1, $M_{\rm BH}=14.8\pm0.1\msun$ ($a_{\rm spin}>0.983$); 
M33 X-7, $M_{\rm BH}=15.7\pm1.5\msun$ ($a_{\rm spin}=0.84$). 
Note that BH spins can be measured by two methods: disk reflection and
disk continuum. For LMC-1 and Cyg X-1 BH spins from both methods are
consistent, while for M33 X-7 BH spin was estimated only through disk
continuum~\citep{Miller2015b}.
There is an apparent tension between the spin estimates of LIGO/Virgo BHs and
BHs in these HMXBs. However, this tension can be possibly avoided if HMXBs and 
LIGO/Virgo systems form through different evolutionary scenarios.

First, none of these three HMXBs is expected to produce a BH-BH merger. Future 
evolution of these systems was studied and it was shown that none of these systems 
will form a BH-BH merger. They will either end up as single objects (in
which BH merges with its massive companion star), or at best they may form BH-NS
systems~\citep{Belczynski2012b}. Note that this conclusion was reached in the 
framework of the classical binary evolution that does not take into account 
homogeneous evolution of rapidly rotating stars that allows for the formation of 
BH-BH mergers in alternative ways~\citep{Marchant2016,Mandel2016a,deMink2016}. 

Second, it was argued that HMXBs and LIGO/Virgo systems may form through different 
evolutionary scenarios. As explained in the past, the classical isolated binary evolution 
channel forms BH-BH mergers from (initially) very wide binaries ($a \gtrsim 1000 \rsun$; 
see Fig.~\ref{fig.evol1} or ~\ref{fig.evol2} or \cite{deMink2015,Belczynski2016b}). 
Therefore, if the tidal spin-up operates, it only acts at the very end of the 
evolution during the BH-WR stage (potentially) spinning up some fraction of the 
second-born BHs in BH-BH mergers (see Sec.~\ref{sec.tides} and ~\ref{sec.xeff}). 
This makes the majority of BHs in BH-BH mergers to have low spins if efficient 
angular momentum transport is adopted for massive stars (as in our MESA or Fuller 
models). 
On the other hand, HMXBs were proposed to form from initially very close binaries 
($a \sim 100 \rsun$, e.g.,~\cite{Valsecchi2010,Qin2019}). In such scenario a system 
with a massive main sequence primary is tidally locked so, through its evolution, 
its spin is kept high at the expense of the orbital angular momentum. At such small orbital 
separations the binary undergoes case A mass transfer and/or homogeneous evolution 
that keeps reducing H-rich envelope of the donor primary and keeps it from expanding. 
After main sequence evolution, the primary is a compact WR star that quickly collapses 
to a BH. The BH is found in the near proximity of its companion star (if natal kick is 
small), naturally producing a (wind-fed) HMXB. Possibly, tidal interactions can keep 
the primary spin high until the end of its nuclear evolution which would lead to the 
formation of a rapidly spinning BH. However, the detailed evolutionary calculations 
have shown that if efficient (Tayler-Spruit dynamo) angular momentum transport is 
assumed for the primary star,  then properties of the three HMXBs considered here 
can not be reproduced~\citep{Qin2019}.
 
Therefore, the tension still exists, albeit it seems like HMXBs and LIGO/Virgo BH-BH 
mergers originate from different initial populations of binaries (close versus wide), 
but specific details of evolution leading to the formation of HMXBs need to be worked 
out.  

Note that formation scenarios that require rapidly spinning stars (whether they 
lead to formation of BH-BH mergers or BH HMXBs) are not available in our study.
Rapid rotation induces efficient mixing and reduces radial expansion of stars in 
homogeneous or semi-homogeneous evolution. Our calculations can only be applied to 
slow- or moderately-rotating stars in classical binary evolution. Both binary 
channels (homogeneous and classical) do not exclude but rather complement 
each other.

Finally one should stress that estimates of BH spins through properties of their 
accretion disks, are not \textsl{direct} spin measurements but are model--dependent, 
so there is still a possibility that the tension is only apparent.

\subsection{Origin of LIGO/Virgo BHs}
\label{sec.origin}

At the moment (only O1/O2: data available) it seems that the classical isolated binary 
evolution formation channel can explain all the basic properties of the LIGO/Virgo
BH-BH mergers. This does not mean than other channels do not contribute to
LIGO/Virgo detections. For example, if we compare the average rate estimates of the 
isolated binary channels (see Tab.~\ref{tab.ratesB} and ~\ref{tab.ratesA}) and 
the dynamical formation channels in globular clusters (see Sec.~\ref{sec.dynamics}) 
it appears that $1$ in $10$ BH-BH mergers might come from a globular cluster. There may 
be yet another dynamical contribution from young open clusters to BH-BH
formations~\citep{Ziosi2014,DiCarlo2019}. Even the homogeneous isolated binary
channel is not excluded by the observed LIGO/Virgo low BH-BH effective spins, as apparently 
BHs with a broad range of spins ($a_{\rm spin} \sim 0.2-1$) can be produced also in this 
scenario (Ph. Podsiadlowski, private communication). In the published models with rather 
sparse metallicity sampling and conservative evolutionary assumptions (e.g., no 
pair-instability pulsation supernova mass loss\footnote{Note that pair-instability 
pulsation supernova mass loss, if taken into account, can reduce spin by about $30\%$
~\citep{Marchant2018}.}) BH spins are found in somewhat narrower range: 
$a_{\rm spin} \sim 0.4-1$ for $M_{\rm BH} \lesssim 60\msun$ (e.g., see Fig.9 of 
\cite{Marchant2016}). 

If a BH-BH merger is discovered with either of the binary components in the mass range 
$70/80 \lesssim M_{\rm BH} \lesssim 135\msun$ with high spin $a_{\rm spin} \approx 0.7$, 
this would strongly suggest a dynamical formation scenario. The lower-mass limit 
corresponds to the pair-instability pulsation supernova effects on the presupernova 
star and on the remnant BH mass~\citep{Woosley2017,Limongi2018,Belczynski2020a}, while 
the upper mass limit corresponds to the end of the pair-instability supernova process, 
that is believed to disrupt the entire star without BH formation~\citep{Fryer2001,
Heger2002}. The lack of BHs in this mass range is referred to as the ``second mass 
gap''~\citep{Belczynski2014,Spera2015,Marchant2016,2017ApJ...851L..25F}. It seems 
unlikely that isolated binaries can fill this gap, but repeated BH-BH mergers in 
dense environments could produce such heavy BHs, and it is expected that they would 
have moderately high spins $a_{\rm spin} \approx 0.7$~\citep{Gerosa2017,Fishbach2017a}. 
However, it cannot be excluded that a BH with high mass and low spin is formed by a 
merger of two BHs~\citep{Belczynski2020b}. 
Alternatively, the detection of a low-spin BH with mass within the second mass gap 
may point to either {\em (i)} inconsistencies in pair-instability supernova theory, 
or {\em (ii)} a primordial BH origin~\citep{Green2017}. 
There is also an issue of the second mass gap width. The lower bound may be as high 
as $70-80\msun$ depending on details of input physics in stellar models
~\citep{Limongi2018,Belczynski2020a}.
The upper bound may change if 
fusion reaction rates of heavy elements that are involved in pair-instability supernovae 
change. Note that the reaction rates are uncertain~\citep{fields_2018_aa}. 
Possibly, the second mass gap is narrower than currently believed.

In the classical isolated binary evolution channel we have shown that, if LIGO/Virgo 
observations of BH-BH mergers remain consistent with small effective spins 
$\chi_{\rm eff} \simeq 0$, this would indicate that low natal spins are common in 
BHs. In this case, our work shows that stellar models with mild rotational coupling 
between the stellar interior (core) and the outer zones (envelope) would be 
disfavored. Our main finding is that angular momentum transport in massive stars (so 
far unconstrained by the electromagnetic observations) is more efficient than 
predicted by the Geneva shellular model. This demonstrates how LIGO/Virgo 
observations can be used to make clear astrophysical inferences and guide stellar 
evolution astrophysics. Our conclusion is not subject to the main known population 
synthesis uncertainties, since we have explored a large range of the key parameters: 
BH natal kicks, initial star rotation, tides, spin-up by accretion onto BHs. 

Note that adjusting the angular momentum transport to produce low spinning BHs in 
BH-BH mergers, to fit LIGO/Virgo observations, leads to several astrophysical 
inferences and constraints on massive binary evolution. {\rm (i)} RLOF between two 
massive stars (see second evolutionary stage in BH-BH formation; Fig.~\ref{fig.evol1} 
or ~\ref{fig.evol2}) cannot effectively spin up the core of the accreting MS star 
(since this would produce a BH with large spin). This can be avoided if the RLOF is 
highly non-conservative and not much mass is accreted onto MS star (this is not 
excluded by any EM observations). Note that we have assumed that $50-80\%$ of the 
mass is lost in RLOF, but in our models this fraction can be easily increased. 
Alternatively, if accretion is significant and if the entire star is spun-up; then 
we obtain an additional constraint on the angular momentum transport. It needs to be 
effective enough to remove most of the angular momentum from the highly spinning 
core to the envelope in about $1$ Myr (time between RLOF and CE which expels the 
envelope from the binary). {\em (ii)} Spin-up of the first-formed BH by accretion 
in the CE phase must be negligible as predicted by recent studies \citep[][see the 
fifth evolutionary stage in BH-BH formation; Fig.~\ref{fig.evol1} and
~\ref{fig.evol2}]{MacLeod2017,Murguia2017,Holgado2017}. {\em (iii)} Tidal torques 
are not effective in BH-WR binaries that form BH-BH mergers, as the WR star spin-up 
would lead to the formation of a highly spinning BH (see the sixth evolutionary 
stage in BH-BH formation; Fig.~\ref{fig.evol1} and ~\ref{fig.evol2}).
In our evolution a small, but significant fraction ($\sim 20-30\%$; see Figs.~\ref{fig.xeff2} 
and ~\ref{fig.xeff3}) of the BH-WR/WR-BH/WR-WR binaries are found on orbits smaller 
than $\sim 10-20\rsun$ (orbital periods $P_{\rm orb}<1.3$\,d) for which the tides 
are expected to be effective for WR stars~\citep{Kushnir2016,Zaldarriaga2017,Qin2018}. 
Therefore if there are no detections of highly spinning BHs in LIGO/Virgo observations 
it will imply that tides are not as effective as argued by recent work, or that BH-BH 
mergers are not produced by the isolated classical binary evolution~\citep{Hotokezaka2017}. 
Note, however, that already one event (GW170729) may have an effective spin as high as 
$\chi_{\rm eff}=0.57$ ($90\%$ credible limits; see Tab.~\ref{tab.ligodata}). On the 
other hand, in ten events one can expect one outlier within the $90\%$ credible limits. 
Therefore at the moment the comparison of models with observations remains inconclusive 
in respect to tides. Yet, the comparison of models predicting tidally spun-up stars, 
producing highly spinning BHs, with the number of LIGO/Virgo high effective spin BH-BH 
mergers may help to constrain the strength of tides in near future.  

Finally, note that our main conclusion that angular momentum transport needs to be 
more effective than predicted by shellular model for massive stars adds support to 
the similar conclusion that was reached for low-mass stars from Kepler asteroseismology 
data~\citep{Cantiello2014}.

\section{Conclusions}
\label{sec.final}

We have updated our method of population synthesis calculations with revised and 
extended input physics that is important for the formation of double compact object 
mergers: BH-BH, BH-NS and NS-NS. New models of pair-instability pulsation supernovae 
that are crucial in mass estimates of heavy black holes have been introduced. We have 
employed the recent estimates of star--formation rate density and cosmic metallicity 
evolution in our calculations. These two factors play an important role in setting the
double compact object merger rates~\citep{Belczynski2010a,Dominik2013,Chruslinska2019,
Neijssel2019}. We have introduced new models for BH natal spin that allow for a direct 
comparison with LIGO/Virgo estimates of the effective spin parameter of BH-BH mergers. 
These models connect gravitational-wave observations with detailed stellar evolution 
calculations of angular momentum transport in massive stars that is otherwise hidden 
from electromagnetic observations. We have also allowed for significantly decreased 
stellar winds with respect to the standard wind prescriptions~\citep{Vink2001} used 
in modeling. Winds are an important factor that sets the shape of black hole mass
distribution. Finally, updated prescriptions of accretion onto compact objects in 
stable Roche-lobe overflow, common envelope, and from stellar winds have been used in 
our calculations. 

In the following we summarize our main findings. 
\begin{enumerate}

\item 
Our study is the first to employ detailed single stellar evolutionary
calculations in binary population synthesis to compare LIGO/Virgo BH-BH
effective spins with those resulting from several angular momentum transport
mechanisms. If LIGO/Virgo BH-BH mergers originate from the classical
isolated binary evolution channel, then their low effective spins inform
about the effective angular momentum transport in massive stars and most
likely disfavor effective tidal interactions in close binaries with WR stars.
According to our evolutionary framework the Tayler-Spruit magnetic dynamo
(as implemented in the MESA or in the Fuller model) reproduces the effective
spin measurements very well, while meridional currents (as implemented in
the Geneva model) are inconsistent with the current LIGO/Virgo data.\\

\item For some of our models the predicted merger rate densities of BH-BH, BH-NS and 
NS-NS systems are within the LIGO/Virgo empirical estimates. The rates
depend strongly on the cosmic metallicity evolution, the choice of NS/BH natal kicks
and the treatment of the common envelope phase. Due to the similar effects of these factors 
on the rates, it is still not possible to derive strong conclusions about
any of these pieces of input physics separately. However, some combinations
of parameters may be already excluded. Inter-parameter degeneracies are an
important factor that cannot be overlooked in deriving astrophysical conclusions from
gravitational-wave observations.\\  

\item The range of the observed BH masses appears to be in agreement with all three 
of our adopted PPSN models: from strong PPSN with maximum BH mass of 
$M_{\rm BH,max} \sim 40\msun$, to moderate PPSN with $M_{\rm BH,max} \sim 50\msun$, 
and to weak PPSN with  $M_{\rm BH,max} \sim 55\msun$. If heavier BHs are observed
it will indicate that, either the pair-instability pulsation supernovae and the 
pair-instability supernovae do not work as predicted, or that BHs with very heavy 
mass originate from other formation channels. The values of BH spins may distinguish  
these two possibilities. If a massive BH ($\sim 100 \msun$) is observed 
with low spin it will indicate a classical isolated binary evolution formation 
channel, but such an observation will cast shadow on PPSN/PSN predictions. If, 
however, such a BH is found to have a large spin as expected from consecutive BH 
mergers, then its presence will point to formation in a dense stellar environment 
(e.g., globular cluster).\\  

\item If, based on our limited set of models and the initial ten O1/O2 detections,
we were to make statements about the physics of massive binaries, we could venture
to say that with our revised cosmic metallicity evolution it seems that to
reproduce simultaneously the observed BH-BH, BH-NS, and NS-NS merger rates one should
use submodels B (no CE allowed with HG donors) if BH kicks are low, but if these kicks
are high, submodels A (CE allowed with HG donors) should be preferred. 
If individual BH masses are in fact as high as LIGO/Virgo reports for the most 
likely values from O1/O2 run (e.g., $50.6\msun$ for GW170729) then we can already 
exclude strong mass ejection during the PPSN.\\ 

\end{enumerate}

Note that similar conclusions were reached by \cite{Bavera2019} and 
\cite{Neijssel2019}, although somewhat different approach to detailed 
stellar evolution models and a different population synthesis code was used.

\begin{acknowledgements}
We have benefited from comments from Selma de Mink, Enrico
Ramirez-Ruiz, Ilya Mandel, Thomas Janka, Serena Repetto, Simon Stevenson,
Sambaran Banerjee, Tom Maccarone, Craig Heinke, Phil Charles, Doron Kushnir,
Tsvi Piran, Miguel Holgado, Antonio Claret, Tassos Fragos, Philipp 
Podsiadlowski, and Matt Benacquista.
We would like to thank thousands of {\tt Universe@home} users that have provided their
personal computers and phones for our simulations, and in particular to Krzysztof
Piszczek (program IT managers).
KB acknowledges support from the Polish National Science Center (NCN) grant  
Maestro (2018/30/A/ST9/00050) and  KB, SM and JPL from the NCN grant OPUS 
(2015/19/B/ST9/01099).
JK and MC acknowledge support from the Netherlands Organisation for 
Scientific Research (NWO).
C.E.F. acknowledges support from a Pre-doctoral Fellowship administered by
the National Academies of Sciences, Engineering, and Medicine on behalf of
the Ford Foundation, an Edward J Petry Graduate Fellowship from Michigan
State University, and the National Science Foundation Graduate Research
Fellowship Program under grant number DGE1424871.
MG ans AA were partially supported by the National Science Centre, Poland,
through the grant UMO-2016/23/B/ST9/02732.
This work has been supported by the EU COST Action CA16117 (ChETEC).
RH, KN, KB acknowledge support from the World Premier International
Research Centre Initiative (WPI Initiative), MEXT, Japan.
D.G. is supported by Leverhulme Trust Grant No. RPG-2019-350 and by NASA
through Einstein Postdoctoral Fellowship Grant No. PF6-170152 awarded by the
Chandra X-ray Center, which is operated by the Smithsonian Astrophysical
Observatory for NASA under contract NAS8-03060.
TB acknowledges support from the Polish National Science Center
(NCN) Grant No. UMO-2014/14/M/ST9/00707.
EB is supported by NSF Grants No.~PHY-1607130, AST-1716715, and by FCT
contract IF/00797/2014/CP1214/CT0012 under the IF2014 Programme.
C.G., G.M., and S.E. acknowledge support from the Swiss National Science
Foundation (project number 200020-160119).
ROS is supported by NSF AST-1412449, PHY-1505629, and PHY- 1607520.
DAB acknowledges support from National Science Foundation Grant No.
PHY-1404395.
G.W. is partly supported by the President’s International Fellowship
Initiative (PIFI) of the Chinese Academy of Sciences under grant
no.2018PM0017 and by the Strategic Priority Research Program of the Chinese
Academy of Science Multi-waveband Gravitational Wave Universe (Grant No.
XDB23040000).
AA is supported by the Carl Tryggers Foundation for Scientific Research
through the grant CTS 17:113.
MCM acknowledges support by NASA grant 80NSSC18K0527.
This research was supported in part by the National Science Foundation 
under Grant No. NSF PHY-1748958 to KITP UC Santa Barbara. Several authors 
(KB, EB, CLF, DAB, DEH, RO, KN, MCM, GM, JPL) were supported by KITP UC
Santa Barbara to work on this project. 
JPL was supported by a grant from the French Space Agency CNES.
S.C.L acknowledges support by the funding HST-AR-15021.001-A.
DEH was partially supported by NSF grant PHY-1708081, and also by the Kavli 
Institute for Cosmological Physics at the University of Chicago through an 
endowment from the Kavli Foundation. He also gratefully acknowledges support 
from a Marion and Stuart Rice Award.
We thank the Niels Bohr Institute for its hospitality while part of this work
was completed, and acknowledge the Kavli Foundation and the DNRF for supporting
the 2017 Kavli Summer Program.
\end{acknowledgements}

\bibliographystyle{aa}
\bibliography{biblio}

\clearpage
\section{Appendix}
\label{sec:appendix}

Data availability: data generated during this project with population synthesis 
code {\tt StarTrack} is available online: \url{www.syntheticuniverse.org}
under the tab "StarTrack models vs. Gravitational Wave Observations".
Code availability: the {\tt StarTrack} code is not an open source code and is 
not publicly available. New models will be calculated upon requests to the first 
author: chrisbelczynski@gmail.com.

\subsection{Geneva Stellar Models}
\label{sec.Geneva}

The physics included in the employed {\tt Geneva} rotating models is described
in detail in \citet{Eggenberger2008,Ekstrom2012}. The models have been computed
with a modest core-overshooting during the core H- and He-burning phase (the
core has been extended beyond the Schwarzschild limit by a length given by
$10\%$ of the local pressure scale height). Mass loss rates by stellar winds are
accounted for (see the above references for the details). The effects of
rotation are included according to the theory by \cite{Zahn1992}. In these
models, the transport of the angular momentum and the mixing of the chemical
species in the radiative zones are due to shear instabilities and meridional
currents. In contrast with many stellar models in which
the effects of meridional currents are accounted for by a diffusion equation,
in the present models the angular momentum transport is accounted for
by solving an advective equation. In essence, transport by meridional currents
is an advective process. Diffusion treatments can lead not only to
wrong estimates of the
amplitude, but also to wrong signs (a diffusive
process always tends to flatten any gradient, while an advective process can
increase gradients in some circumstances).
Note also that shear instabilities are the main drivers for the
transport of the chemical species, while angular momentum is mainly transported
by meridional currents. The angular momentum transport by meridional currents
mildly couples the core rotation with that of the envelope. In the convective
zones, the transport of angular momentum and chemical species is assumed to be
extremely efficient: convective zones are assumed to rotate as a solid body, and
the chemical composition is homogenized.

In Table~\ref{tab.spin1} we present a suite of {\tt Geneva} evolutionary models
~\citep{Ekstrom2012,Georgy2013,Groh2019} and Eggenberger et al. (in preparation) 
with $M_{\rm zams}=9$--$120\msun$ for a wide range of chemical compositions: 
$Z=0.014$--$0.0004$.  Specifically, we are listing CO core mass for each 
evolutionary model along with our estimate of the associated BH spin magnitude 
(see Fig.~\ref{fig.bhspin1} for a plot of the relation between the two).

The spin magnitude is very sensitive to the BH mass
($\propto 1/M_{\rm BH}^2$; see eq.~\ref{eq.bhmag}). The CO core mass
may be taken as a proxy for BH mass (cf. Appendix~\ref{sec.COBH} and
Figure~\ref{fig.mbhmco}). Below we present the behavior of the CO
core mass within the {\tt Geneva} models. This discussion explains the
non-monotonic dependence of BH spin on metallicity.

As naturally expected, the CO core mass increases with the initial
stellar mass within models of the same metallicity. This intuitive
trend is reversed only at the highest metallicity considered here
($Z=0.014$), at which stellar winds are most efficient in mass
removal. In particular, this is true for very massive stars
($M_{\rm zams} \gtrsim 100\msun$).

Another intuitive expectation is not supported by these stellar
models.  For a given initial stellar mass, one may naively expect that
the CO core mass would increase with decreasing metallicity, as winds
are getting weaker and star remains more massive. However, for example
for $M_{\rm zams}=85\msun$ we find
$M_{\rm CO,2}=26.4,\ 35.8,\ 27.4,\ 44.2\msun$ for
$Z=0.014,\ 0.006,\ 0.002,\ 0.0004$, respectively. In the following, we
explain this non-monotonic behavior in terms of our adopted stellar
evolution model.

Stellar winds are weaker at low metallicity thus this leads to increase of 
the CO core mass with decreasing metallicity. This general trend is mitigated 
by two other physical processes. Decreasing metallicity leads to the formation 
of extended convective H-burning shells that tend to slow down the growth
of the He core (and subsequently the CO core) mass. At high $Z$
massive stars lose most of their envelopes, so there is no vertical
structure available for extended convective H-burning shell. The
H-burning shell is compact and moves outwards (once H is totally
depleted) through the envelope, adding mass to the underlying He
core. At low $Z$, massive stars not only retain their envelopes
longer, but they are also more compact, and increased density helps to
form extended convective zones. Within the extended convective shell
reaching far above H-burning (which occurs only at the shell base),
intensive mixing keeps bringing new fuel to the H-burning zone
(keeping it in the same position), and keeps redistributing newly
formed He across this large shell.  Instead of adding He to the core,
the newly formed He is redistributed through the material of the
extended convective shell. Another process that leads to CO core mass
decrease with metallicity is connected to the diffusion of elements
due to the action of meridional currents and turbulence. With
decreasing metallicity stars are more compact and the vertical scale
of diffusion decreases, leading to less effective mixing of fresh fuel
into burning zones. This, in turn, lowers the CO core mass.

For the particular set of assumptions used in the {\tt Geneva} code,
the non-monotonic behavior of the CO core mass is the effect of the
complex and metallicity-dependent interplay of strength of stellar
winds, the H-burning shell extent, and the model for the efficiency of
element diffusion within the meridional current. This explains the
non-monotonic metallicity dependence of our BH natal spin model (see
Fig.~\ref{fig.bhspin1}).

\begin{table*}
\caption{Geneva stellar models\tablefootmark{a}.}
\centering
\begin{tabular}{c r c r c r c c c c c}
\hline\hline
Z & $M_{\rm ZAMS}$ & $V_{\rm rot,i}$ & $M_{\rm He}$ & $M_{\rm CO,1}$ & $M_{\rm CO,2}$ & $P_{NS},a_{\rm spin}$ & $M_{\rm NS/BH}$ & $M_{\rm star,g}$ & $M_{\rm star,st1}$ & $M_{\rm star,st2}$ \\
  & [$\msun$]      & [km s$^{-1}$]   & [$\msun$]    & [$\msun$]      & [$\msun$]    &  [ms,]                & [$\msun$]       & [$\msun$]        & [$\msun$]          & [$\msun$] \\ 

\hline
\hline
   {\bf Z=0.014} &   9 &  248 & 2.9  & 1.61 & 2.40 & 0.6  & 1.18 &  8.5 &  8.5 &  8.9  \\
                 &  12 &  262 & 3.8  & 2.26 & 3.40 & 0.6  & 1.19 & 10.2 & 11.0 & 11.7  \\
                 &  15 &  271 & 5.1  & 3.07 & 4.59 & 0.5  & 1.35 & 11.1 & 11.3 & 13.9  \\
                 &  20 &  274 & 7.1  & 4.50 & 6.86 & 0.5  & 1.76 &  7.2 & 12.3 & 17.6  \\
                 &  25 &  295 & 9.7  & 6.69 & 9.38 & 0.90 & 9.66 &  9.7 &  9.2 & 20.4  \\
                 &  32 &  306 & 10.1  & 7.45 & 10.1 & 0.88 & 6.24 & 10.1 & 10.5 & 23.4  \\
                 &  40 &  314 & 12.3  & 9.29 & 12.3 & 0.87 & 7.88 & 12.3 & 11.6 & 22.1  \\
                 &  60 &  346 & 18.0  & 14.2 & 18.0 & 0.62 & 14.2 & 18.0 & 13.5 & 36.0  \\
                 &  85 &  368 & 26.4  & 21.6 & 26.4 & 0.29 & 21.6 & 26.4 & 18.1 & 52.6  \\
                 & 120 &  389 & 19.0  & 15.2 & 19.0 & 0.13 & 15.2 & 19.0 & 20.7 & 66.9  \\

\hline
   {\bf Z=0.006} &  15 &  271 & 5.1  & 2.87 & 4.71 & 0.5  & 1.30 & 14.0 & 13.5 & 14.5  \\
                 &  20 &  292 & 7.2  & 4.41 & 7.02 & 0.5  & 1.72 & 13.9 & 16.7 & 19.0  \\
                 &  25 &  301 & 9.6  & 6.50 & 9.55 & 0.80 & 9.62 & 10.5 & 17.8 & 22.8  \\
                 &  32 &  334 & 13.5  & 9.98 & 13.4 & 0.90 & 11.6 & 13.6 & 17.5 & 27.5  \\
                 &  40 &  334 & 18.9  & 14.9 & 18.7 & 0.90 & 18.9 & 18.9 & 14.8 & 21.3  \\
                 &  60 &  378 & 32.8  & 28.3 & 32.8 & 0.90 & 28.2 & 32.8 & 18.9 & 38.6  \\
                 &  85 &  410 & 35.8  & 30.2 & 35.8 & 0.37 & 30.1 & 35.8 & 28.8 & 61.5  \\
                 & 120 &  435 & 52.5  & 45.1 & 52.4 & 0.25 & 42.3 & 52.5 & 31.4 & 77.9  \\

\hline
   {\bf Z=0.002} &   9 &  255 & 2.8  & 1.55 & 1.59 & 0.5  & 1.20 &  8.9 &  8.4 &  8.9  \\
                 &  12 &  271 & 3.9  & 2.01 & 3.17 & 0.5  & 1.20 & 11.8 &  9.4 & 11.3  \\
                 &  15 &  303 & 5.0  & 2.76 & 4.65 & 0.5  & 1.26 & 14.7 & 12.8 & 14.6  \\
                 &  20 &  305 & 7.2  & 4.36 & 7.04 & 0.5  & 1.73 & 18.7 & 17.8 & 19.3  \\
                 &  25 &  319 & 9.6  & 6.29 & 9.42 & 0.90 & 9.56 & 21.8 & 19.8 & 23.4  \\
                 &  32 &  338 & 13.1  & 9.32 & 13.0 & 0.85 & 10.2 & 24.1 & 18.9 & 28.0  \\
                 &  40 &  358 & 17.6  & 13.3 & 17.5 & 0.90 & 17.6 & 27.4 & 14.5 & 17.1  \\
                 &  60 &  400 & 31.6  & 26.7 & 27.4 & 0.001 & 31.6 & 39.1 & 22.8 & 39.3  \\
                 &  85 &  435 & 45.3  & 26.7 & 27.4 & 0.28 & 43.0 & 74.8 & 35.7 & 66.6  \\
                 & 120 &  438 & 85.6  & 76.4 & 83.2 & PSN  &  0.0 & 85.6 & 46.6 & 91.6  \\  

\hline
  {\bf Z=0.0004} &   9 &  270 & 1.6  & 1.54 & 1.59 & 0.6  & 1.20 & 8.90 &  8.9 &  9.0  \\
                 &  12 &  285 & 3.9  & 2.02 & 3.28 & 0.6  & 1.19 & 11.9 & 11.8 & 11.9  \\
                 &  15 &  319 & 5.0  & 2.87 & 4.81 & 0.5  & 1.28 & 14.9 & 14.8 & 14.9  \\
                 &  20 &  314 & 7.8  & 4.84 & 7.35 & 0.4  & 1.86 & 19.4 & 19.7 & 19.9  \\
                 &  25 &  343 & 10.0  & 6.60 & 9.81 & 0.88 & 10.0 & 24.0 & 24.4 & 24.8  \\
                 &  32 &  366 & 12.6  & 8.71 & 12.5 & 0.80 & 8.86 & 30.5 & 30.0 & 31.4  \\
                 &  40 &  393 & 16.9  & 12.5 & 16.9 & 0.82 & 12.5 & 34.6 & 15.6 & 23.4  \\
                 &  60 &  435 & 28.4  & 28.4 & 27.8 & 0.79 & 28.4 & 49.9 & 24.9 & 41.7  \\
                 &  85 &  469 & 47.0  & 40.1 & 44.2 & 0.26 & 44.6 & 57.8 & 37.9 & 68.5  \\
                 & 120 &  463 & 70.2  & 62.1 & 65.3 & PSN  & 0.0  & 92.5 & 66.6 &  104  \\ 

\hline
\hline
\end{tabular}
\tablefoot{\\
\tablefootmark{a}{For all models we list:
$Z$: metallicity, 
$M_{\rm ZAMS}$: initial star mass,
$V_{\rm rot,i}$: initial rotation at equator, 
$M_{\rm He}$: He core mass,
$M_{\rm CO,1}$: CO core mass defined by $<1\%$ He (used in MESA),
$M_{\rm CO,2}$: CO core mass defined by $>20\%$ CO (adopted in our study),
$P_{\rm NS},a_{\rm spin}$: for NSs, the NS spin period in milliseconds, for BHs, the
dimensionless spin magnitude, 
$M_{\rm NS/BH}$: remnant mass,
$M_{\rm star,g}$: final mass of a star in Geneva code,
$M_{\rm star,st1}$: final mass of a star in {\tt StarTrack} with standard winds,
$M_{\rm star,st2}$: final mass of a star in {\tt StarTrack} with reduced ($30\%$) winds.}
}
\label{tab.spin1}
\end{table*}

\begin{table*}
\caption{MESA stellar models with $40\%$ critical initial rotation\tablefootmark{a}.}
\centering
\begin{tabular}{c r c r r r c r r r r}
\hline\hline
Z & $M_{\rm ZAMS}$ & $V_{\rm rot,i}$ & $M_{\rm He}$ & $M_{\rm CO,1}$ & $M_{\rm CO,2}$ & $P_{NS},a_{\rm spin}$ & $M_{\rm NS/BH}$ & $M_{\rm star,m}$ & $M_{\rm star,st1}$ & $M_{\rm star,st2}$ \\
  & [$\msun$]      & [km s$^{-1}$]   &[$\msun$]    & [$\msun$]      & [$\msun$]    &     [ms,]             &       [$\msun$] & [$\msun$]        & [$\msun$]          & [$\msun$] \\ 
\hline
\hline
   {\bf Z=0.014} &  10 &  257 & 2.9  &  1.2 &  1.5 &   8.0 & 1.20 &   3.4 &  9.4 &   9.8 \\
                 &  15 &  270 & 4.5  &  2.2 &  2.7 &   6.4 & 1.27 &   5.7 & 11.3 &  13.9 \\
                 &  20 &  274 & 6.7  &  3.6 &  4.6 &   4.4 & 1.79 &  10.8 & 12.3 &  17.6 \\
                 &  25 &  276 & 7.8  &  4.2 &  6.2 & 0.094 & 6.16 &  19.9 &  9.2 &  20.4 \\
                 &  32 &  272 & 12.0  &  7.9 &  9.1 & 0.107 & 9.23 &  25.3 & 10.5 &  23.4 \\
                 &  40 &  267 & 16.6  & 11.8 & 13.2 & 0.105 & 16.8 &  27.8 & 11.6 &  22.1 \\
                 &  60 &  254 & 27.6  & 21.1 & 22.6 & 0.083 & 28.0 &  46.5 & 13.5 &  36.0 \\
                 &  85 &  239 & 41.6  & 33.4 & 35.2 & 0.046 & 42.6 &  56.2 & 18.1 &  52.6 \\
                 & 120 &  220 & 60.8  & 49.9 & 52.8 & 0.037 & 55.0 &  80.0 & 20.7 &  66.9 \\
\hline
   {\bf Z=0.006} &  10 &  268 & 2.9  &  1.0 &  1.6 &   6.7 & 1.20 &   3.9 &  9.6 &   9.9 \\
                 &  15 &  279 & 4.7  &  2.2 &  2.9 &   5.9 & 1.32 &  10.0 & 13.5 &  14.5 \\
                 &  20 &  293 & 6.8  &  3.8 &  4.6 &   9.7 & 1.79 &  18.3 & 16.7 &  19.0 \\
                 &  25 &  296 & 8.6  &  5.1 &  6.0 & 0.101 & 8.67 &  23.0 & 17.8 &  22.8 \\
                 &  32 &  295 & 11.8  &  7.9 &  8.9 & 0.100 & 8.87 &  29.3 & 17.5 &  27.5 \\
                 &  40 &  289 & 14.4  &  8.9 & 10.3 & 0.099 & 16.1 &  31.3 & 14.8 &  21.3 \\
                 &  60 &  277 & 26.6  & 20.3 & 21.9 & 0.093 & 26.8 &  52.4 & 18.9 &  38.6 \\
                 &  85 &  266 & 39.9  & 31.2 & 32.9 & 0.081 & 47.1 &  69.0 & 28.8 &  61.5 \\
                 & 120 &  251 & 61.8  & 50.6 & 53.3 & 0.072 & 55.0 &  91.7 & 31.4 &  77.9 \\

\hline
   {\bf Z=0.002} &  10 &  285 & 3.0  &  1.1 &  1.6 &  7.1  & 1.20 &   6.5 &  8.4 &   9.6 \\
                 &  15 &  298 & 4.9  &  2.5 &  3.0 &  7.7  & 1.35 &  14.7 & 12.8 &  14.6 \\ 
                 &  20 &  305 & 6.6  &  3.7 &  4.3 &  7.6  & 1.72 &  19.5 & 17.8 &  19.3 \\
                 &  25 &  310 & 8.2  &  4.9 &  7.7 & 0.096 & 8.29 &  23.9 & 19.8 &  23.4 \\
                 &  32 &  312 & 12.0  &  8.1 &  9.0 & 0.104 & 9.11 &  30.4 & 18.9 &  28.0 \\
                 &  40 &  313 & 15.4  & 10.9 & 12.1 & 0.142 & 15.4 &  37.3 & 14.5 &  17.1 \\
                 &  60 &  304 & 26.9  & 20.8 & 22.4 & 0.108 & 27.0 &  55.1 & 22.8 &  39.3 \\
                 &  85 &  286 & 39.5  & 31.6 & 33.6 & 0.090 & 39.7 &  76.9 & 35.7 &  66.6 \\
                 & 120 &  268 & 60.9  & 50.1 & 54.1 & 0.065 & 55.0 &   105 & 46.6 &  91.6 \\

\hline
  {\bf Z=0.0004} &  10 &  302 & 3.1  &  1.4 &  1.7 &  7.1  & 1.20 &   9.8 &  9.9 &  10.0 \\
                 &  15 &  315 & 4.9  &  2.5 &  3.1 &  8.5  & 1.36 &  14.9 & 14.8 &  14.9 \\
                 &  20 &  323 & 6.8  &  3.9 &  4.6 &  5.2  & 1.79 &  19.7 & 19.7 &  19.9 \\
                 &  25 &  329 & 8.7  &  5.4 &  6.2 & 0.138 & 8.81 &  24.5 & 24.4 &  24.8 \\
                 &  32 &  335 & 12.4  &  8.3 &  9.3 & 0.110 & 9.83 &  31.0 & 30.0 &  31.4 \\
                 &  40 &  338 & 14.3  &  9.7 & 10.7 & 0.109 & 13.7 &  38.3 & 15.6 &  23.4 \\
                 &  60 &  341 & 40.0  & 32.3 & 34.1 & 0.058 & 43.1 &  55.7 & 24.9 &  41.7 \\
                 &  85 &  340 & 43.3  & 34.7 & 36.9 & 0.097 & 41.7 &  79.9 & 37.9 &  68.5 \\
                 & 120 &  334 & 63.1  & 47.3 & 61.1 & 0.072 & 28.4 &   112 & 66.6 &   104 \\

\hline
\hline
\end{tabular}
\tablefoot{\\
\tablefootmark{a}{For all models we list:
$Z$: metallicity, 
$M_{\rm ZAMS}$: initial star mass,
$V_{\rm rot,i}$: initial rotation at equator,
$M_{\rm He}$: He core mass,
$M_{\rm CO,1}$: CO core mass defined by $<1\%$ He (used in MESA),
$M_{\rm CO,2}$: CO core mass defined by $>20\%$ CO (adopted in our study),
$P_{\rm NS},a_{\rm spin}$: for NSs, the NS spin period in milliseconds, for BHs, the
dimensionless spin magnitude, 
$M_{\rm NS/BH}$: remnant mass,
$M_{\rm star,m}$: final mass of a star in MESA,
$M_{\rm star,st1}$: final mass of a star in {\tt StarTrack} with standard winds,
$M_{\rm star,st2}$: final mass of a star in {\tt StarTrack} with reduced ($30\%$) winds.}
}
\label{tab.spin2}
\end{table*}

\subsection{MESA Stellar Models}
\label{sec.MESA}

The \texttt{MESA} stellar models were evolved using \texttt{MESA} revision {\tt 10398} 
\citep{paxton_2011_aa,paxton_2013_aa,Paxton2015,paxton_2018_aa,paxton_2019_aa}.
The models are evolved from the pre-main sequence to core He-depletion. We use 
temporal and spatial parameters similar to those used in \citet{farmer_2016_aa} 
and \citet{fields_2018_aa} that provide convergence to the $\approx$ 10$\%$ level.
The \texttt{MESA} models use the \texttt{mesa-49.net} network that follows $49$ 
isotopes from $^{1}$H to $^{34}$S. We include mass loss using the `\texttt{Dutch}` 
wind scheme with an efficiency value of 0.8 \citep{nieuwenhuijzen_1990_aa,
nugis_2000_aa,vink_2001_ab,glebbeek_2009_aa}. We use the Ledoux criterion for 
convection with an efficiency parameter of $\alpha_{\rm{MLT}}=1.6$, and the
\texttt{mlt++} approximation for convection \citep{paxton_2013_aa}. 

Additional mixing processes due to convective boundary mixing are included using the 
exponentially decaying diffusion coefficient framework of \citet{Herwig2000} based on 
hydrodynamic simulations of \citet{Freytag1996}. The following values of $f$ were used: 
$f=0.014$ above H- and He-burning regions,  $f=0.001$ below H- and He-burning regions, 
$f=0$ elsewhere. Note that the additional parameter, $f_0=0.001$ was used, where 
$f_0 H_P$ is the distance from the boundary inside the convective zone where the 
exponential decay starts. \citet{Jones2015} compare \texttt{GENEC} models to 
\texttt{MESA} models and find that using $f=0.022$ on top of convective core H and He 
burning in \texttt{MESA} matches \texttt{GENEC} models with a penetrative overshoot, 
$\alpha_{\rm ov}=0.2 H_P$. Given that the \texttt{GENEC} models used in this paper use 
$\alpha_{\rm ov}=0.1 H_P$, using $f=0.014$ in \texttt{MESA} yields similar core masses 
and lifetimes.

In \texttt{MESA}, rotation is implemented using the shellular 
approximation \citep{Zahn1992,meynet_1997_aa}. In our models, we initialize solid 
body, uniform rotation at the zero age main-sequence as a fraction of the Keplerian 
critical rotation rate, in most of our models this value is $40\%$. We use the 
suggested values from \citep{heger_2000_aa} for the diffusion coefficients for the 
transport of angular momentum and material due to various instabilities. Of the
most influential of these mechanisms included in the \texttt{MESA} models is the 
Tayler-Spruit dynamo. The modeling of this dynamo in the stellar models increases 
the efficiency in angular momentum transport and can lead to a significantly 
different angular momentum profiles and core rotation rates \citep[][see also 
Sec.~\ref{sec.ang_mom_transport}]{heger_2005_aa}. 

We find that, at the essence, the Taylor-Spruit dynamo dominates the transport of 
angular momentum in our \texttt{MESA} models and is very efficient at transporting 
the angular momentum from the core to the upper layers, from which it is lost in 
winds. As a result, all our models end with a very similar (small) amount of angular 
momentum and thus small spin parameter $a<0.15$ , see Figure~\ref{fig.bhspin2} and also 
Table.~\ref{tab.spin2}. Given that stronger stellar winds lead to a more efficient 
loss of angular momentum, one would expect a dependence of the final spin on mass 
and metallicity. We find a hint of such a dependence in our models: the more massive 
as well as the higher metallicity models tend to end with systematically lower final 
spins (Fig.~\ref{fig.bhspin2}). However, this is not very clear in our models, and 
the impact of mass and metallicity is sub-dominant to the fact that the efficient 
transport of angular momentum through the Taylor-Spruit dynamo leads to small spins. 

\begin{figure}
\includegraphics[width=\columnwidth]{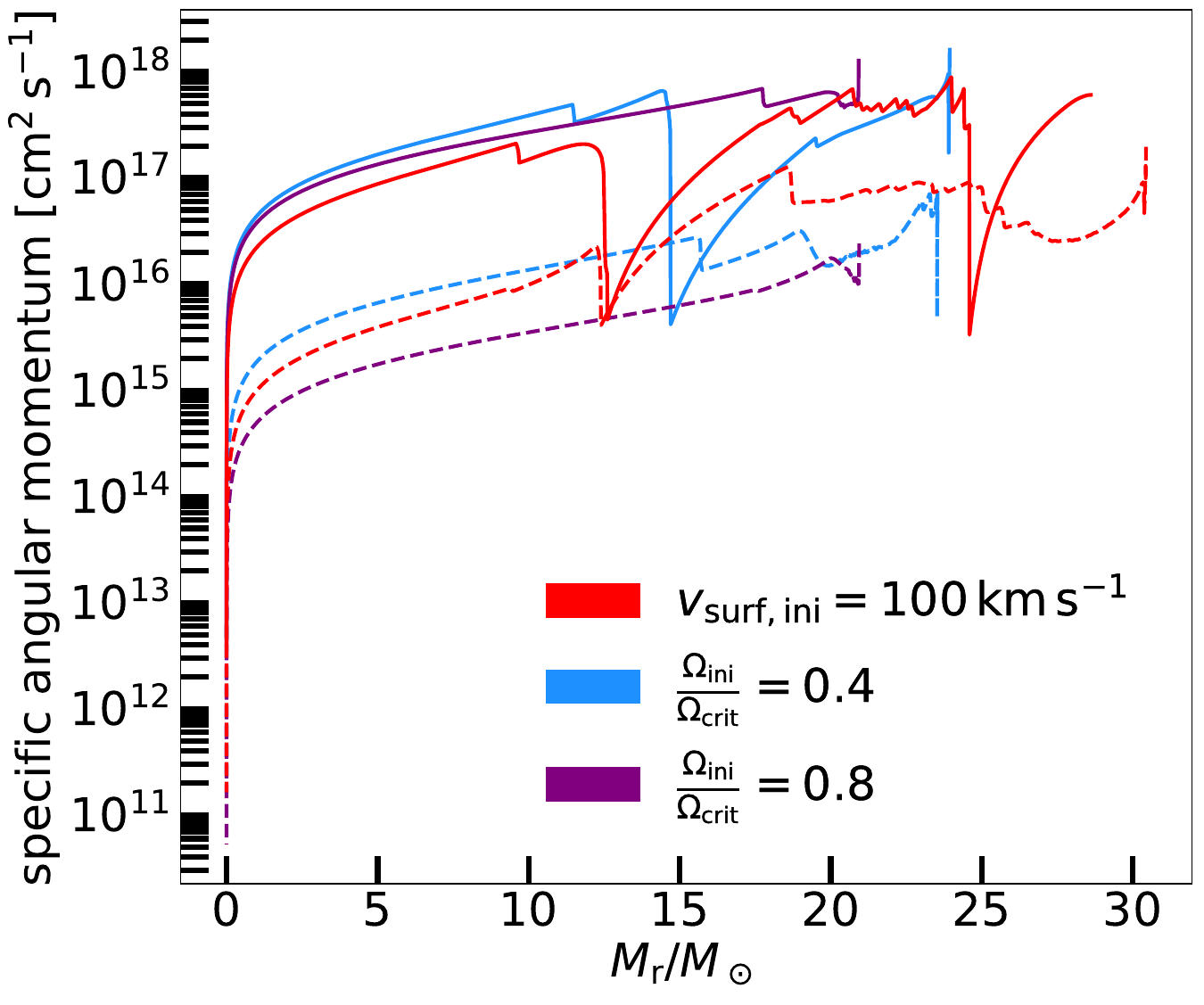}
\caption{Specific angular momentum profile at the end of core helium burning for the 
$32\msun$ \texttt{MESA} models at $Z=0.002$. The models were calculated with different 
initial rotation rates; slow ($100\kms$, red), fast ($40\%$ of critical rotation, blue) 
and very fast ($80\%$ of critical rotation, purple). Magnetic (dashed line) and 
non-magnetic models (solid line) are considered. 
Large dips in the curve indicate the base of a convective region, where angular 
momentum transport is very efficient. The maximal extent of the curves to the right 
indicate the mass of the star. We can see from these curves that the faster rotating 
models lose more mass in winds due to rotation-enhanced mass loss.
}
\label{fig.jrot_endHe_MESA}
\end{figure}

A star might spin up during its main-sequence evolution due to mass accretion 
from a companion or be born as a very fast rotator. We mimic this scenario with 
a single, very fast rotating \texttt{MESA} model ($80 \%$ of critical rotation) 
and compare it to the slow and fast rotating $32 \msun$ \texttt{MESA} models plotted 
in Figure~\ref{fig.jrot_endHe} for both, magnetic and non-magnetic angular momentum 
transport. The specific angular momentum profile at core helium depletion of the very 
fast rotating models is presented in Figure~\ref{fig.jrot_endHe_MESA}. It clearly 
shows, as already pointed out earlier, that the rotation rate at core helium depletion 
depends mainly on the angular momentum transport mechanism (the difference between 
the models using the Taylor-Spruit dynamo will be further reduced during the advanced 
phases of stellar evolution) and not the initial rotation rate. This is a result of 
the evolution after the main-sequence where the Taylor-Spruit dynamo extracts most of 
the angular momentum from the core. The core contracts due to the missing energy 
generation and the envelope expands as a consequence of the virial theorem and energy 
conservation. The contracting core spins up faster in the models with faster initial 
rotation, creating a stronger shear between core and envelope. The stronger shear leads 
to a stronger angular momentum transport of the magnetic dynamo 
($\nu_{\rm ST} \sim -(\frac{d \ln \Omega}{d \ln r})^2$) in a region where stratification 
is dominated by the chemical composition gradient \citep{Spruit2002}, which is the case 
for the region above the helium core where the shear develops after the MS evolution in 
a massive star), resulting in similar specific angular momentum profiles. Therefore, a 
star that is either spun up by a binary companion or is born fast will still end its life 
with a slow spinning core when an efficient angular momentum transport mechanism is active.

In the non-magnetic models presented in Figure~\ref{fig.jrot_endHe_MESA}, most of the 
angular momentum is transported in the short phase between core hydrogen and helium 
burning (as in the magnetic models) but the transport is much less efficient. 
Therefore, the core of models excluding magnetic fields will not be able to slow 
down and will have a high final rotation rate. 

There are dips in the angular momentum profiles in Figure~\ref{fig.jrot_endHe_MESA}, 
One of the red curves (slowest rotation) has two dips, because when helium is exhausted 
in the core there is a convective hydrogen zone (starting at $12\msun$) and the surface 
convective zone (starting at $25\msun$) present. This is not the case in the other models 
because {\em (i)} they evolve more to the blue (to higher temperature), {\em (ii)} faster 
rotating models tend to smooth temperature and chemical composition, leading to a shorter 
time with convective hydrogen core and {\em (iii)} some models lose nearly their entire 
envelope, hence, there is no surface nor hydrogen convective zone. 
Models with high rotation end evolution with lower total mass as visible in 
Figure~\ref{fig.jrot_endHe_MESA}. Rotation enhances mass loss in general in several ways: 
directly by reducing the effective gravity, indirectly by rotation-induced mixing leading 
to increased luminosities and sometimes leading to quicker evolution to the red supergiant 
or WR phases. For the very fast rotating models evolving quasi-chemically homogeneously, 
the dominant effects are rotation-induced mixing leading to increased luminosities, which 
in turn enhances the wind mass loss and reaching the WR phase earlier in the evolution. 

We note that apart from the spin values, there are also differences between the final 
CO core masses between \texttt{GENEC} and \texttt{MESA} models (Table~\ref{tab.spin1} 
and \ref{tab.spin2}). Some level of discrepancy is not unexpected solely due to the 
fact that those sets of models were run with slight differences in the efficiency of 
chemical mixing, criteria for convection, and treatment of convective boundary 
regions. More importantly, even though both sets of modes were computed using similar 
prescriptions for mass loss, differences in evolution in the HR diagram alone can 
lead to significant differences in the total amount of mass lost in winds and 
therefore the final CO core masses. This can be most clearly seen when comparing the 
most massive Solar metallicity models ($Z = 0.014$, initial masses $60$, $85$, and 
$125\msun$), in which case the  \texttt{MESA} models end their evolution with 
noticeably higher CO core masses. We caution that in the case of such massive stars 
and especially at higher metallicity, the difficulties in numerical treatment of 
radiation-dominated, super-adiabatic envelope layers can lead to significant 
differences in the HR diagram evolution between different codes \citep{paxton_2013_aa}. 
Given that such envelopes are likely dynamically unstable and highly turbulent, any 
1D models of late evolution of massive stars should be consider highly uncertain. 	

All \texttt{MESA} inlists used to produce these models are publicly available at 
\href{http://mesastar.org}{mesastar.org}.

\subsection{CO Core Mass versus BH Mass}
\label{sec.COBH}

\begin{figure}
    \includegraphics[width=\columnwidth]{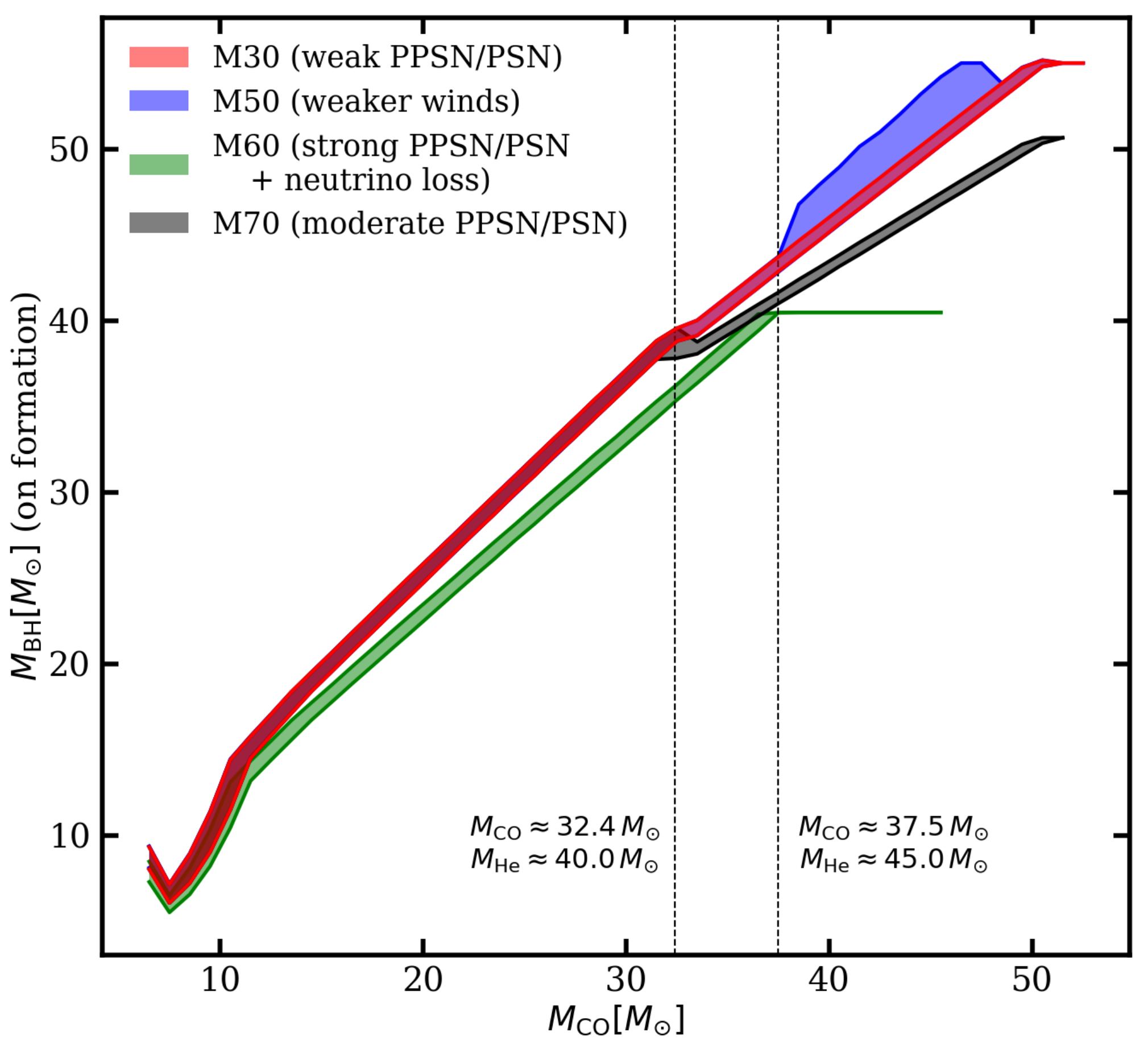}
    \caption{
Final CO core masses of BH progenitors and their corresponding BH formation masses for 
the components (both primary and secondary) of detectable merging BH-BH systems in four 
of our models: M30, M50, M60, and M70 (see Table.~\ref{tab.models} for an overview of 
all the model ingredients). For readability, the filled areas encapsulate $96\%$ of all 
the BHs, excluding few outliers and showcasing the $M_{\rm CO}$ to $M_{\rm BH}$ relation 
for the representative majority. The presented models are different in ways that affect 
the BH formation masses, i.e. the PPSN/PSN prescription (see Sec.~\ref{sec.bh_mass}) as 
well as the assumed fraction of mass lost in neutrinos during a BH formation ($1\%$ in 
M30, M50, M70 and $10\%$ in M60). Notice that model M70 only differs from M30 and M50 for 
$M_{\rm CO} \gtrsim 32.4 \msun$, whereas model M50 only differs from M30 for 
$M_{\rm CO}>37.5\msun$. Those threshold core masses correspond to the threshold helium 
core masses $M_{\rm He}$ in different models PPSN/PSN prescriptions. 
}
\label{fig.mbhmco}
\end{figure}

The relation between the BH mass at formation and the final CO core mass of its 
progenitor in our simulations is set by formulae based on supernova modeling
~\citep[the 'Rapid' engine of][]{Fryer2012}, together with the pre-supernova mass 
of the star that is set by stellar and binary evolution. In Figure~\ref{fig.mbhmco} 
we show the $M_{\rm CO}$--$M_{\rm BH}$ relation for BH-BH mergers detectable in the 
O1/O2 LIGO runs within models M30, M50, M60, and M70. The presented models are 
different in ways that affect the BH formation masses, i.e. the PPSN/PSN prescription 
(see Sec.~\ref{sec.bh_mass}) as well as the assumed fraction of mass lost in 
neutrinos during a BH formation (1\% in M30, M50, M70 and 10\% in M60). Note that 
even though we plot $M_{\rm BH}$ at the moment of BH formation, it is very similar to 
the final BH mass. We note that the CO core masses presented in Figure~\ref{fig.mbhmco} 
are the final CO core masses (just before the BH formation), which in some cases have 
already been reduced due to PPSN mass loss. Those are the $M_{\rm CO}$ masses that we 
use in order to assign the newly formed BHs with natal spins (see Sec.~\ref{sec.spins}).

In general, the $M_{\rm CO}$ to $M_{\rm BH}$ relation is quite similar between our 
models and, for the most part, almost linear. The initial decrease in $M_{\rm BH}$ (for 
$M_{\rm CO} < 7.5 \msun$), and the subsequence increase followed by a change in slope 
at around $M_{\rm CO} = 11 \msun$ are the results of changes in the fraction of material 
$f_{\rm fb}$ that falls back onto the proto-NS in the 'Rapid' supernovae engine 
\citep[see Eq.~16 in][]{Fryer2012}. The maximum fall-back ($f_{\rm fb} = 1.0$, a direct 
collapse formation of a BH) is expected for the CO core masses $M_{\rm CO}$ either in 
range between $6$ and $7\msun$ or for $M_{\rm CO} > 11 \msun$. Partial fall-back and the 
ejection of some of the pre-SN mass for progenitors with $M_{\rm CO}$ within $7$ and 
$11\msun$ is what is responsible for the dip in $M_{\rm BH}$ in Figure~\ref{fig.mbhmco} 
at the lower $M_{\rm CO}$ end. 

The fact that BH masses are systematically smaller in model M60 compared to other models 
is a direct consequence of a higher mass fraction being lost in neutrinos during a BH 
formation (10\% in M60 compared to 1\% in other models).

The influence of PPSN/PSN kicks in above $M_{\rm CO} \approx 32.4 \msun$ for models M30, 
M50, M70 (which corresponds to $M_{\rm He} \approx 40 \msun$) and above 
$M_{\rm CO} \approx 37.5 \msun$ in the case of M60 ($M_{\rm He} \approx 45 \msun$). Above 
this core mass, PPSN removes outer part of the envelope and therefore reduces the final 
pre-SN mass with respect to evolution without PPSN (see Fig.~\ref{fig.ppsn}). This leads 
to a smaller mass of the BH formed in direct collapse of the remaining post-PPSN star. As 
expected, Figure~\ref{fig.mbhmco} reveals that the more mass is lost in PPSN (depending 
on the model) the smaller the final BH mass. Finally, in the case of most massive stars 
($M_{\rm He} > 65 \msun$) all the models assume that a PSN disrupts the entire star and 
that no remnant remains. 

The impact of weaker stellar winds in model M50 ($30\%$ of mass loss in winds at any 
evolutionary stage compared to other models) on the $M_{\rm CO}$ to $M_{\rm BH}$ relation 
is relatively small, only differentiating models M30 and M50 in any way for 
$M_{\rm CO}>40\msun$. This is because, in most cases of the BH-BH merger formation, both 
and primary and secondary are going to lose their entire envelope prior to core collapse 
anyway, due to RLOF. In fact, as many as $90\%$ of all the BH formed from $M_{\rm CO}>40\msun$ 
progenitors in model M50 fall into the same area in Figure~\ref{fig.mbhmco} as BHs in 
model M30.

\subsection{Making GW170104}
\label{sec.GW170104}

Figure~\ref{fig.evol1} shows the formation and evolution of a binary which results in 
a BH-BH merger with properties similar to GW170104. This particular system was evolved 
within our model M20 (see Sec.~\ref{sec.calcul} for an overview of our models and 
Sec.~\ref{sec.Binary} for an overview of our binary evolution calculations) under the 
assumptions of Geneva-based BH natal spins (Fig.~\ref{fig.bhspin1}) and no WR spin-up 
during the BH-WR stage (Sec.~\ref{sec.tides}. The progenitor binary was formed in a 
low-metallicity environment $Z=0.001$ ($5\% \zsun$) as a pair of MS stars with masses 
of $94.6\msun$ and $62.5\msun$ on a wide orbit, with pericenter distance of 
$\sim 2000\rsun$. As the primary ends its MS evolution and rapidly expands as a HG star, 
it initiates a stable case-B mass transfer.  In model M20 we assume that $20\%$ of the 
mass is accreted by the MS companion, while the other $80\%$ is lost from the system. 
During mass transfer the primary is stripped of almost all of the H-rich envelope, and 
it becomes a naked helium WR star with mass of $41.7\msun$. Only $0.3$ Myr later it 
finishes its nuclear evolution and forms a BH ($M_{\rm BH1}=32.9\msun$) in a direct 
collapse event (no natal kick), losing $10\%$ of its mass in neutrino emission. The 
mass of the CO core of the BH progenitor is $M_{\rm CO}=29.6\msun$, which means that
the first BH is formed with zero natal spin ($a_{\rm spin1} = 0.0$, see 
Fig.~\ref{fig.bhspin1} for $Z = 0.002$ (used as representative for $Z = 0.001$).

The companion, on the other hand, has increased its mass during the mass
transfer to $71.1\msun$. It has also been rejuvenated (which we model
according to eq.~58 of \cite{Belczynski2008a}) and spun up. At the offset
of mass transfer, it is about $80\%$ of its way through the MS phase. During
the HG stage it expands to $1330\rsun$, which is, however, not enough to
cause a Roche lobe overflow (RLOF). It further increases its radius to
$1650\rsun$ as a core-He-burning (CHeB) supergiant, at which point it
initiates a dynamically unstable mass transfer and the common envelope (CE)
phase. As a result, the wide binary orbit ($a=3678\rsun$) decays to $a=34\rsun$,
and the secondary is left without any envelope as a naked WR star.  We
assume a $5\%$ Bondi-Hoyle accretion rate onto a compact object during
CE, which leads the first BH to only a slight increase in mass (by
$0.6\msun$) and spin ($a_{\rm spin1}=0.05$). Shortly after ($0.3$ Myr
after the CE) the secondary forms a BH with mass
$M_{\rm BH2}=24.7\msun$ in a direct collapse with no natal kick and
only $10\%$ neutrino mass loss. Because the secondary accretes a
significant fraction of its mass ($\sim 20\%$) while still on the MS,
we assume that its increased rotation leads to a $\sim 20\%$ increase
of the CO core mass with respect to the non-rotating stellar models
(see Sec.~\ref{sec.Binary}).  With the increased CO core mass of
$M_{\rm CO}=26.0\msun$, the second BH is assigned an initial spin of
$a_{\rm spin2} = 0.14$.

The BH-BH is formed after $4.9$ Myr of binary evolution on a close
($a=37.4\rsun$), almost circular ($e=0.05$) orbit. The time to
coalescence via gravitational wave emission is $6.1$ Gyr. For this
particular evolution/model we assume no natal kicks at the formation
of the massive BHs, so the BH spin vectors are aligned with the
binary angular momentum ($\Theta_1=\Theta_2=0^\circ$). This produces
an upper limit on the effective spin parameter (see
eq.~\ref{eq.xeff}). For these particular BH masses, spins and spin
tilts, we obtain a rather low effective spin $\chi_{\rm eff}=0.09$.
The progenitor binary forms at $z=1.2$, so $\sim 5$ Gyr after the Big
Bang (close to a peak in star formation: $z\approx2$ means $3.2$ Gyr
after the Big Bang), and the BH-BH merger takes place at $z=0.2$
($\sim 11$ Gyr after the Big Bang). The gravitational waves from the
BH-BH merger propagate for $\sim 2.5$ Gyr to reach the LIGO detectors
at the present time. All of the system properties are within $90\%$ of
the credible limits of GW170104 \citep[][see also Sec.~\ref{sec.Xeff} 
for a revised limits on $\chi_{\rm eff}$]{Abbott2017a}.

\subsection{The Effective Spin Parameter: $\chi_{\rm eff}$}
\label{sec.Xeff}

The Bayesian analysis of GW170104 reported in \cite{Abbott2017a}
adopts prior assumptions about the relative likelihood of different
spin magnitudes (uniform) and directions (isotropic). These
assumptions are not suitable for comparison to the binary evolution
model we adopt, which requires both individual spins to be initially
aligned and then only mildly (if at all) misaligned by natal
kicks. Recent detailed analysis addressing the impact of prior
assumptions showed that they can indeed impact the inferred parameters
\citep{Vitale2017a,Williamson2017}.

We therefore reassess the reported limit, concluding that
$\chi_{\rm eff} < 0.2$ at approximately $90\%$ confidence in the
context of our model.  We can justify this reanalysis using only the
reported LIGO result on $\chi_{\rm eff}$, restricted to
$\chi_{\rm eff}>0$.  Approximating the LIGO distribution as nearly
normal with mean $\mu =-0.21$ and width $\sigma_{\chi} \simeq 0.155$,
we construct a truncated normal distribution
$\propto \theta(\chi) \exp - (\chi+\mu)^2/2\sigma^2$, which has a
$90\%$ upper limit at $x\simeq 0.2$.

We arrive at a similar result by reanalyzing the underlying LIGO data
using the same model and techniques (including the prior), then
restricting to configurations with positive individual spins
$\chi_{1,z},\chi_{2,z}>0$. Our revised upper limit is consistent with
the range of plausible $\chi_{\rm eff}$, as reported by
\cite{Abbott2017a} (see Fig.~5 of the supplementary material),
corresponding to approximately a $98$--$99\%$ confidence limit within
the strong assumptions of their original analysis.

\subsection{NS Spins}
\label{sec:neuspin}

\subsubsection{Observed Pulsar Spin Period Distribution}
\label{sec.nsspins1} 
 
Although astronomers have amassed a large sample of pulsar spin-period
measurements, extrapolating from observed periods to birth periods is an
open area of research~\citep[for a review, see][]{miller15}. Because pulsar
emission spins down the neutron star with time, any estimate of the birth
period requires an estimate of the pulsar age, and an understanding of the
rate of spin-down from pulsar emission (including an understanding of
magnetic field evolution in neutron stars).  The age of a pulsar ($t_{\rm
pulsar}$) can be estimated assuming the angular momentum is lost through
electromagnetic radiation from a pulsar with dipole magnetic
fields~\citep{manchester77}:
\begin{equation}
t_{\rm pulsar} = \frac{P}{(n-1)\dot{P}} (1-(P_0/P)^{n-1}),
\end{equation}
where $n$ is the pulsar braking index ($n=3$ for magnetic dipole radiation),  
$P_0$, $P$ are respectively the initial and current pulsar spin periods. If 
the age of the pulsar is known by other means (e.g., the supernova remnant age), 
measurements of the spin period along with constraints on the braking index 
provide an estimate of the birth spin period. Clearly the pulsar age 
depends on the choice of $n$. In addition, other sources of angular momentum 
loss exist: e.g. nonsphericities of the neutron star can cause gravitational 
wave emission, lowering the spin period.
 
Despite these uncertainties, astronomers have estimated the neutron star
birth spin period distribution. The fastest pulsars could be born spinning less
than 10\,ms (the Crab pulsar is believed to have been born spinning at
17\,ms). \cite{popov12} found that a Gaussian with an average spin period 
of 100\,ms (with a $1-\sigma$ deviation of 100\,ms) fits the observations. 
\cite{faucher06} found a slightly higher average, 300\,ms ($1 \sigma$ 
deviation of 150\,ms). \cite{igoshev13} argue that the differences between 
these two studies could be explained by the choice of magnetic field 
evolution and either distribution could be made consistent with  the data. 
\cite{noutsos13} also obtained a distribution of periods peaking below 125\,ms, 
but found an additional set of long-period birth spins ($>0.5$\,s). They 
found that poor age estimates limit determinations of the birth pulsar spin 
distribution and described methods to estimate pulsar ages kinematically.
 
\begin{figure}
\includegraphics[width=\columnwidth]{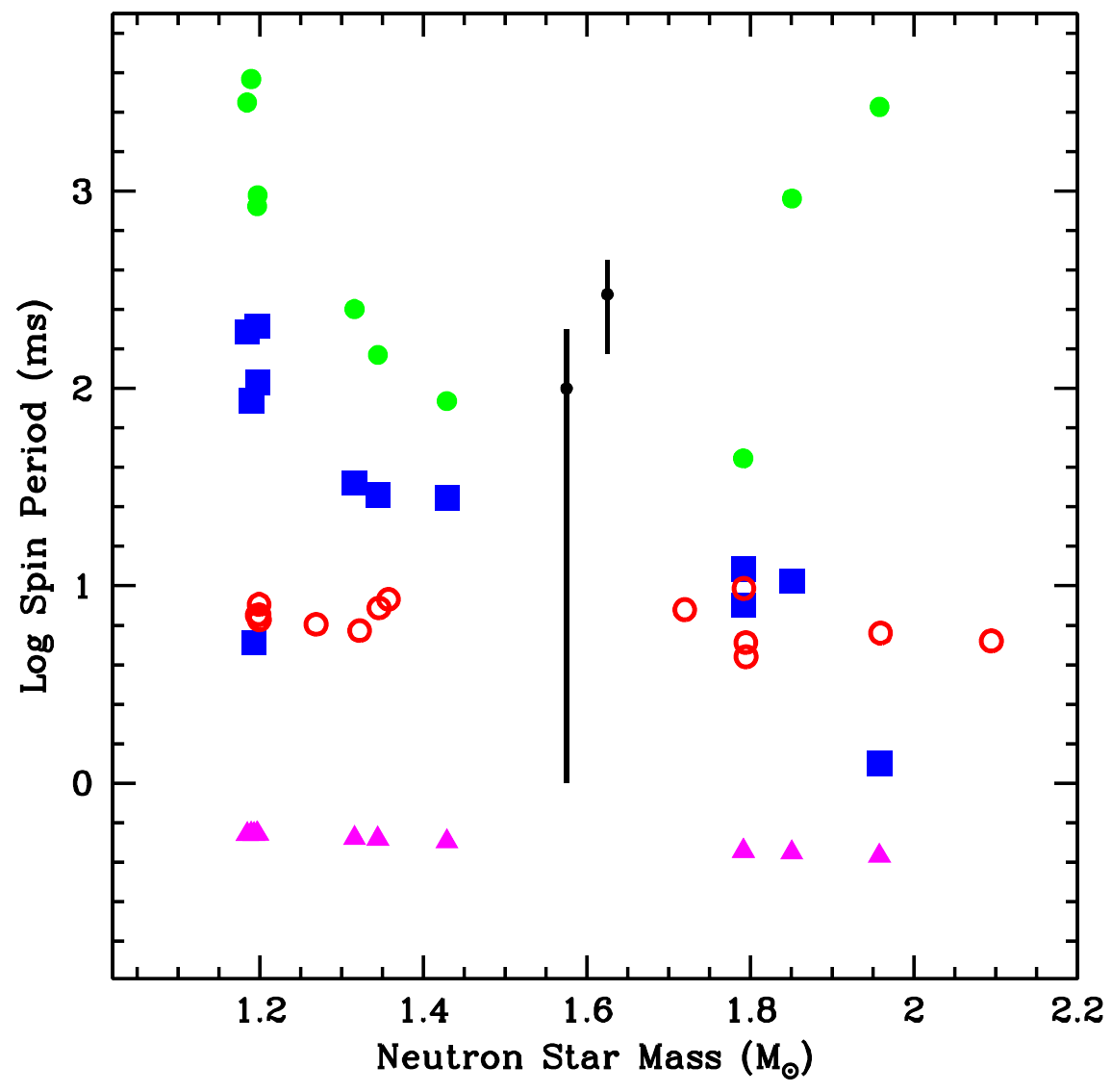}
\caption{
Neutron star spins from our progenitors as a function of neutron star mass. 
We assume any remnant from our suite of models with a mass below $2.5\msun$ 
is a neutron star and consider only these remnants. 
With the Geneva models, we have studied the angular momentum coupling of 
different burning layers and their effect on the neutron star spin. The 
Geneva models with the original mild coupling produce spin periods shorter than 
$1$\,ms (not shown here, but very close to magenta triangles; see below). Some 
angular momentum loss in the supernova engine (e.g. magnetic coupling such as 
a magnetar engine) would be required to slow neutron stars down for such 
models to match the data. The spins are nearly the same if the coupling 
extends through the CO core (magenta triangles). Coupling through the helium 
layer (blue squares and empty red circles; different helium core definitions) 
and hydrogen layer (green solid circles) produces slower neutron stars. The 
helium-coupled models match the data (but realize that we are using rapidly 
rotating progenitors with no angular momentum loss during NS formation). 
The MESA models (not shown here) with the Tayler-Spruit dynamo produce 
$\sim 10$\,ms pulsars; generating spins that match fastest-spinning pulsars 
with rapidly spinning progenitor stars, an indication that the MESA models 
produce reasonable coupling results. The Fuller models (not shown here) do not 
produce rotating neutron stars, requiring spin-up mechanisms in the supernova 
engine to match the data. For comparison we also mark (with black lines) the
range of two observational estimates. More details on models and observations 
are given in Sec.~\ref{sec.nsspins1}. 
}
\label{fig.spin}
\end{figure}

With these uncertainties in mind, we can now compare stellar models to the 
observed spin distribution.  As with the stars forming black holes, we can 
estimate the birth spins of of neutron stars from their massive star progenitors 
by assuming that the angular momentum of the collapsing core is accreted onto
the proto-neutron star along with its mass.  We limit the accreted angular
momentum ($j_{\rm acc}$) to the centrifugally supported value:
\begin{equation}
j_{\rm acc} = min (j_{\rm shell}, \sqrt{r_{\rm NS} G M_{\rm encl}}) 
\end{equation}
where $j_{\rm shell}$ is the angular momentum of the accreting shell,
$r_{\rm NS}$ is the neutron star radius, $G$ is the gravitational constant
and $M_{\rm encl}$ is the mass enclosed in the shell. For the neutron star 
radius, we assume a $10$\,km.  We consider only compact remnants with masses 
below $2.5\msun$ as this is our adopted maximum NS mass. By summing up the
angular momentum from the accreted material, we obtain the total angular momentum 
of the compact remnant. The moment of inertia of a neutron star ($I_{\rm NS}$) 
depends upon the equation of state~\citep{worley08} and can vary by roughly 
$50\%$ with the choice of the equation of state but it is roughly linear with 
neutron star mass.  For our calculations, we assume 
\begin{equation}
I_{\rm NS} = 1.5 \times 10^{45} (M_{\rm NS}/M_\odot) \ {\rm g \, cm^2}
\end{equation}
where $M_{\rm NS}$ is the neutron star mass.  With the total angular
momentum and moment of inertia, we can determine the birth spin period
of the neutron stars from our progenitors (Figure~\ref{fig.spin}).
The spins from these models assume stars born rotating at $40\%$ breakup
velocity, producing core spins near the maximum allowed by a given angular
momentum transport mechanism. We have not included any angular momentum loss 
mechanisms, but there are several possibilities of extracting rotational 
energy from the protoneutron star. For example, if the rotational energy is 
tapped to help driving the explosion (e.g., by interaction with a disk or when 
forming a magnetar), the total angular momentum of the system will be reduced. 
Therefore the rotation rates produced in our models are only upper limits on 
the real spin rates. 

The results of Figure~\ref{fig.spin} demonstrate that some magnetic braking  
is needed to reduce the angular momentum of the core, confirming decades-old 
arguments for magnetic breaking~\citep{heger00, fryer00}. The Geneva models 
with the original distribution of angular momentum through a star produce only 
sub-ms pulsars. While the spins from these models without magnetic coupling 
have too much angular momentum, they are ideally suited to determine the 
amount of coupling needed to produce the correct spin periods. We assume the 
angular momentum is distributed with a constant angular velocity (solid body 
rotation) across the mass of different cores. If we assume the angular momentum 
to be conserved, we can  calculate the angular velocity of the core and 
recalculate the spin of the pulsar produced. For example, if we sum up the 
angular momentum of the CO core and divide it by its moment of inertia, we 
get an average angular velocity. If we use this constant angular velocity to 
redistribute the angular momentum in the CO core, we revise our estimate of 
the spin of the compact remnant. For this redistribution, the spin periods 
remain below a ms (magenta triangles). To truly slow down the birth spin periods 
of neutron star, we must couple the angular momentum through the helium core 
(meaning that one assumes a constant angular velocity from star center to
the outer boundary of the helium core).  
Figure~\ref{fig.spin} shows the resultant spins if we assume coupling through 
the helium core with a helium core definition of $X_{\rm He}>0.4,\ X_{\rm H}<0.01$ 
(blue squares), and $X_{\rm He}>0.5$ (red empty circles). This produces maximum birth 
spin periods of between $\sim 10-1000$\,ms. Further coupling, the H-rich 
layers, would produce spin periods (solid green circles) that are too slow to 
match the observed pulsar distribution. Given that these spins are maximum spin 
values (recall that we are using fast rotating progenitors and assuming no 
angular momentum loss in the explosion mechanism), our pulsar spin observations 
argue against angular-momentum coupling beyond the helium core.
 
If we compare the spin periods produced using our MESA models using a prescription 
similar to \cite{heger00}, we find spin periods of roughly $\sim 6-10$\,ms, 
matching the fastest-spinning non-recycled pulsars and the results of 
\cite{heger00,fryer00}. These periods are produced using rapidly spinning 
progenitors, so we would expect their periods to match the fastest-spinning 
systems and this result is a relatively good match to the observed pulsar 
distribution. For the Fuller models, the birth spin period would be too slow 
to match the data, arguing for some mechanism to spin-up the neutron star.  
Simulations of the asymmetries in the engine do show spin-up in the core. 
The amount of spin-up, and whether it is sufficient to solely explain the pulsar 
spin periods, remains a matter of debate~\citep{Blondin2007,Fryer2007,Foglizzo2009,
Rantsiou2011,Kazeroni2016}. 

\subsubsection{Rotational Explosion Properties}
\label{sec.nsspins2}
  
It has been argued that fast spinning magnetars can drive a subset of
supernova explosions.  Here we determine the role spins can play in
the supernova explosion itself. With strong (magnetar-strength: $\sim
10^{15} {\rm \,Gauss}$) magnetic fields, a pulsar can quickly release
the rotational energy of a newly formed neutron star.  To calculate
the energy available for such a model, we need to estimate the
rotational energy of the neutron star:
\begin{equation}
E_{\rm rot} = 1/2 I_{\rm NS} \omega^2 = 5\times10^{50} (\omega/1000Hz)^2 \,
{\rm erg}
\end{equation}
where $\omega$ is the angular velocity.  If the neutron star is
spinning with a ms period (e.g., original Geneva models), it can
produce a $10^{51}$\,erg explosion if it can tap $10\%$ of the
rotational energy to drive a jet. With the $6-10$\,ms periods 
(e.g., MESA models with rapidly rotating stars) we find that, even 
if all of the rotational energy is tapped to drive an explosion, 
rotation is unable to produce a normal-energy supernova.  If the 
strong magnetic fields are formed quickly, a spin-powered magnetar 
engine will deposit its energy in the slowly-moving ejecta,  
accelerating this innermost material (and adding additional heating)  
but not contributing significantly to the total energy budget of the 
explosion. 

Similarly, because the angular momentum is lowest in the cores of 
these stars, such models are unable to form a disk around the neutron  
star. Hence, engines that invoke jets produced by magnetic fields 
generated in a disk will not work. Note, however, that a disk can
form after the formation of a $3\msun$ remnant (presumably a black
hole) as then more angular momentum is trapped in the remnant. It is
likely that any model with Tayler-Spruit efficient angular
momentum transport (e.g., MESA or Fuller models) invoking these
engines will have to rely upon some means to spin up the star prior to
collapse ~\citep[see review by][]{fryer99b}. Note that this statement
does not depend on the initial stellar spin of models as independent
of adopted initial spin, both MESA and Fuller models end up with very
small angular momentum in the core (see
Fig.~\ref{fig.jrot_endHe}). Such rare spin-up events would be able to
explain rare outbursts such as gamma-ray bursts but, if these spins
are correct, it is unlikely that magnetars or disks play a big role in
explaining supernovae~\citep[for a review, see][]{Fryer2019}.

\subsection{Binary Evolution Calculations}
\label{sec.Binary}

We employ the {\tt StarTrack} population synthesis code~\citep{Belczynski2002,
Belczynski2008a}. The existing improvements relevant for massive star evolution
include updates to the treatment of CE evolution~\citep{Dominik2012}, the compact
object masses produced by core collapse/supernovae~\citep{Fryer2012,Belczynski2012a}
including the effect of pair-instability pulsation supernovae and pair-instability
supernovae~\citep{Belczynski2016c}, stellar binary initial conditions set by
observations~\citep{deMink2015}, and observationally constrained star formation
rate and metallicity evolution over cosmic time~\citep{Madau2014,Belczynski2016b}.
The code adopts by default the fallback-decreased natal kick prescription
(see below). Additionally, we explore three different models for the BH natal spins 
proposed in the current upgrade (Sec.~\ref{sec.spins}) as well as three different 
models for PPSN/PSN (Sec.~\ref{sec.bh_mass}).

In our population synthesis calculations we evolve stars on a finite grid of metallicity:
$Z=0.0001$, $0.0002$, $0.0003$, $0.0004$, $0.0005$, $0.0006$, $0.0007$, $0.0008$, $0.009$,
$0.001$, $0.0015$, $0.002$, $0.0025$, $0.003$, $0.0035$, $0.004$, $0.0045$, $0.05$,
$0.006$, $0.0065$, $0.007$, $0.0075$, $0.008$, $0.0085$, $0.009$, $0.0095$,
$0.01$, $0.015$, $0.02$, $0.025$, $0.03$.
If in our population synthesis model the metallicity is
$Z<0.00089$ we adopt the BH spin model corresponding to $Z=0.0004$;
if $0.00089 \leq Z < 0.00346$ we adopt the BH spin model corresponding to $Z=0.002$;
if $0.00346 \leq Z < 0.00916$ we adopt the BH spin model corresponding to $Z=0.006$;
and if $Z \geq 0.00916$ we adopt the BH spin model corresponding to $Z=0.014$.
The limits are half points in decimal logarithm between the four metallicities of
the BH natal spin model ($Z=0.014,\ 0.006,\ 0.002,\ 0.0004$).

For the initial orbital period distribution of massive binaries we use
$f_p (\log p/{\rm day}) \propto ( \log p/{\rm day} )^{-0.5}$ in range $[0.15, 5.5]$ 
\citep{Sana2012}. For the initial eccentricity distribution we use 
$f_e (e) \propto e^{-0.42}$ in the range $[0.0, 0.9]$. The initial mass of the 
primary star is taken from a broken power law IMF: 
$\propto M^{-1.3}$ for $0.08 \leq M<0.5\msun$, $\propto M^{-2.2}$ for 
$0.5 \leq M<1.0\msun$, and $\propto M^{-2.3}$ for $1.0 \leq M \leq 150\msun$ 
\citep{Kroupa2001,Bastian2010,Duchene2013}. The initial secondary mass is taken from 
a uniform mass ratio distribution $f_q (q)  \propto q^0$ in the range $q\in [0.1, 1]$.

We adopt maximum binary fraction: $f_{\rm bi}=1.0$ for stars of any mass and any  
metallicity. This is a sound assumption for massive stars of non-negligible metallicity
~\citep{Raghavan2010,Chini2012,Moe2013,Moe2016}. However, the binarity may be smaller 
for low mass stars, e.g., $f_{\rm bi}=0.5$ ($2/3$ of stars in binaries). 
For each metallicity we evolve $N=2 \times 10^6$ massive binaries (primary mass 
$>5\msun$, secondary mass $>3\msun$) and this corresponds to the total simulation
stellar mass of $M_{\rm sim}=1.9\times 10^8\msun$ (all stars over entire IMF) for 
$f_{\rm bi}=1.0$. Note that had we assumed $f_{\rm bi}=0.5$ for all stars then 
$M_{\rm sim}=2.8\times 10^8\msun$. Therefore, such a change would decrease of all 
our rate predictions for double compact objects by $\sim 30\%$ (see eq.7 of 
\cite{Belczynski2016a}). 

The initial distributions described above assume that all the binary parameters are 
independent from each other. However, various correlations between those parameters 
have long been hinted by the observations \citep[eg.][]{Abt1990,Duchene2013}. It was 
only recently that \citet{Moe2016}, using results from more than 20 surveys of massive 
binary stars, were able to fit analytic functions to the correlated distributions and 
obtain a join probability density function $f(M_1, q, P, e) \ne f(M_1)f(q)f(P)f(e)$. 
In some pockets of the entire parameter space they found differences larger than an order 
of magnitude with respect to the typically used method of combining several independent 
distributions. Similarly, even though a conclusive evidence for significant IMF variations 
with environmental conditions is still lacking, there is an increasing amount of results 
suggesting such departures from the IMF universality \citep{Bastian2010, Kroupa2013}. 
Notably, both theoretical arguments \citep[see ][for an overview]{Klencki2018} and 
observations of some GCs in the Milky Way \citep{Marks2012a,Marks2012b} seem to point 
towards a top-heavy IMF\footnote{With respect to the IMF assumed in this study.} in 
low-metallicity galaxies, which are likely an important formation site of massive BH-BH 
mergers. Recently, \citet{Klencki2018} analyzed the significance of the inter-correlations 
in initial distributions quantified by \citet{Moe2016} as well as of the possible variations 
in the IMF slope for the massive stars on the formation rate and properties of compact 
binary mergers. They found that the effect of those factors is very small compared to other 
uncertainties (eg. rates affected by less than a factor of 2). Their result holds even for 
very significant changes in the IMF slope due to the coupling of the IMF and the cosmic SFRD. 
This justifies the simplified assumptions of the universal IMF and non-correlated initial 
binary parameter distributions used in this study.

For old models (M10 and M13) we have redone the calculations with the same input physics, 
but with the addition of the new distribution of natal BH spins.  Additionally, we have 
updated the calibration for all models (decreasing rates by a factor of $0.926$) to account 
for small inconsistencies in our previous estimates \citep{Klencki2018}. Below we comment 
on factors introduced in our new models (M20, M23, M25, M26).

The fraction of mass retained in the binary ($f_{\rm a}$) during stable RLOF is not well 
established, and could be fully conservative ($f_{\rm a}=1$), fully non-conservative 
($f_{\rm a}=0$), or anywhere in between (e.g., \cite{Meurs1989}). Donor stars are typically 
the more massive components, as they are the ones that evolve (and typically expand) more 
quickly. Consequently, more massive donors often have a much shorter thermal timescale than 
their companions. For that reason, in the case of a thermal-timescale mass transfer when 
the mass transfer rate is related to the thermal-timescale of the donor star, the less 
massive companions will not have enough time to thermally readjust in order to  accommodate 
all of the transferred mass. This likely leads to a large fraction of the transferred mass 
being ejected from the system. Even in the case of a slower, nuclear-timescale mass transfer, 
accretion is typically expected to be small because the accretor quickly becomes spun up to 
its critical surface rotation velocity. We note that the picture is possibly more complicated 
due to the uncertain efficiency of stellar winds in carrying away the angular momentum from 
the accretor surface \citep[eg.][]{Vanbeveren2018}.

Nonetheless, we decided to revise and test our assumption on the accretion efficiency. 
In the previous work (as well as the remaining models in this study) we adopted 
$f_{\rm a}=0.5$. Recent estimates of mass transferring BH-BH progenitors resulted 
(typically) in $f_{\rm a}<0.5$~\citep{Stevenson2017}, so in models M20-M26 we adopt
$f_{\rm a}=0.2$. As a consequence, the secondary stars (accretors) remain less massive 
than in our previous models, and this generates a wider BH-BH binary mass ratio 
distribution than reported in our earlier studies. The efficiency of accretion in the 
first episode of mass transfer was noted to have possible impact on the BH-NS formation 
\citep{Kruckow2018}. Note that ~\cite{Stevenson2017} used not a single value for 
$f_{\rm a}$, but estimated the accretion efficiency from the relative thermal timescales 
of the donor and accretor for a given binary. 

Even a small amount of accretion during RLOF ($\sim$ few percent of
the accretor's mass) may effectively spin up accreting
stars~\citep{Packet1981}. With our adopted RLOF retention fraction of
$f_{\rm a}=0.2$, accretors in BH-BH progenitor binaries typically gain
about $10\msun$, which is enough to spin up even very massive
stars. This accretion usually happens around the middle (or shortly
thereafter) of the accretor's main sequence life, and therefore it
allows for effective rotational mixing and the formation of more
massive He and CO cores. {\tt Geneva} stellar evolution models
indicate that the CO core masses in rotating stars ($40\%$ critical
velocity) are $20\%$ more massive than in non-rotating models. So far
all CO core masses calculated in our binary evolution models were
obtained from non-rotating models~\citep{Hurley2000}.  Here we
increase CO core mass of accreting low-metallicity ($Z<0.002$) MS
stars by $20\%$. For high-metallicity stars, the effects of rotation
on the CO core mass are suppressed due to angular momentum loss
through stellar winds~\citep{Georgy2012}. This change may increase
the mass of the second BH, and also lower its spin magnitude.
In all models we assume that material is lost from a binary in RLOF
with specific angular momentum $dJ/dt= j_{\rm loss}
[J_{\rm orb}/(M_{\rm don}+M_{\rm acc})] (1-f_{\rm a}) dM_{\rm RLOF}/dt$,
with $j_{\rm loss}=1.0$~\citep{Podsiadlowski1992}.

Our criteria for the mass transfer stability and the occurrence of CE are described 
in detail in Sec.5 of \citet{Belczynski2008a}. To treat CE evolution we assume energy 
balance with fully effective conversion of orbital energy into envelope ejection 
($\alpha=1.0$), while the envelope binding energy for massive stars is calibrated 
using a parameter $\lambda$ that depends on stellar radius, mass, and metallicity 
for all models. For massive stars we assume $\lambda \approx 0.1$~\cite{Xu2010}.
Additionally, we either allow or do not allow (submodels A and B) for HG stars to 
initiate and survive CE evolution~\cite{Belczynski2007,Pavlovskii2017}. In submodels 
A only stars with well developed core-envelope boundary (beyond HG) can successfully 
initiate and survive CE, depending on the energy balance. In submodels B we
also allow for HG stars to initiate and survive CE.
Recent calculations show that the accretion rates onto compact objects in CE 
inspiral can be reduced even by a factor of $10^{-2}$ with respect to the rates 
resulting from the Bondi-Hoyle approximation when the structure of the envelope 
(in particular, the density gradients around the inspiraling object) are taken 
into account~\citep{Ricker2008,MacLeod2017,Murguia2017,Holgado2017}.
\citet{MacLeod2015a} argue that accretion structures forming around
compact objects embedded in the CE may span a large fraction of the
envelope radius, and so traverse substantial density gradients.
Introducing gradients in the CE structure leads to net non-zero
angular momentum of the flow around an accreting object (which is not
the case in the standard Hoyle formalism), and by doing so limits
accretion: steeper density gradients correspond to smaller accretion.
The typical values of the density gradients found by these authors
introduce a considerable perturbation to the flow. For most density
gradients considered by \cite{MacLeod2017}, the accretion rate is well
below $10\%$ of Bondi-Hoyle accretion rate. Based on these findings,
we adopt $f_{\rm bond}=5\%$ of the Bondi-Hoyle accretion rate onto a
BH in CE in our current simulations. Therefore massive BHs
($M_{\rm BH} \sim 30\msun$) accrete $\sim 0.5\msun$ in a typical CE
event, as opposed to $\sim 1.0\msun$ in our earlier calculations
($f_{\rm bond}=10\%$). To assess the Bondi-Hoyle accretion rate we 
follow the approach presented in~\cite{Belczynski2002}. 

Spectroscopic analysis of OB stars in the Local Group 
\citep[MW, SMC, LMC, eg.][]{Oskinova2013,Hainich2018,Ramachandran2019} as well 
as other low-metallicity dwarf galaxies \citep[eg.][]{Bouret2015} has shown a 
systematic offset in the wind mass loss rates between theoretical predictions 
\citep[][assumed in most our models]{Vink2001} and the empirical 
log$\dot{M} - $log$L$ relation. Namely, the theoretical models seem to 
overestimate the actual wind mass loss rates of hot stars by at least a factor 
of a few. The most likely cause for this discrepancy is clumping of the wind, 
which is not accounted for in the standard models of radiatively driven smooth 
winds. For that reason, in model M50 we reduce all the wind mass loss rates down 
to $30\%$ of the usually assumed values. 

Compact remnants formed in supernovae can receive proper motions via two
classes of engine-driven natal kicks: asymmetric matter ejecta or asymmetric
neutrino emission. For BHs, asymmetric matter ejection mechanisms only work
when matter is ejected, as opposed to prompt collapse or complete recapture
of all ejected material (``fallback''). In our calculations, we expect that
only a small fraction of systems eject a substantial amount of  matter,
enabling a substantial BH natal recoil kick. In contrast, asymmetric
neutrino emission mechanisms operate even without any mass ejection, and
thus they can affect any model of BH formation.
Although neutrino mechanisms have been invoked to explain recoil velocities
of pulsars and X-ray binaries \citep{Lai1998,Repetto2015}, the proposed kick
models all require strong magnetic fields. Models without strong magnetic
fields are unable to produce significant neutrino kicks \citep{Tamborra2014}.

The sterile neutrino oscillation model~\citep{Kusenko1996,Fryer2006b} argued
that neutrinos produced in the core could oscillate to sterile neutrinos and
escape the core. Large magnetic fields align the ions and electrons, forcing
both the neutrino scattering and absorption cross sections to be anisotropic.
To ensure asymmetric neutrino emission, these strong magnetic fields must be
at the last scattering surface for the neutrinos. If the magnetic field in
the core is high enough to align the ions and electrons, the neutrinos in the
core will be anisotropic. If these neutrinos oscillate into sterile neutrinos,
they can escape, retaining their anisotropies and generating large natal kicks.

Alternatively, the neutrino bubble instability \citep{Socrates2005}
argues that magnetic-acoustic instabilities develop, transporting
neutrino radiation to the photosphere. These instabilities carry
neutrinos, and the luminosity escaping the neutrinosphere will be
enhanced at these "bubbles". If the magnetic-acoustic bubbles are
globally asymmetric, the neutrino emission will also be asymmetric,
producing a neutrino-driven kick. Current supernova calculations have
several limitations: (i) they do not model high magnetic fields, (ii)
they do not sufficiently resolve the hydrodynamics, and (iii) they do
not include the neutrino oscillation physics necessary to produce
these kicks. So high, neutrino-driven BH natal kicks cannot be
ruled out.

In models M10 and M20 we test asymmetric mass ejection kicks, as we
employ fallback-decreased natal kicks~\cite{Fryer2012}. To mimic
asymmetric neutrino emission mechanisms, we explore an alternative
phenomenological prescription for BH natal kicks in models M13, M23,
M25, M26, M33, M35, and M43 where we impart kicks which are random in 
direction, with magnitude drawn from a Maxwellian with a given 1-dimensional
$\sigma$, independent of the BH mass or its progenitor history (see
Tab.~\ref{tab.models}).

\subsection{Detectabilty of Mergers in Gravitational Waves}
\label{sec.GW}

All the compact object mergers are redistributed according to star
formation history across cosmic time ($z\approx0$--$15$ for Population
I and II stars; e.g., \cite{Madau2014}), taking into account the time
delay between binary formation at Zero Age Main Sequence and the merger.

For each merger at redshift $z$ we use phenomenological
inspiral--merger--ringdown waveforms (IMRPhenomD; \cite{Khan2016}) to
calculate the signal-to-noise ratio in the O1/O2 LIGO runs. A given
merger is considered detectable (depending on its random sky location
and orbital orientation with respect to the detectors) if the
signal-to-noise ratio in a single detector is greater than $8$. Only
detectable mergers are used in our comparisons with O1/O2 data (e.g., rates in
Tables~\ref{tab.ratesB},~\ref{tab.ratesA}, or effective spins in Sec.~\ref{sec.xeff}).

With this method we obtain a self-consistent redshift distribution of
mergers in the local Universe, and we also account for LIGO/Virgo
detectability of our synthetic mergers (in particular, more massive mergers
can be detected at larger redshifts). A more detailed description of
detectability criteria can be found in \cite{Belczynski2016a,Belczynski2016b,
Belczynski2016c}.

\subsection{BH-NS merger rate limits from the ongoing O3 run}
\label{sec.app_BHNS_limits}

Here, we analyze the constraint on the rate density of BH-NS mergers that the 
recent LIGO/Virgo candidate \citep[S190814bv]{GCN_324,GCN_333}\footnote{see 
https://gracedb.ligo.org/superevents/S190814bv/} entails.

We do not know the mass of the object but we assume that the mass of the NS 
$M_{\rm NS}$ must be in the range from $1.3$ to $3\msun$, while the BH mass 
$M_{\rm BH}$ is in the range from $5$ to $50\msun$. We denote the chirp mass 
of the system as $\cal M$. The O3 has lasted for about $4.5$ months with the 
$80\%$ uptime which given the effective observation time $t_{\rm obs}$ of
$150$ days, or $0.41$ years for O1/O2 and $t_{\rm obs}$ of $3.6$ months, or 
$0.30$ years. For the range of the search we assume that the sensitivity to 
NS-NS mergers was $r_{\rm NSNS}=80$\,Mpc (O1/O2) and $r_{\rm NSNS}=135$\,Mpc 
(O3). We can estimate the range to the BH-NS merges from 
\begin{equation}
r_{\rm BHNS} = r_{\rm NSNS} \times ({\cal M}/1.2\msun)^{5/6}, 
\label{eq.bhnsrange}
\end{equation}
where we assumed that the fiducial chirp mass of a NS-NS system is $1.2\msun$. 
The observed volume is then $V_{\rm BHNS}= (4/3) \pi r_{\rm BHNS}^3$. The rate 
density of BH-NS mergers, estimated from this one event, can be estimated as
\begin{equation}
{\cal R} = {1 \over (t_{\rm obs} V_{\rm BHNS})_{\rm O1/O2} + (t_{\rm obs} V_{\rm BHNS})_{\rm O3}}=1.6 - 60 \gpy, 
\label{eq.bhnsrate}
\end{equation}
and the range corresponds to the limits in which we have allowed NS and BH 
mass to vary. This estimate will go down by approximately a factor of two if 
no similar objects will be detected in the remainder of O3 ($0.7$\,yr). On the 
other hand, the range will become narrower once we know the mass estimates of
NS and BH. Note however, that including the effect of Poisson distribution 
width will broaden the result again. Thus we are confident that the above 
estimate is approximately accurate under the assumption that the detected
system was in fact BH-NS merger and not BH-BH merger with a very light 
secondary BH. We confront the merger rate limit estimated above with the 
results from our models in Figure~\ref{fig.rates2}.

\end{document}